\documentclass[twocolumn]{aastex631}
\usepackage{graphicx,times,epsf}       
\usepackage{soul,xcolor}
\usepackage{longtable}
\usepackage{natbib}
\usepackage{amssymb}
\usepackage{amsmath}
\shortauthors{Shan et al.}
\begin{document}

\title{TESS Timings of 31 Hot Jupiters with Ephemeris Uncertainties}
\email{Fan Yang: sailoryf@nao.cas.cn; sailoryf1222@gmail.com}
\email{Ji-Feng Liu: jfliu@nao.cas.cn}

%\affil{Department of Astronomy, Beijing Normal University, Beijing 100875, People's Republic of China\\}
\author[0000-0002-5744-2016]{su-su shan}
\affil{National Astronomical Observatories, Chinese Academy of Sciences, 20A
Datun Road, Chaoyang District, Beijing 100101, China\\}
\affil{School of Astronomy and Space Science, University of Chinese Academy of Sciences,
Beijing 100049, China\\}

\author[0000-0002-6039-8212]{fan yang}
\affil{National Astronomical Observatories, Chinese Academy of Sciences, 20A
Datun Road, Chaoyang District, Beijing 100101, China\\}
\affil{Department of Astronomy, Beijing Normal University, Beijing 100875,
People's Republic of China\\}
\affil{School of Astronomy and Space Science, University of Chinese Academy of Sciences,
Beijing 100049, China\\}

\author[0000-0002-1310-4664]{You-Jun Lu}
\affil{National Astronomical Observatories, Chinese Academy of Sciences, 20A Datun Road, Chaoyang District, Beijing 100101, China\\}
\affil{School of Astronomy and Space Science, University of Chinese Academy of Sciences,
Beijing 100049, China\\}

\author{Xing Wei}
\affil{Department of Astronomy, Beijing Normal University, Beijing 100875,
People's Republic of China\\}

\author{Wen-Wu Tian}
\affil{National Astronomical Observatories, Chinese Academy of Sciences, 20A Datun Road, Chaoyang District, Beijing 100101, China\\}
\affil{School of Astronomy and Space Science, University of Chinese Academy of Sciences,
Beijing 100049, China\\}

\author{Hai-Yan Zhang }
\affil{National Astronomical Observatories, Chinese Academy of Sciences, 20A Datun Road, Chaoyang District, Beijing 100101, China\\}
\author{Rui Guo}
\affil{Department of Astronomy, School of Physics and Astronomy, Shanghai Jiao Tong University, 800 Dongchuan Road, Shanghai 200240, China}

\author{Xiao-Hong Cui}
\affil{National Astronomical Observatories, Chinese Academy of Sciences, 20A Datun Road, Chaoyang District, Beijing 100101, China\\}
\author{Ai-Yuan Yang}
\affil{Max Planck Institute for Radio Astronomy, Auf dem Hügel 69, 53121 Bonn, Germany\\}
\author[0000-0002-6434-7201]{Bo zhang}
\affil{Department of Astronomy, Beijing Normal University, Beijing 100875,
People's Republic of China\\}

\author[0000-0002-2874-2706]{ji-feng liu}
\affil{National Astronomical Observatories, Chinese Academy of Sciences, 20A Datun Road, Chaoyang District, Beijing 100101, China\\}
\affil{School of Astronomy and Space Science, University of Chinese Academy of Sciences,
Beijing 100049, China\\}
\affil{WHU-NAOC Joint Center for Astronomy, Wuhan University, Wuhan, China\\}

\begin{abstract}
A precise transit ephemeris serves as the premise for follow-up exoplanet observations. We compare TESS Objects of Interest (TOI) transit timings of 262 hot Jupiters with the archival ephemeris and find 31 of them having TOI timing offsets, among which  WASP-161b shows the most significant offset of -203.7$\pm$4.1 minutes. The median value of these offsets is 17.8 minutes, equivalent to 3.6 $\sigma$. We generate TESS timings in each sector for these 31 hot Jupiters, using a self-generated pipeline. The pipeline performs photometric measurements to TESS images and produces transit timings by fitting the light curves.
We refine and update the previous ephemeris, based on these TESS timings (uncertainty $\sim$ 1 minute) and a long timing baseline ($\sim 10$ years). Our refined ephemeris gives the transit timing at a median precision of 0.82 minutes until 2025 and 1.21 minutes until 2030. We regard the timing offsets mainly originating from the underestimated ephemeris uncertainty. All the targets with timing offset larger than 10$\sigma$ present earlier timings than the prediction, which cannot be due to underestimated ephemeris uncertainty, apsidal precision, or R$\o$mer effect as those effects should be unsigned. For some particular targets, timing offsets are likely due to tidal dissipation. Our sample leads to the detection of period decaying candidates of WASP-161b and XO-3b reported previously.
\end{abstract}

\section{Introduction}

Transit ephemeris is crucial for exoplanet follow-up investigations, e.g., atmosphere analysis \citep{Berta2012, Deming2013, YangLD} and orbital evolution \citep{Lendl2014, Dawson2018, Millholland2018,2020wasp12b}. 
The newly commissioned Transiting Exoplanet Survey Satellite \citep[TESS]{Ricker2015} provides precise timings in a long baseline when combined with previous works, which enables us to obtain a better transit ephemeris.

The observed transit timing could deviate from the ephemeris prediction due to either underestimation of ephemeris uncertainties \citep{Mallonn2019}, or physical processes \citep{Agol2018}. The TTV could originate from tidal dissipation, orbital precession, R$\o$mer effect, mass loss and multi-planets \citep{2017wasp12b,2020wasp12b,2021wasp12bTESS,wasp-4b2020, Ragozzine2009, Lai2010,Mazeh2013, Agol2018}. For hot Jupiters, the interactions of planet companions are usually not massive and close enough to generate significant TTVs \citep{Huang2016, Dawson2018}.

TTV provides direct evidence of tidal dissipation that likely drives hot Jupiter migration \citep{Dawson2018}. WASP-12b has been reported to undergo tidal dissipation by observational TTVs \citep{2017wasp12b,2020wasp12b,2021wasp12bTESS}. The TTVs are at the level of $\sim$ 5 minutes in a 10-year baseline compared to the ephemeris obtained from a constant period \citep{2020wasp12b}. Apsidal precession is reported to be the major arguing explanation and seems to be ruled out with more than 10-year observations, including most recent TESS timings \citep{2017wasp12b,2020wasp12b,2021wasp12bTESS}. The referred works also discuss and exclude the other possible effects, including R$\o$mer effect and mass loss \citep{Ragozzine2009, Lai2010}.

The R$\o$mer effect, i.e. the acceleration towards the line-of-sight probably due to stellar companions, has been reported to dominate the TTV of WASP-4b \citep{wasp-4b2020}. Using TESS light curves, \cite{WASP-4b} present a period decreasing at -12.6 $\pm$ 1.2 ms yr$^{-1}$. Further radial velocity (RV) monitoring indicates the Doppler effect contributes most of the period decreases \citep{wasp-4b2020}. For another example, WASP-53b and WASP-81b should harbor brown dwarf companions which could cause TTVs $\sim$ 30s, according to the calculation of \cite{Triaud2017}.

We compare TESS timings and archival ephemeris predictions\footnote{Exoplanet Archive: \url{https://exoplanetarchive.ipac.caltech.edu/index.html}}, and
report transit timing offsets of 31 hot Jupiters in this work. The paper is organized as follows. We present the sample selection and data reduction in Section 2. In Section 3, transit timings and offsets compared to the previous ephemeris are shown. The ephemeris refinement is also shown in this section. In Section 4, we discuss the possible physical origin of the timing offsets. We briefly summarize the work in Section 5.

\section{Sample Selection and TESS Timing}

The exoplanet sample in this work are hot Jupiters identified from previous work and have access to transit timings from the TESS Objects of Interest (TOI) Catalog \citep{TOIcatalog}. The archival data is extracted from the NASA Exoplanet Archive \citep{ExoplanetArchive} 10.26133/NEA12 table as of August 2021.
The sample selection requires an orbital period of fewer than 10 days, a planet mass larger than 0.5 $M_{J}$, and a planet radius larger than 0.5 $R_{J}$. 
These criteria leave 421 hot Jupiters. After crossmatched this sample and the TOI catalog, we find out TESS observed and reported new transit timing for 262 hot Jupiters.

\subsection{TESS Photometry and TOI Catalog}

\begin{figure}
  \centering
  \includegraphics[width=3.2in]{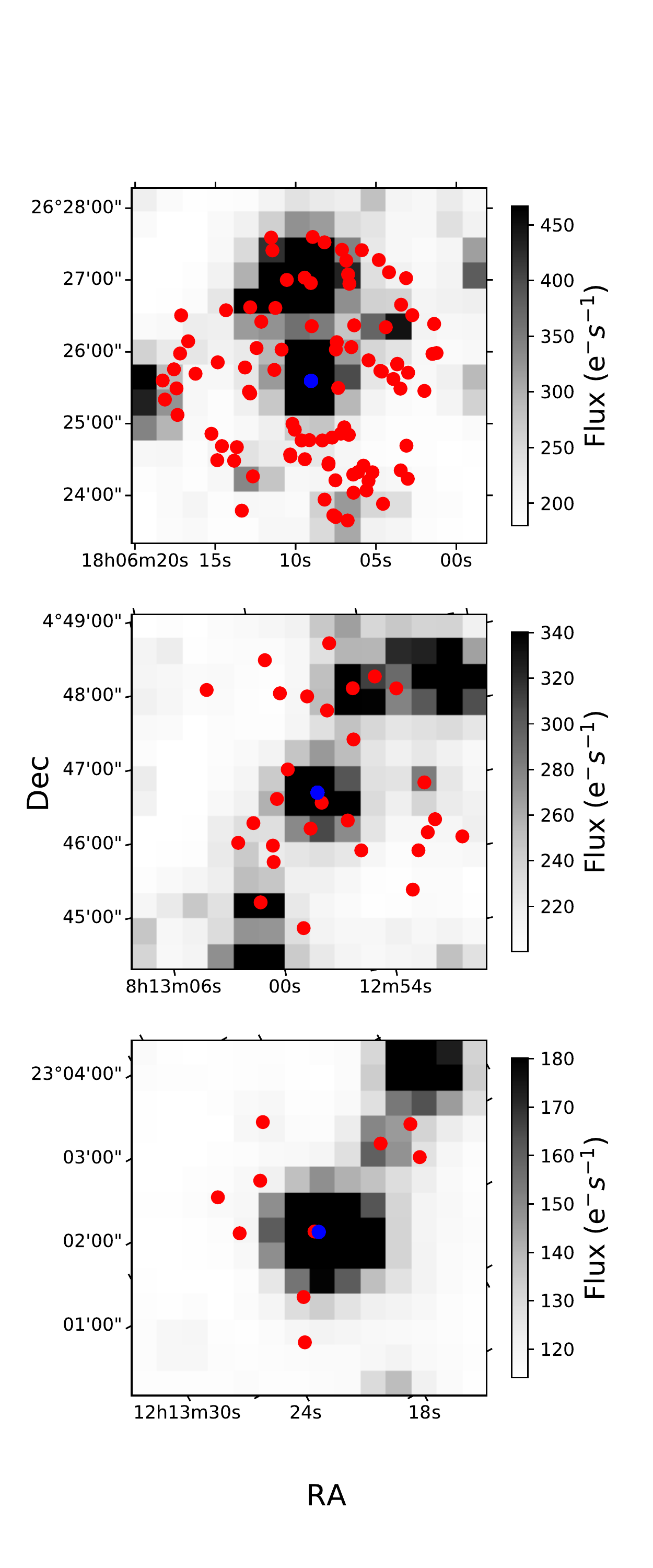}
  \caption{TESS example images of 14$\times$14 pixels. The images correspond to HAT-P-31b, HAT-P-35b, and WASP-56b, from top to bottom. The blue points refer to the planet position in the Gaia catalog \citep{GaiaDr2} while red points present nearby source positions.}
  \label{image:example}
\end{figure}

TESS is launched in 2018, possessing four cameras with a total field of view (FOV) of 24×96 square degrees, equivalent to a pixel resolution of 21 arcseconds \citep{Ricker2015}. The full frame image (FFI) covering FOV is released in a cadence of 30 minutes (as shown in Figure \ref{image:example}) while $\sim$ 200,000 targets are recorded with 11 $\times$ 11 pixel cut-off images in a cadence of 2 minutes (known as target pixel file; TPF). The TESS data is available in MAST: \dataset[10.17909/t9-yk4w-zc73, 10.17909/t9-nmc8-f686]..

The TOI catalog is built based on the light curves obtained from TESS image products, including both 2 minute and 30 minute frames \citep{TOIcatalog}. The 2-minute cadence light curve is generated by the Science Processing Operations Center (SPOC) pipeline
and the 30 minute light curves by the Quick Look Pipeline (QLP) \citep{Twicken2016, Huang2020}. \citet{TOIcatalog} generate an automated pipeline to derive transit parameters and thereby identify planet candidates with the method referred to the Kepler Robovetter \citep{Thompson2018}. More than 2000 planet candidates (continuously updating) are identified in the TOI catalog including both newly discovered and previously known planets.

The timing from the TOI catalog provides a long time baseline when compared with the previous ephemeris. The median timing baseline of the 262 exoplanets is 2368 days, while the median uncertainties of timings from archival data and from the TOI catalog are 0.59 and 0.84 minutes. The median uncertainty of archival periods is 4$\times$10$^{-6}$ days. 159 of 262 hot Jupiters show consistent TESS timings within 1 $\sigma$ when compared to the previous ephemeris predictions. This circumstantially demonstrates the accuracy of TOI timings. We neglect the difference between the Barycentric Julian Date (BJD) and Heliocentric Julian Date (HJD) in this work. The difference is within $\pm$4s, beyond the timing precision discussed.

The TOI catalog has been well utilized for exoplanet research, including TTV analysis which uses the data in a similar condition to this work \citep{TOIforTTV, Martins2020TOI1,2021HowardTOI2}.

\subsection{TESS Transit Timing Acquirement}
\begin{figure}
  \centering
  \includegraphics[width=3.3in]{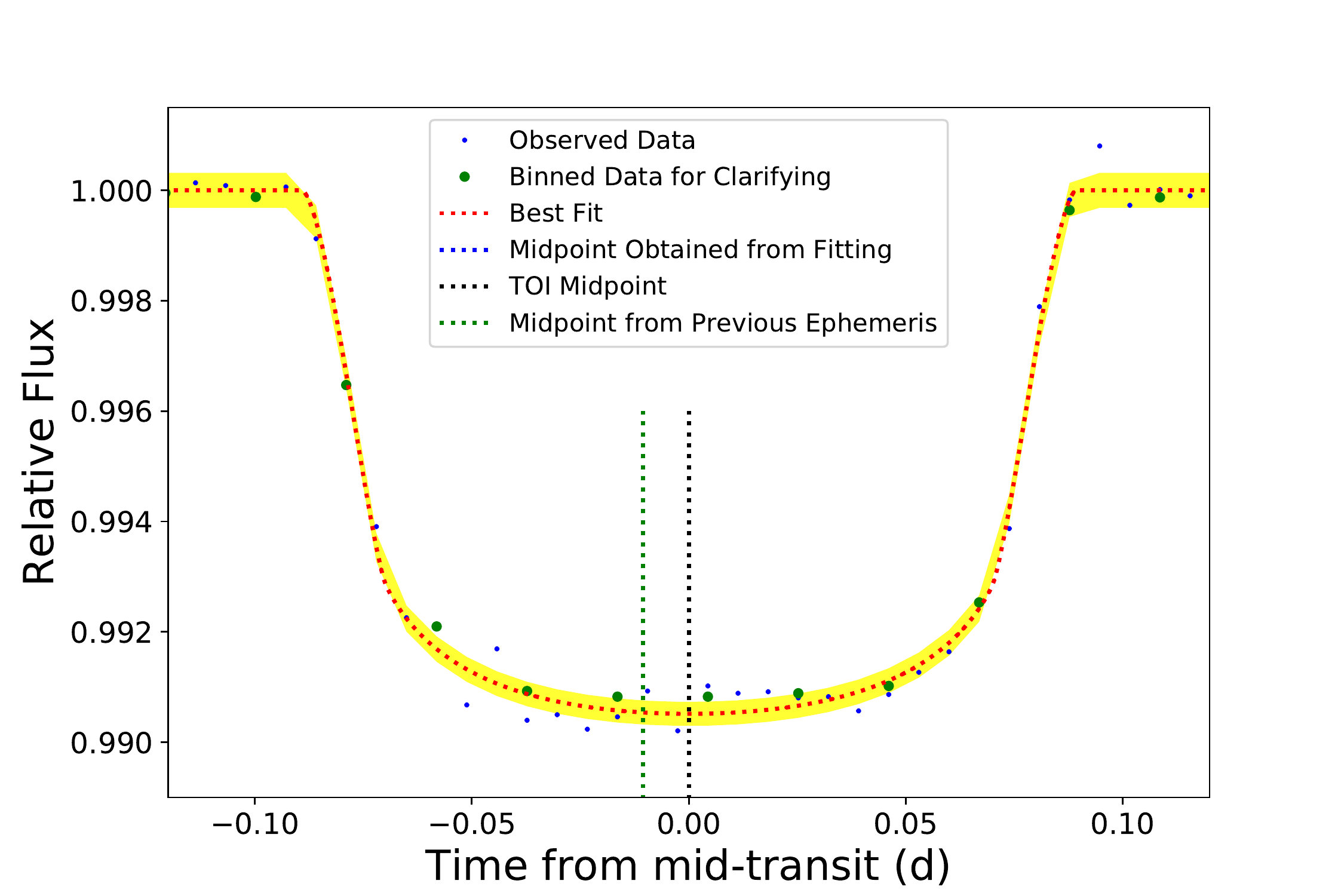}
  \includegraphics[width=3.3in]{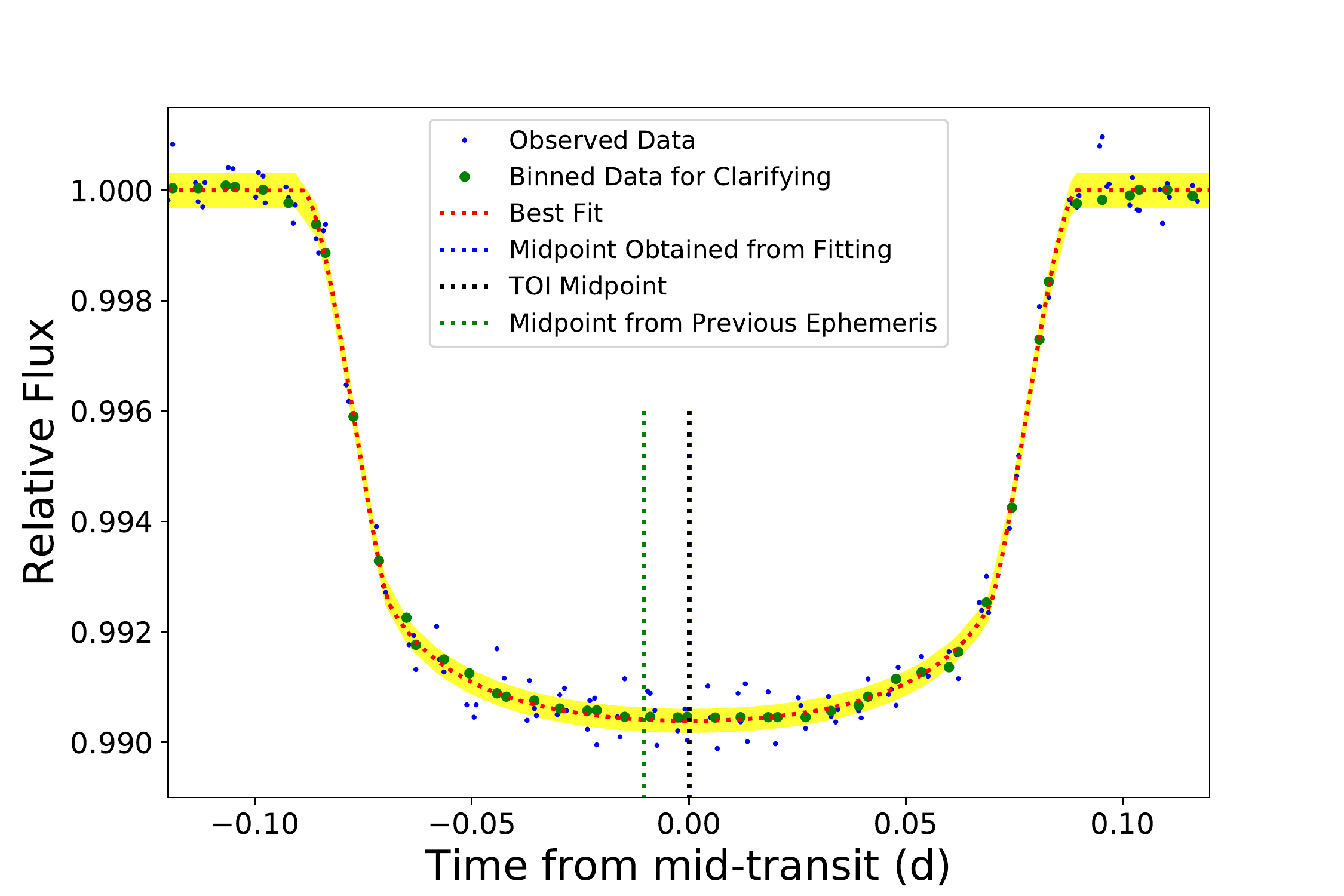}
  \caption{Light curves of KELT-19Ab as an example: single epoch around TOI timing (top panel), folded multiple visits at reference epoch (bottom panel). The blue points present observations (10-minute cadence) while the green points are bins of every three points for clarity. The red line gives the transit model fit with the yellow region indicating 1 $\sigma$ confidence region. The vertical blue line gives the fitted timing; the black vertical line, TOI timing; the green vertical line, previous ephemeris prediction. The timings from single epoch fitting, folded epoch fitting are only 0.14 minutes earlier, 0.20 minutes later than TOI, corresponding to a negligible difference as shown in the image (overlapped blue and black line). The observed TESS timings show an offset of $\sim$ 15 minutes, compared to the previous ephemeris prediction as shown in the vertical green line. The fitting uncertainty is 0.54 minutes for a single epoch, and 0.23 minutes for folded epochs. 
  }
  \label{image:lc}
\end{figure}

\begin{figure*}
   \centering
  \includegraphics[width=7.5in]{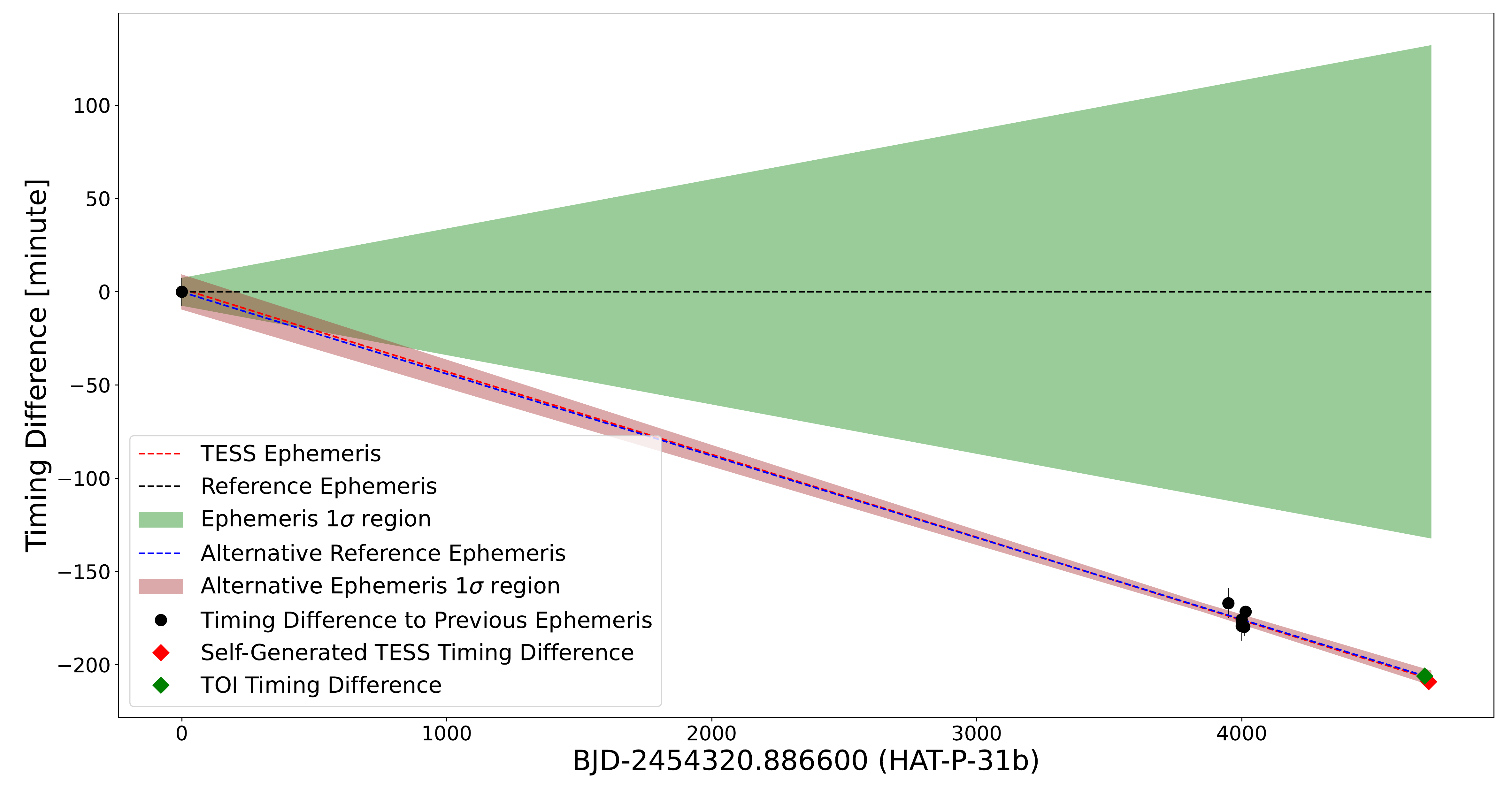}
\caption{The timing difference of HAT-P-31b.  The timing difference is the observed mid-transit times minus the ephemeris predictions.  The red point refers to TESS timing difference; black points refer to timing differences of other observations from literature paper \citep{Kipping2011, Mallonn2019}; the black dashed line is reference ephemeris; blue line alternative reference ephemeris; the red line is the refined ephemeris derived by combining TESS observation; the green region is 1 $\sigma$ significant region of reference ephemeris; the brown region is 1 $\sigma$ significant region of alternative reference ephemeris. We note that our refined ephemeris overlaps the alternative reference ephemeris, indicating the consistency of the two ephemerides. 
}
\label{image:timing}
\end{figure*}

A precision validation of TOI timing is necessary, for the purpose of the study on timing offsets to the previous ephemeris. A majority of the hot Jupiters (159 of 262) present consistent TOI timings which could be circumstantial evidence. Direct evaluation is performed by independently reducing the data and obtaining the TESS timings. We generate a half-automatic pipeline to obtain and fit the light curve from TESS images \citep{Yangatmos, YangLD}.

The pipeline includes two parts, i.e., a photometric pipeline and transit modeling. The photometric pipeline works on TESS image products (as shown in Figure \ref{image:example}) and includes modules: e.g., astrometry checking, aperture photometry, deblending of the nearby contamination flux, and light curve detrending. The photometric pipeline generates light curves from raw images of both 30-minute cadence (FFI) and 2-minute cadence (TPF). 
During TESS extended mission, the 30-minute cadence FFI is updated with a 10-minute cadence. The 10-minute FFI is used in our pipeline. 
Currently, we do not search for the recently released 20-second cadence data.

The photometric data reduction starts at finding if the TPF is available for the source. We would use 2-minute cadence TPF for light curve generation and 30-minute cadence FFI cut-off as a substitution when TPF is not available. The astrometry would then be checked and corrected if there was any pointing jitter \citep{Yangatmos, Yanghats5b}. The astrometry check is based on the comparison between the nominal position reported by Gaia \citep{Gaia2018} and the target center in the TESS image. 
Circular aperture photometry is performed with a radius of 63 arcsecs. 
The background is estimated as the median value of the lowest fifth percentile of pixel fluxes in the vicinity of the target. The photometry error is the quadratic sum of the Poison error and the standard deviation of the background.

The flux contamination from nearby sources is modeled and removed using the flux profile as a function to the given center \citep{Yangatmos}. The detrending for long-term structure removing is performed by modeling the light curve of 0.6 days centering at the transit mid-point after masking the planetary transit. We use a linear function for modeling the long-term structure. We have tested with high order polynomial functions (up to 10 orders) as well as a cubic spline function which gives negligible differences for five validation targets in this work and the exoplanets investigated in previous work \citep{Yangatmos, Yanghats5b, YangLD}.

The detrended light curve is performed with a stellar activity check from archival data and TESS photometry to avoid possible timing bias. The strong stellar activity would be taken into consideration. We note that the starspot perturbation is more significant to brightness than to the shape of the light curve unless the transit comes across the starspot \citep{Makarov2009, Agol2010}. Among the comparison sample in this work, we do not find any significant transit across starspot. Empirically, a sun-like star would hold a variability at a level of $\sim$ 10 ppm on plant transit timescale \citep{Jenkins2002}. In addition, the bias of timing estimation caused by starspot would be weakened by 
detrending process.

More details and evaluations of the pipeline are referred to in previous work \citep{Yangatmos, YangLD}. From the tests and applications so far, the derived transit parameters are within 1 $\sigma$ when we apply the same fitting to TPF light curves.

We derive timings of 31 hot Jupiters using our self-generated pipeline. And we check if the timing obtained from our pipeline is consistent with TOI timing and find the difference is commonly within 2 minutes. Comparison details are present in Section \ref{sect: result}.

%The comparison is performed on the planets WASP-161b with the largest time difference earlier than archival ephemeris prediction, WASP-17b with the largest time difference later than archival ephemeris prediction, WASP-58b that is consistent TESS timing compared to the previous ephemeris \citep{Anderson2010, wasp161, Mallonn2019}. The comparison also includes WASP-121b, KELT-19Ab which have been applied for analysis of the transit depth in previous work \citep{Yangatmos, YangLD}.

Applying Markov Chain Monte-Carlo (MCMC) \citep{pymc,pya}, the light curve is fitted with a planet transit model \citep{Mandel_Agol2002,Eastman2013}. The choice of `circular orbit' or `Keplerian orbit' is consistent with the archival reference work. We briefly describe the transit modeling here with more details available in \citep{Yangatmos, YangLD}.

For `circular orbit', the free parameters during our fitting are transit mid-point ($T0$), the radius ratio of planet to star ($R_{p}/R_{\ast}$), the semi-major axis ( $a/R_{\ast}$), and the limb darkening parameters. For `Keplerian orbit', the model has extra free parameters (during our fitting) of the longitude of the ascending node, the orbital eccentricity ($e$), the ascending node, the periapsis passage time, and the periapsis argument ($\omega$). The fitting model as well as parameterization are taken from \citep{Eastman2013}, e.g., formats of $e$ and $\omega$ are $\sqrt{e}$sin$\omega$ and $\sqrt{e}$cos$\omega$.

The MCMC fitting runs 50,000 steps after the first 50,000 steps as initialization. All the priors are uniform, except for the limb darkening which applies Gaussian prior interpreted in limb darkening table \citep{TESSLD}. We apply quadratic law \citep{Sing2010, KippingLD} for parameterizing limb darkening in this work. Specifically, the format is standard parameterization of $u1$, $u2$. The uncertainty of the light curve applied for obtaining MCMC final result is the standard deviation of the light curve residual from an initial fitting. We do not apply a time-dependent term nor a jitter term for the uncertainty given that no significant evidence of time-dependent and jitter structures has been found during the previous TESS research experiences \citep{Yangatmos, YangLD, Yanghats5b,yangxo3b}. We note that extra free parameters during fitting may reduce the fitting $\chi^2$ which might be considered in future applications.

We apply the transit model to both the light curve of a single epoch and the light curve folded from one TESS sector (examples as shown in Figure \ref{image:lc}). The folding is based on the archival ephemeris and we evaluate the fitting parameter bias if folding an inappropriate period \citep[using the same method as in][]{Yanghats5b}. For TESS one sector, the timing bias is $\sim$ 4 minutes if the period is biased at 0.0004 days. Such large period bias would cause significant TESS timing offsets when compared to ephemeris prediction and thereby are flagged. The fold-and-check method has been well utilized in period searching researches \citep{PDM1989,yangbinary,yang2021ltd064402245919}. In this work, we utilize and present the final timings obtained from folded light curves in one TESS sector. TOI timings are used for sample selection.

The oversampling technique is applied to mitigate influences caused by the sampling rate of TESS 30 minute data. \citep{Kipping2010} reports transit parameter bias caused by under-sampling and proposes the oversampling technique using a numerical solution to Kepler's equation. In previous work, we discuss the sampling influence on inclination and transit depth with and without oversampling technique \citep{Yangatmos}. 
In this work, we check if timing precision could improve by oversampling technique.
The median timing uncertainty is $\sim$ 4 minutes for modeling to 30-minute cadence light curve without oversampling. The timing uncertainty is $\sim$ 1 minute for the 2-minute cadence light curve. Applying oversampling routine from \citep{Kreidberg2015}, we resample the 30-minute light curve to the cadences of 1-minute, 2-minute, and 10 minutes.
The timing uncertainties obtained from fitting the resampled light curves are the same as the result when not applying the oversampling technique. We also test the oversampling to the 2-minute cadence light curve obtained from TPF. The resampling rate is set to be 0.5 and 1 minute. The test also yields a negligible timing difference. We note that the oversampling is particularly effective in estimating inclination as described in \citep{Yangatmos}.

An extra timing uncertainty would be induced to be up to a few 0.1 minutes for the 30-minute light curve. The FFI cut-off we use sets the time stamp as the same as the time of the FFI center. The timing difference during BJD to JD switching can be as large as 0.5 minutes for sources on CCD center and corner that TESS CCD is at 12 degrees $\times$ 12 degrees size. This extra uncertainty is negligible considering the uncertainty of 4 minutes for timings obtained from the FFI cut-off. 2-minute light curves do not suffer such an issue that the time correction has been performed to TPF.

The median timing offset between our results and TOI timings is 1.43 minutes among test sample. The median TOI timing uncertainty is 0.83 minutes. We conclude that it is reasonable to use TOI timings. And the TOI timing offset to the previous ephemeris is regarded as significant if the offset is larger than 4 minutes which is $\sim$ 3 times the median difference. We also require the timing offset to be larger than 1 combined $\sigma$ which is the square root of the quadratic sum of archival ephemeris uncertainty and TESS timing uncertainty. These criteria lead to a final sample of 31 hot Jupiters.

\section{Hot Jupiters with TESS Timing Offsets}
\label{sect: result}

We obtain a sample of 31 targets with TOI timing offsets compared to the previous ephemeris prediction. An example is shown in Figure \ref{image:timing} with the whole sample as shown in Figure \ref{image:timing appendix}. The parameters are present in Table \ref{table:timing}, including planet ID, TESS time minus the predicted time from the previous ephemeris ($\Delta T_{C}$), transit midpoint $T_{C}$, orbital period $P$, reduced chi-squared statistic ($\chi^2_{red}$) of linear period fitting, category flag, parameter reference. We take TOI timings as TESS timings when calculating $\Delta T_{C}$ and replace them with self-generated timings for WASP-173Ab, TOI-1333b, TOI-628b, KELT-21b, KELT-24b, and WASP-187b.

In our sample, the median $\Delta T_{C}$ is 17.8 minutes while the median combined uncertainty is 4.9 minutes. Therefore the signal-to-noise ratio (SNR) is 3.6. Among 31 Jupiters, WASP-161b presents the earliest offset timing of -203.7$\pm$4.1 minutes. WASP-17b gives the latest offset timing of 70.8$\pm$11.7 minutes. The timing uncertainty is derived as the quadratic sum of uncertainties of previous ephemeris and TESS timing.

We classify the sources into three categories, according to the potential properties implied by the timings. Type I target refers to a source of which timings are modeled with a linear function. The timing inconsistency could be either due to systematic error underestimation or some physical process. The linear function indicates a model with a constant derivative, referring to a constant period. Type II refers to the targets of which the timing differences can not be modeled by a linear function but by a quadratic function. The quadratic function can be due to abnormal points or physical processes which lead to a constant period derivative. 
We identify the targets as type III if the timings can not be fitted with any linear or quadratic functions.
The possible physical origin of the timing offsets is discussed in Section.4.

Specifically, we present reduced chi-squared ($\chi^2_{red}$) statistic for type I targets if the data set number is larger than 2 (as shown in Table \ref{table:timing}). We note that the limited amount of data sets induces large uncertainty when calculating $\chi^2_{red}$ \citep[see details in][]{Andrae2010}. The Bayesian Information Criterion \citep[BIC, details in][]{BICmostcited} difference for XO-3b is larger than 383, preferring a quadratic function to a linear fit \citep{yangxo3b}. We note that some hot Jupiters classified as type I may be better fitted with a quadratic function \citep[e.g., WASP-161;][]{Yang1612022}, though the significance is not as high as XO-3b. These tentative signals need careful following up investigation and are not highlighted in this work.

%\clearpage
%\newpage
{
\setlength{\tabcolsep}{2pt}
\setlength\LTcapwidth{\textwidth} 
%\begin{landscape}
\begin{longtable*}{|l|l|l|l|l|l|l|}
\caption{Exoplanet parameters. `1' in column `Planet ID' indicates to the reference ephemeris in Figure \ref{image:timing} and \ref{image:timing appendix}, while `2' presents the alternative ephemeris. The TESS timings derived by our pipeline are flagged as this work. The table is sorted by the significance of $\Delta $$T_{C}$. Sources with earlier TESS timing are listed before the targets with later TESS timings.} 
%Category flag \uppercase\expandafter{\romannumeral1} 
\label{table:timing}\\

\hline

Planet ID   &  $\Delta $$T_{C}$   &      $T_{c}$   &  $P$     & $\chi^2_{red}$ &Category flags  & Reference  \\
%\hline
              &         minutes           &     BJD                 &       days         &              &   &               \\

\endfirsthead
\multicolumn{6}{c}%
{\tablename\ \thetable\ -- \textit{Continued from previous page}} \\
\hline

Planet ID   &  $\Delta $$T_{C}$   &      $T_{c}$   &  $P$     & $\chi^2_{red}$ & Category flags     & Reference  \\
%\hline
              &         minutes           &     BJD                 &       days         &     &            &               \\
\hline
\endhead
\hline \multicolumn{6}{c}{\textit{Continued on next page}} \\
\endfoot
\hline
\endlastfoot
\hline % inserts single horizontal line
WASP-161b   &   &2458492.286050$\pm$0.00265&   5.405366$\pm$0.0000039     &    2.0419  &I&   This work; \citet{Yang1612022} \\
&-203.7$\pm$4.1  &  2459249.035676$\pm$0.000594  &       &  &   &   TOI timing   \\
  1          &                  &  2457416.5289$\pm$0.0011      &
5.4060425$\pm$0.0000048   &   &    &   \cite{wasp161}   \\
\hline
  XO-3b   &    &  2458819.06428$\pm$0.00035 &      &     & II  &  This work; \citet{yangxo3b}  \\ 
   &-17.8$\pm$1.2  &  2458819.064098$\pm$0.000279  &      &     &   &  TOI Timing \\ 
1   &                  &      2455292.43266$\pm$0.00015 &
     3.19153285$\pm$0.00000058 &  &   &   \cite{Wong2014}    \\
2   &                  &     2454449.86816$\pm$0.00023     &
   3.1915239$\pm$0.0000068 &    & &  \cite{Winn2008}\\
    &                  &     2456419.04365$\pm$0.00026     &
   3.19153247$\pm$0.00000055&  &  &  \cite{Wong2014}\\
     \hline  
KELT-18b  & & 2458734.280341$\pm$0.000335  & 2.871698$\pm$0.0000004     &    2.7489    &I&    This work \\
  & & 2458748.637347$\pm$0.000331  & &&&         \\
  & & 2458906.582255$\pm$0.000325  & &&&         \\
  & & 2458932.425443$\pm$0.000353  & &&&         \\
  & & 2459624.505991$\pm$0.000214  & &&&         \\
  & & 2459684.811712$\pm$0.000239  & &&&         \\
 &  -26.8$\pm$2.3  &  2458714.181140$\pm$0.000380  &       &&  & TOI timing   \\
1   &                  &     2457542.52504$\pm$0.00039    &
    2.8717518$\pm$0.0000028 &   &  &  \cite{McLeod2017}  \\
2   &                  &     2457542.52463$\pm$0.00057     &
 2.8716992$\pm$0.0000013  &   &  &  \cite{Maciejewski2020}\\
    \hline
    WASP-54b& & 2458949.705160$\pm$0.001171  & 3.693599$\pm$0.0000006     &    0.8468 &I&    This work \\
  & & 2459573.923274$\pm$0.000538  & &&&         \\
  & & 2459669.955842$\pm$0.000641  & &&&         \\  
  &  -55.9$\pm$8.6  &  2458931.236409$\pm$0.000435  &       & &   &  TOI timing    \\
     1       &                  &      2455518.35087$\pm$0.00053     &
    3.6936411$\pm$0.0000059  &   &    &\cite{Bonomo2017}       \\
  \hline  
 K2-237b  & & 2458642.067579$\pm$0.001163  &2.180535$\pm$0.0000006     &    7.8139   &I&    This work \\
  & & 2459387.806193$\pm$0.000636  & &&&         \\
     &  -15.5$\pm$3.9  &  2458626.800781$\pm$0.000869  &       &&   &  TOI timing   \\
1   &                  &  2457656.4633789$\pm$0.0000048   &  2.1805577$\pm$0.0000057    &    &  &    \cite{Smith2019}  \\
\hline
WASP-76b   & & 2459133.976069$\pm$0.000147  & 1.809881$\pm$0.0000002     &    1.8398 &I&    This work \\
  & & 2459472.424209$\pm$0.000121  & &&&         \\
  & & 2459485.093248$\pm$0.000136  & &&&         \\ 
&  -11.9$\pm$2.9  &  2459117.687201$\pm$0.000119  &       &  &    & TOI timing   \\
 1  &                  &   2456107.85507$\pm$0.00034  &  1.809886$\pm$0.000001   &   &   &    \cite{West2016}  \\
 \hline
WASP-95b    & & 2458328.690567$\pm$0.000287  &2.184667$\pm$0.0000002     &    1.1205 &I&    This work \\
  & & 2459075.846388$\pm$0.000132  & &&&         \\
&  -10.7$\pm$2.9  &  2459084.585010$\pm$0.000110  &       &&      &  TOI timing  \\
1   &    &  2456338.458510$\pm$0.000240  &  2.184673$\pm$0.0000014     &  &    &  \cite{Hellier2014}     \\
\hline
WASP-101b   & & 2458481.061101$\pm$0.000185  &3.585708$\pm$0.0000003     &    4.4892 2&I&    This work \\
  & & 2459216.130731$\pm$0.000145  & &&&         \\
&  -17.3$\pm$5.2  &  2459223.302264$\pm$0.000132  &       & &   &  TOI timing     \\
1   &                  &     2456164.6934$\pm$0.0002  &
     3.585722$\pm$0.000004   &   &   & \cite{Hellier2014}    \\
     \hline
WASP-35b & & 2458459.092397$\pm$0.000244  & 3.161569$\pm$0.0000002     &    0.2758&I&    This work \\
  & & 2459179.930001$\pm$0.000126  & &&&         \\
&  -9.5$\pm$3.5  &  2459176.768453$\pm$0.000197  &       & &   &  TOI timing   \\
1   &    &  2455531.479070$\pm$0.000150  &  3.161575$\pm$0.0000020     &   &   &  \cite{Enoch2011}     \\
\hline
TOI-163b& & 2458350.038867$\pm$0.000908  &4.231119$\pm$0.0000016     &    3.2752   &I&    This work\\
  & & 2458371.194185$\pm$0.001123  & &&&         \\
  & & 2458392.345356$\pm$0.001497  & &&&         \\
  & & 2458421.969304$\pm$0.000841  & &&&         \\
  & & 2458451.585295$\pm$0.001530  & &&&         \\
  & & 2458481.204317$\pm$0.000995  & &&&         \\
  & & 2458557.363318$\pm$0.001196  & &&&         \\
  & & 2458574.287291$\pm$0.000781  & &&&         \\
  & & 2458612.367794$\pm$0.000928  & &&&         \\
  & & 2458629.292355$\pm$0.001017  & &&&         \\
  & & 2458671.603439$\pm$0.000874  & &&&         \\
  & & 2459039.709508$\pm$0.000636  & &&&         \\
  & & 2459073.556184$\pm$0.000635  & &&&         \\
  & & 2459069.328010$\pm$0.000735  & &&&         \\
  & & 2459098.946501$\pm$0.000611  & &&&         \\
  & & 2459132.796272$\pm$0.000621  & &&&         \\
  & & 2459175.105352$\pm$0.000618  & &&&         \\
  & & 2459204.723621$\pm$0.000557  & &&&         \\
  & & 2459234.341314$\pm$0.000575  & &&&         \\
  & & 2459263.960120$\pm$0.000622  & &&&         \\
  & & 2459331.657993$\pm$0.000505  & &&&         \\
  & & 2459348.581531$\pm$0.000414  & &&&         \\
  & & 2459373.968750$\pm$0.000331  & &&&         \\  
        &  -57.2$\pm$22.0  &  2459310.502979$\pm$0.000817  &       & &   &  TOI timing  \\
   1   &                  &    2458328.87970$\pm$0.00063     &
    4.231306$\pm$0.000063 &   &  & \cite{Kossakowski2019}   \\
    \hline
  KELT-14b& & 2458493.272296$\pm$0.000185  &1.710054$\pm$0.0000001     &    4.9021    &I&    This work \\
  & & 2459202.944408$\pm$0.000109  & &&&         \\
  & & 2459235.648287$\pm$0.000969  & &&&         \\
  & & 2459235.435128$\pm$0.000110  & &&&         \\ 
&  -10.7$\pm$5.2  &  2459252.535529$\pm$0.000108  &    &  &    &  TOI timing    \\
1   &    &  2457091.028632$\pm$0.000470  &  1.710059$\pm$0.0000025     &   &   & \cite{Rodriguez2016}      \\
2   &    &  2456665.224010$\pm$0.000210  &  1.710057$\pm$0.0000032     &   &   &    \cite{Turner2016}   \\
\hline
KELT-7b   & & 2458816.518431$\pm$0.000430 & 2.734765$\pm$0.0000002     &    3.3771  &I&    This work \\
  & & 2459492.005468$\pm$0.000231  & &&&         \\
  & & 2459519.352936$\pm$0.000238  & &&&         \\
  & & 2459533.027055$\pm$0.000234  & &&&         \\
&  -12.4$\pm$5.4  &  2458819.253410$\pm$0.000240  &       & &     & TOI timing   \\
1   &    &  2456355.229809$\pm$0.000198  &  2.734775$\pm$0.0000039     &   &   &  \cite{Bieryla2015}     \\
\hline

HAT-P-31b &   &  2459025.840900$\pm$0.001368&5.005269$\pm$0.0000056     &    0.9519 &I & This work \\
&  -206.0$\pm$131.6  &  2459010.826736$\pm$0.001149  &   &  &   &   TOI timing   \\
1  &                  &   2454320.8866$\pm$0.0052 &5.005425$\pm$0.000092 &  &   &\cite{Kipping2011}       \\
2&                  &      2458169.9410$\pm$0.0017 &
5.0052724±0.0000063       &    &  &\cite{Mallonn2019} \\  
\hline
 KELT-1b  &   & 2458765.534321$\pm$0.000717 &1.217494$\pm$0.0000003     &    7.4954    &I & This work\\
      &  -67.4$\pm$53.9  &  2458765.533813$\pm$0.000299  &     & &   &  TOI timing   \\
1   &                  &    2455914.1628$\pm$0.0023   &
     1.217514$\pm$0.000015  &   &  &   \cite{Siverd2012}    \\
2   &                  &  2456093.13464$\pm$0.00019         &
    1.21749448$\pm$0.00000080  &   &  &  \cite{Baluev2015}\\
\hline
KELT-21b    & -0.59$\pm$2.5 & 2458690.462229$\pm$0.000704  &3.612769$\pm$0.0000008     &    3.4411  &I&   This work \\
  & & 2458719.364524$\pm$0.000912  & &&&         \\
  & & 2459420.242127$\pm$0.000267  & &&&         \\
&  -9.8$\pm$2.4  &  2458686.841940$\pm$0.000580  &       &  &   &  TOI timing   \\
1   &    &  2457295.934340$\pm$0.000410  &  3.612765$\pm$0.0000030     &    &  &  \cite{Johnson2018}     \\
\hline
 HAT-P-69b   & & 2459247.345980$\pm$0.000274  &     &     &III&    This work \\
  & & 2458510.155715$\pm$0.000546  & &&&         \\ 
&  9.7$\pm$1.5  &  2459242.559429$\pm$0.000245  &       & &     &  TOI timing   \\
1   &    &  2458495.788610$\pm$0.000720  &  4.786949$\pm$0.0000018     &   &   &  \cite{Zhou2019}     \\
\hline
WASP-17b   & & 2458638.332379$\pm$0.000340  &3.735485$\pm$0.0000003     &    5.5856    &I&    This work\\
  & & 2459340.602164$\pm$0.000403  & &&&         \\
&  70.8$\pm$11.7  &  2458627.126221$\pm$0.000584  &      & &   &  TOI timing   \\
1   &    &  2454559.181020$\pm$0.000280  &  3.735442$\pm$0.0000072     &    &  &  \cite{Anderson2010}     \\
         &    &  2454577.85806$\pm$0.00027   &    3.7354380$\pm$0.0000068   &    &  &  \cite{Anderson2011}     \\
         &    &  2454592.80154$\pm$0.00050  &   3.7354845$\pm$0.0000019    &   &   &  \cite{Southworth2012}     \\
                 &    &  2457192.69798$\pm$0.00028  &   3.735438    &  &    &  \cite{Sedaghati2016}     \\
\hline
WASP-178b & & 2458609.523699$\pm$0.000421  &3.344839$\pm$0.0000007     &    3.4716   &I&    This work \\
  & & 2459352.077016$\pm$0.000181  & &&&         \\
&  12.9$\pm$3.1  &  2458602.836430$\pm$0.001860 &       &&     &  TOI timing    \\
&   &  2459358.7671460$\pm$0.0003877 &      & &       &  TOI timing    \\
1   &    &  2456927.068390$\pm$0.000470  &  3.344829$\pm$0.0000012     &   &   & \cite{Hellier2019}      \\
2   &    &  2458321.867240$\pm$0.000380  &  3.344841$\pm$0.0000033     &  &    &    \cite{Rodr2020}   \\
\hline

WASP-33b & &  2458814.59179$\pm$0.000193   &1.219871$\pm$0.0000001     &    4.8662 &I& This work        \\
&  22.4$\pm$6.9  &  2458791.414307$\pm$0.000169  &       & &    & TOI timing   \\
1   &    &  2454163.223730$\pm$0.000260  &  1.219867$\pm$0.0000012     &   &   & \cite{Collier2010}      \\
2   &    &  2455507.522200$\pm$0.000300  &  1.219868$\pm$0.0000011     &   &   &    \cite{von2014}   \\
\hline

KELT-23Ab & & 2458701.953602$\pm$0.000164  &      &     &III&    This work\\
  & & 2458719.996122$\pm$0.000187  & &&&         \\
  & & 2458758.335680$\pm$0.000196  & &&&         \\
  & & 2458765.102458$\pm$0.000197  & &&&         \\
  & & 2458895.908656$\pm$0.000170  & &&&         \\
  & & 2458934.248842$\pm$0.000181  & &&&         \\
  & & 2459443.943092$\pm$0.000132  & &&&         \\
  & & 2459599.557734$\pm$0.000131  & &&&         \\
  & & 2459613.090493$\pm$0.000169  & &&&         \\
  & & 2459651.430016$\pm$0.000135  & &&&         \\
  & & 2459669.472623$\pm$0.000142  & &&&         \\
  & & 2459793.513154$\pm$0.000138  & &&&         \\
&  23.8$\pm$7.7  &  2458683.911214$\pm$0.000056  &  &     &      &  TOI timing   \\
1   &    &  2458140.379200$\pm$0.002700  &  2.255251$\pm$0.0000110     &    &  & \cite{Johns2019}      \\
2   &    &  2458140.386980$\pm$0.000200  &  2.255288$\pm$0.0000007     &    &  &    \cite{Maciejewski2020}   \\
\hline

HAT-P-6b & & 2458759.452299$\pm$0.000648  &3.852999$\pm$0.0000004     &    7.5980&I&   This work      \\
  & & 2458774.864299$\pm$0.000681  & &&&         \\
&  26.3$\pm$9.2  &  2458740.188710$\pm$0.000360  &      &  &   &  TOI timing  \\
1   &    &  2454035.675750$\pm$0.000280  &  3.852985$\pm$0.0000050     &   &   &  \cite{Noyes2008}     \\
\hline

KELT-19Ab & & 2459222.7898588$\pm$0.00020 &  4.611736$\pm$0.0000009     &    4.1958 &I&    This work\\
  & & 2458507.971344$\pm$0.0002751  & &&&         \\ 
&  15.2$\pm$5.9  &  2459222.789720$\pm$0.000183  &      & &      &  TOI timing  \\
1   &    &  2457281.249537$\pm$0.000361  &  4.611709$\pm$0.0000088     &   &   &  \cite{Siverd2018}     \\
\hline

WASP-94Ab  & & 2458352.000206$\pm$0.000642  & 3.950201$\pm$0.0000006     &    0.5179 &I&    This work \\
  & & 2459039.335697$\pm$0.000323  & &&&         \\
&  10.2$\pm$4.0  &  2459039.335846$\pm$0.000386  &       &  &     &  TOI timing  \\
1   &    &  2456416.402150$\pm$0.000260  &  3.950191$\pm$0.0000037     &  &    &  \cite{Bonomo2017}     \\
\hline
WASP-58b  & & 2458695.984265$\pm$0.000376  & 5.017215$\pm$0.0000005     &    2.7651&I&    This work \\
  & & 2458706.018411$\pm$0.000428  & &&&         \\
  & & 2458991.998531$\pm$0.000371  & &&&         \\
  & & 2459017.084690$\pm$0.000379  & &&&         \\
  & & 2459413.444427$\pm$0.000157  & &&&         \\
  & & 2459734.547119$\pm$0.000164  & &&&         \\
  & & 2459764.650223$\pm$0.000159  & &&&         \\
&  37.4$\pm$13.5  &  2458986.981902$\pm$0.000409  &      &&     &  TOI timing   \\
1   &    &  2455183.933500$\pm$0.001000  &  5.017180$\pm$0.0000110     &  &    & \cite{Hebrard2013}      \\
2   &    &  2457261.059700$\pm$0.000620  &  5.017213$\pm$0.0000026     &    &  &    \cite{Mallonn2019}   \\
\hline
WASP-99b   & & 2458393.713195$\pm$0.000480  &5.752591$\pm$0.0000022     &    4.2045 &I&    This work\\
  & & 2459112.785723$\pm$0.000271  & &&&         \\
  & & 2459141.548814$\pm$0.000244  & &&&         \\
&  61.6$\pm$31.2  &  2459135.796019$\pm$0.000239  &      & &     &     TOI timing   \\
1   &    &  2456224.983200$\pm$0.001400  &  5.752510$\pm$0.0000400     &    &  &  \cite{Bonomo2017}     \\
\hline
TOI-1333b   & 2.67$\pm$1.4&  2458715.1230$\pm$0.0010 &  4.720171$\pm$0.0000204     &    0.0318 & I &  This work     \\
&    &  2458752.884599$\pm$0.000828  &      &    &  &      \\
   &   -5.7$\pm$1.5  &  2458715.117140$\pm$0.000550  &       &  &   &  TOI timing    \\
1   &    &  2458913.370330$\pm$0.000450  &  4.720219$\pm$0.0000110     &    &  &  \cite{Rodriguez2021}     \\
\hline

WASP-78b & & 2458446.902114$\pm$0.000472  &2.175185$\pm$0.0000003     &    3.9626&I&    This work \\
  & & 2459162.537455$\pm$0.000227  & &&&         \\
  & & 2459192.991391$\pm$0.000379  & &&&         \\ 
&  18.8$\pm$11.1  &  2459175.589610$\pm$0.000863  &      & &    &  TOI timing  \\
1   &    &  2455882.359640$\pm$0.000530  &  2.175176$\pm$0.0000047     &    &  & \cite{Bonomo2017}      \\
2   &    &  2456139.030300$\pm$0.000500  &  2.175173$\pm$0.0000030     &  &    &    \cite{Brown2017}   \\
\hline

WASP-173Ab   & 1.2$\pm$0.9  &  2458355.195662$\pm$0.00047  & 
 1.386654$\pm$0.0000006     &    0.3322 &  I   &  This work     \\
  &  -30.4$\pm$1.1  &  2458355.173660$\pm$0.000620  & 
    &   &  &  TOI timing    \\
1   &                  &    2457288.8585$\pm$0.0002     &
    1.38665318$\pm$0.00000027&   &  &  \cite{Hellier2019}  \\
 2    &                  &      2458105.59824$\pm$0.00090   &
 1.3866529$\pm$0.0000027  &  &   &  \cite{Labadie2019}  \\
 \hline
 TOI-628b   &  3.8$\pm$3.4  &  2458469.232700$\pm$0.002220  &  3.409512$\pm$0.0000335     &     &   I &  This work \\
          &  7.4$\pm$1.2 &  2458469.235200$\pm$0.000430  &      &   &   &   TOI timing  \\
1   &    &  2458629.479720$\pm$0.000390  &  3.409568$\pm$0.0000070     &   &   &  \cite{Rodriguez2021}     \\
\hline
 KELT-24b  & 1.0$\pm$0.9 & 2458695.919325$\pm$0.000633  & 5.551490$\pm$0.0000011     &    1.7072&I&    This work\\
  & & 2458868.015428$\pm$0.000148  & &&&         \\
  & & 2458895.773212$\pm$0.000162  & &&&         \\
  & & 2459412.061344$\pm$0.000102  & &&&         \\
  & & 2459423.164772$\pm$0.000217  & &&&         \\
  & & 2459606.363810$\pm$0.000204  & &&&         \\
  & & 2459617.467116$\pm$0.000223  & &&&         \\ 
&  7.9$\pm$0.9  &  2458684.821890$\pm$0.000320  &  
&  &      &  TOI timing   \\
1   &    &  2458540.477590$\pm$0.000360  &  5.551493$\pm$0.0000081     &   &   & \cite{Rodriguez2019}      \\
2   &    &  2458268.454590$\pm$0.000870  &  5.551492$\pm$0.0000086     &    &  &    \cite{Maciejewski2020}   \\
\hline
WASP-187b   & 7.3$\pm$8.7 &  2458785.428921$\pm$0.001771  & 5.147885$\pm$0.0000027     &    &I&    This work \\
&  34.5$\pm$8.7  &  2458764.856300$\pm$0.002600  &      & &  &  This work     \\
1   &    &  2455197.352900$\pm$0.002000  &  5.147878$\pm$0.0000050     &    &  &  \cite{Schanche2020}     \\
\hline
%\multicolumn{this is a try of the version}
\end{longtable*}
%\end{landscape*}
}

\begin{figure}
  \centering
  \includegraphics[width=3.3in]{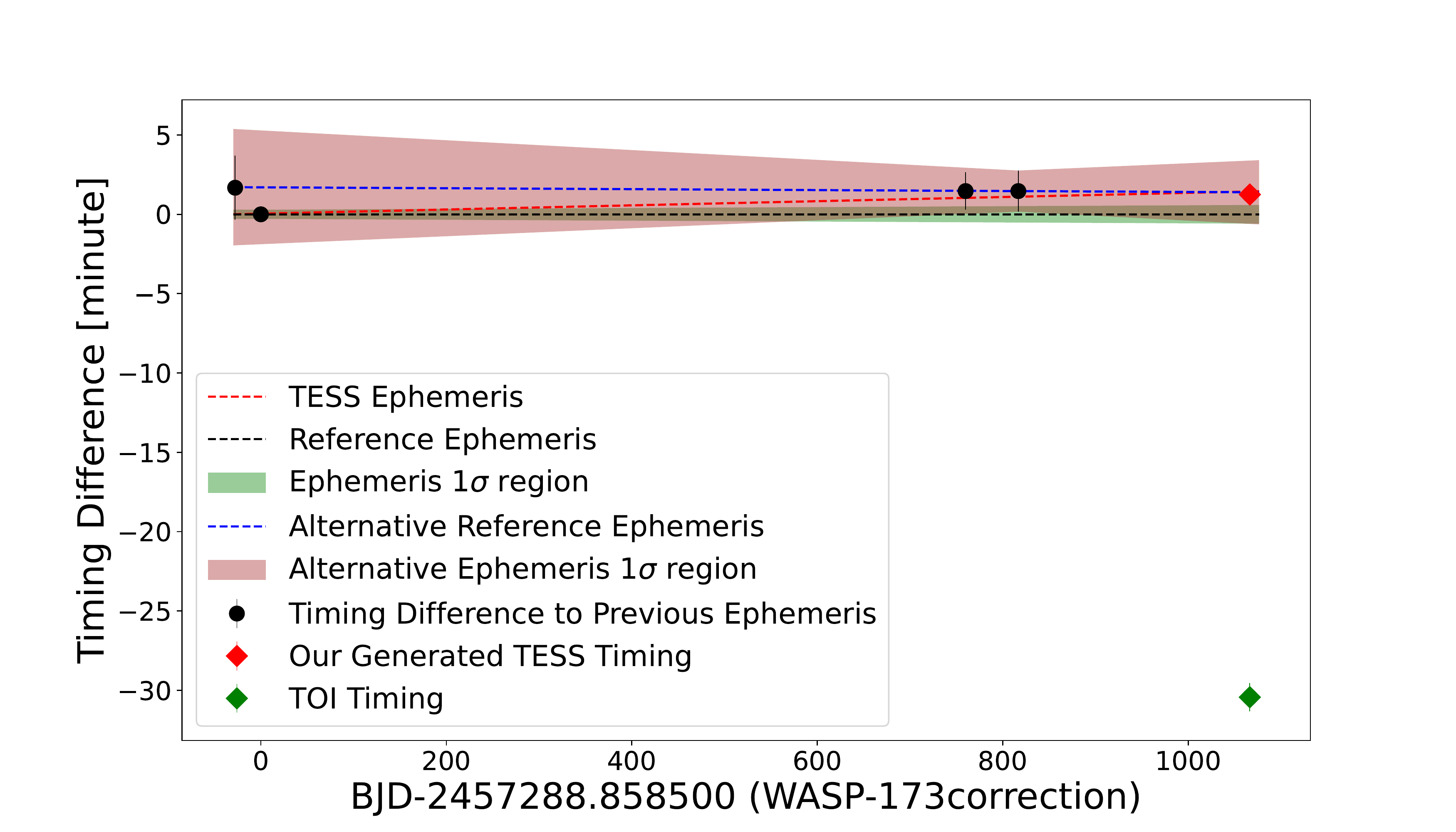}
\includegraphics[width=3.3in]{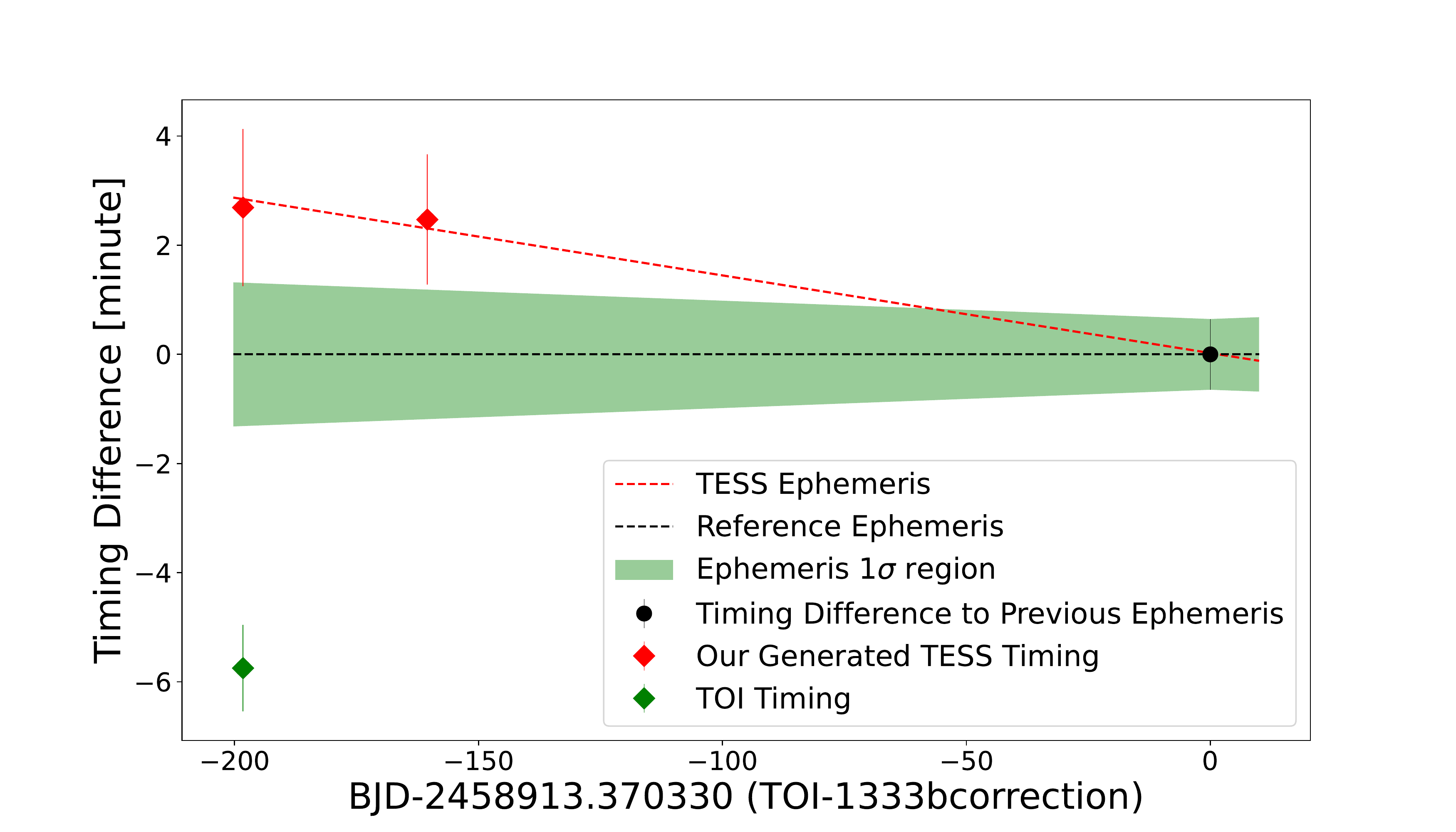}  
\includegraphics[width=3.3in]{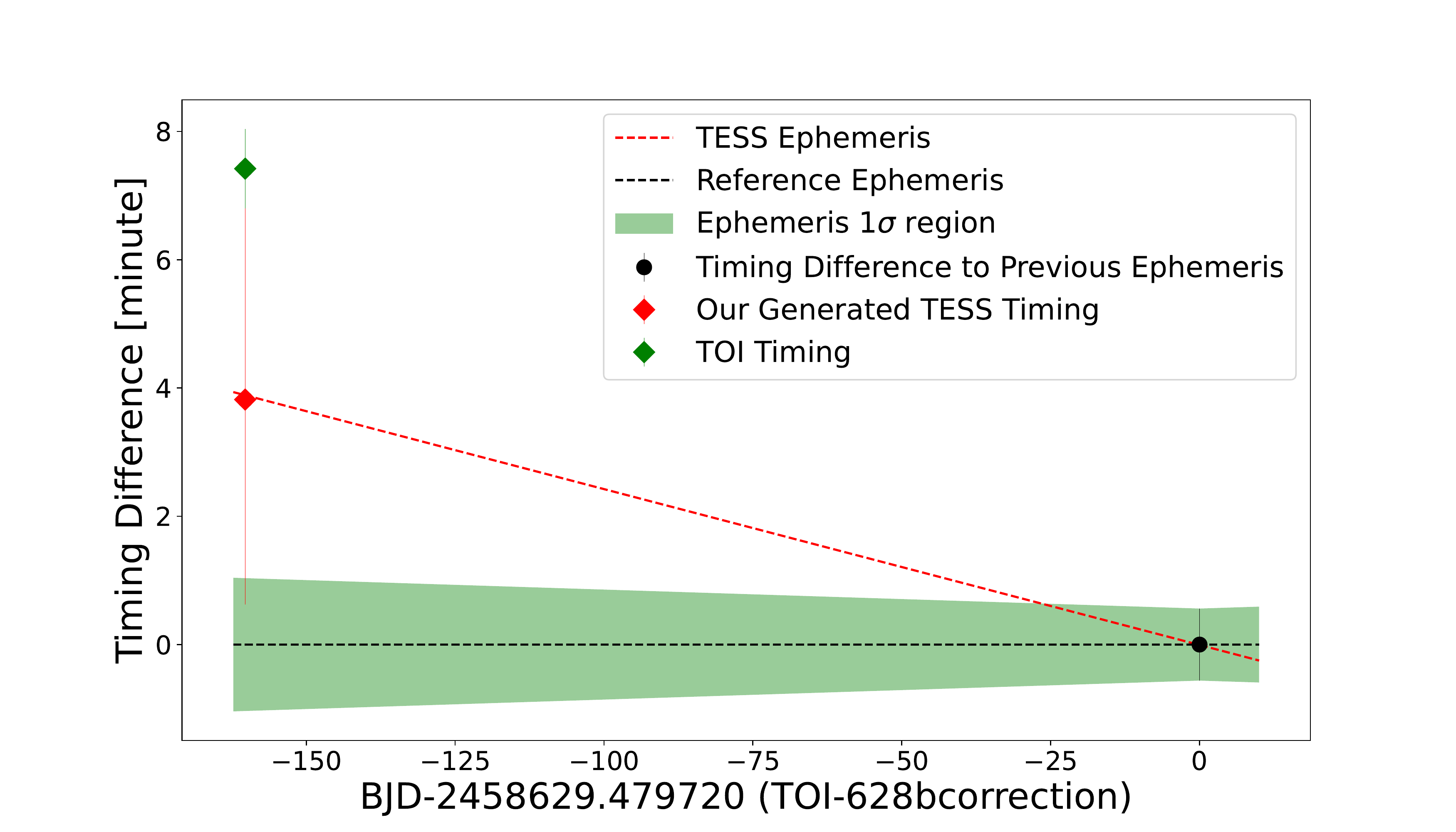}
\caption{The timing differences with corrected timings for WASP-173Ab, TOI-1333b, TOI-628b. The symbols are similar to Figure \ref{image:timing}. The green diamond indicates TOI timing, the red diamond gives the timing generated from TESS raw images.
}
    \label{image:timing_corre}
\end{figure}

We  manually verify the TOI timings of 31 hot Jupiters  among which WASP-173Ab, TOI-1333b, and TOI-628b need timing re-calibration. We check the TESS raw data (2-minute cadence) of WASP-173Ab and find an abnormal data point around a transit at 2468356.564637 (BJD). The abnormal data biases the modeling if not clipped when performing an automatic pipeline. The points should be clipped if excess 10 $\sigma$ to the residual of successful transit fitting.
We refit the TESS light curve with abnormal data clipped. The timing is 2458355.195662$\pm$0.00047 (BJD) when we fit one transit visit and 2458355.195907$\pm$0.0001 (BJD) when fitting visits folded through the whole sector. These two results are consistent within 0.35 minutes and are different from TOI timing at 29 minutes. The refitted TESS timing is consistent with the previous ephemeris (as shown in Figure \ref{image:timing_corre}).

TOI-1333b timing derived by refitting TESS light curve is 2458715.1230$\pm$0.0010 (BJD) which is 8.4 minutes later than TOI derived timing (as shown in Figure \ref{image:timing_corre}). The TESS 30 minute data (available for the TOI-1333b) has some abnormal points around transits which would bias the timings if not applying the sigma clipping process. Removing the abnormal data points, we refit the light curve for the timing. The timings derived from single transit and combined transits have a difference of 1.8 minutes (within 0.3 combined $\sigma$). The timing is close ($\sim$ 1 $\sigma$) to the prediction of previous ephemeris (Figure \ref{image:timing_corre}).

 We derive a combined timing of 1469.23270$\pm$0.00222 (BJD) for TOI-628b while a single transit visit obtains a midpoint at 1469.2332$\pm$0.0074 (BJD). The value is $\sim$ 1 $\sigma$ earlier than TOI timing and is consistent with the previous ephemeris.

Comparing with our generated TESS timings, TOI timings of KELT-21b, KELT-24b, and WASP-187b present differences of 10, 8, and 35 minutes, respectively. We note that TOI timings are highly reliable given that only 5 sources among 262 TOI hot Jupiters are found with possible issues, referring to a possibility of less than 2$\%$.

\subsection{Ephemeris Refinement}

We refine the ephemeris of type I targets in our sample. We do not apply any ephemeris refinements to type II and III sources. The new ephemeris consists of TESS timings and a refined period (as shown in Table \ref{table:timing}). The period is obtained from a linear fit of TESS timings and timings taken from archival papers (as listed in Table \ref{table:timingsupplement}) as well as Exoplanet Archive \citep{ExoplanetArchive}. The refinement has a median precision of 0.82 minutes until 2025 and 1.21 minutes until 2030. The largest uncertainties are 34 minutes in 2025 and 61 minutes in 2030, coming from TOI-628b, due to the shortest baseline. Other than TOI-628b and TOI-1333b, all the refined timing uncertainties are within 5 minutes.

The ephemeris precision depends on the length of the time baseline and transit timing precision. The timing uncertainties could be underestimated due to the techniques in light curve generation and high dimension model fitting \citep{Yangatmos, YangLD}. Combined timing derived from multi-visits based on a constant period assumption might be biased if the folding period is not precise, especially when the light curves partially cover the transits. Correcting the timing biases in archival papers (if present) is beyond the scope of this work.

The period could be updated when more observations are available \citep{Mallonn2019, Edwards2021, Wang2021}. The periods from the previous works are significantly different from the periods derived in our refinement. We note that these period differences might origin from physical processes which make the refinement inappropriate (as discussed in Section~\ref{sec:physics}).

\begin{figure}
  \centering
  \includegraphics[width=3.3in]{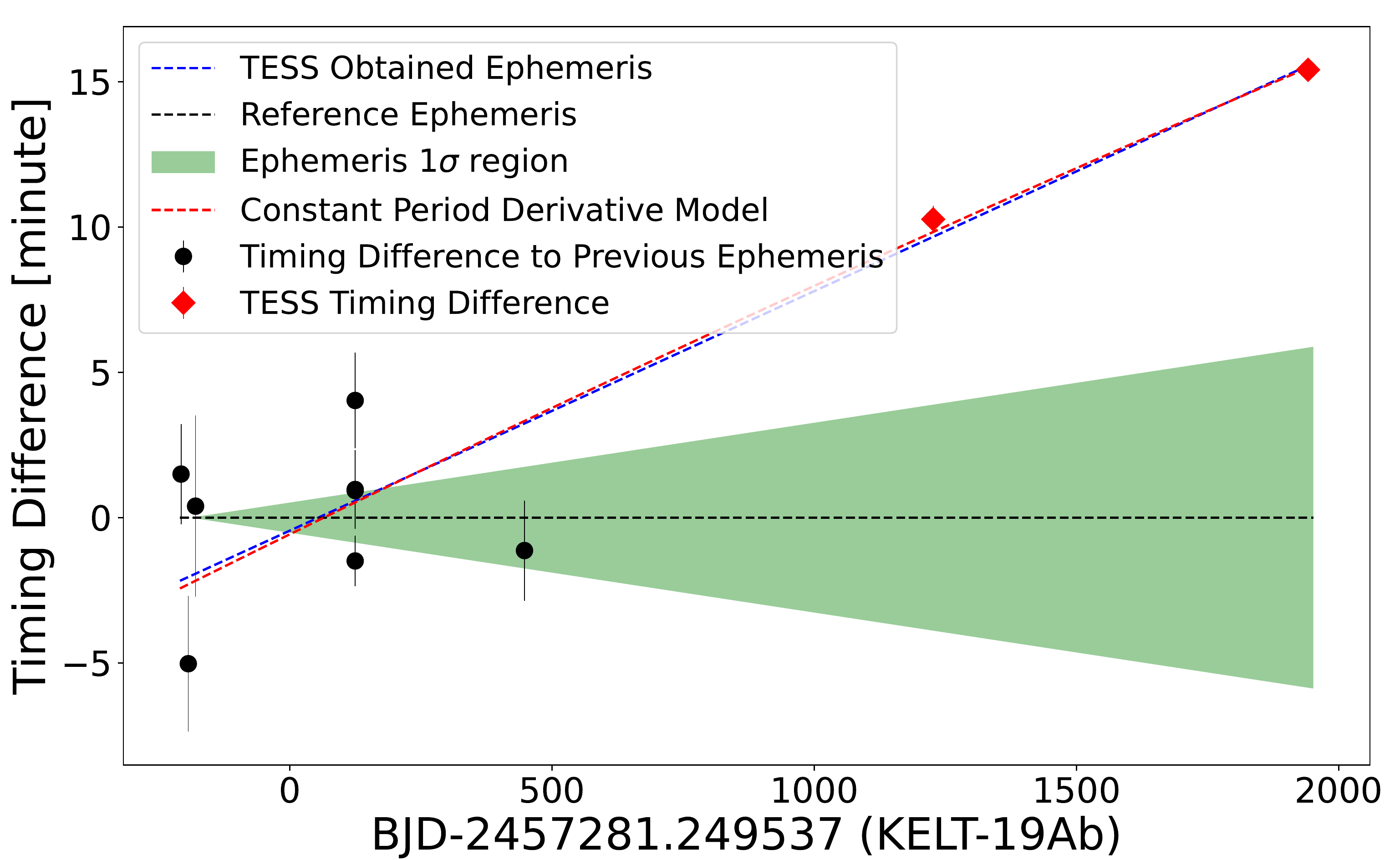}
\caption{KELT-19Ab timings fitted with a quadratic function. The symbols are similar to Figure \ref{image:timing}. The red line shows the quadratic function model.}
    \label{image:timingbinary}
\end{figure}

\section{Discussion: Possible Physical Origin}
\label{sec:physics}

Some targets in our sample present very significant period differences when compared to former results. It might not be a good hypothesis to regard all the differences originating from the underestimation of archival period uncertainties. Period bias caused by a timing shift of 2 minutes would be only $\sim$ 10$^{-5}$ days when the time baseline is 1 year.

We argue that a very significant period difference might be attributed to physical period-changing processes. We find in our sample that the targets with offset SNR larger than 10 all present earlier observation timings. These sources are WASP-161b, XO-3b, and KELT-18b, among which WASP-161b and XO-3b are detected with clues of TTVs in our following work \citep{Yang1612022,yangxo3b}. The period difference caused by systematic underestimation should be unsigned which is not the case. The tidal dissipation could explain the observational phenomenon.

The tidal torque transfers the energy between the star-planet orbit and the rotation of the star and planet \citep{Goldreich1966, Lin1996, Naoz2011, Wu2011, Dawson2018, RodetandLai}. The process could cause the period decay and the apsidal procession \citep{Hut1981, Ragozzine2009}. The induced TTV has been discovered in WASP-12b at $\sim$ a few minutes \citep{2011wasp12b,2017wasp12b}. And TESS provides the most recent evidence for WASP-12b TTV \citep{2021wasp12bTESS}.

%\begin{figure*}
%  \centering
%  \includegraphics[width=7.5in]{xo-3bqua.pdf}
%\caption{The TTV of XO-3b. The symbols are similar to Figure %\ref{image:timing} while the red dashed line presents the fitted quadratic function as described in the text.}
%    \label{image:xo-3}
%\end{figure*}

We report WASP-161b, which shows the most significant TESS timing offsets in this sample, presenting a period derivative ($\dot{P}$) of -1.16$\times$10$^{-7}\pm$2.25$\times$10$^{-8}$ \citep[as details described in][]{Yang1612022}. WASP-161b possibly is undergoing tidal dissipation. We have approved CHEOPS \citep{Benz,pycheops} for two visit observations in 2022 for further investigation. WASP-161b is regarded as a type I target in this work.

The period of XO-3b has been reported differently in previous works \citep[][and references therein]{Winn2008, Winn2009, Johns-Krull2008, Wong2014, Bonomo2017}. TESS timing presents an offset of -17.8$\pm$1.2 minutes (14.8 $\sigma$) to the newest archival ephemeris from \citet{Bonomo2017}. The timing generated by our pipeline is consistent within 0.3 minutes to TOI timing. And the uncertainties are similar ($\sim$ 0.45 minutes). \citet{yangxo3b} reports the XO-3b as a tidal dissipation candidate by joint analyzing archival timings and TESS timing.

The $\dot{P}$ is -6.2$\times$10$^{-9}$$\pm$2.9$\times$10$^{-10}$ days per orbit per day which relates to a timescale of orbital decay of 1.4 Myr. Applying equilibrium tide \citep{Hut1981,Themodel}, \citet{yangxo3b} obtain a modified tidal quality factor $Q'_\star$ as 1.5$\times$10$^{5}$$\pm$6$\times$10$^{3}$ if assuming the period decaying is due to the stellar tide. $Q'_p$ is 1.8$\times$10$^{4}$$\pm$8$\times$10$^{2}$ under the assumption that period decaying is due to the planetary tide.

The number and properties of the detected dissipating planets would calibrate a series of crucial models in the planet formation theory, e.g., the dissipation as well as circularization timescale, and the possibility of capturing a floating planet or interacting with a stellar companion \citep{Dawson2018}.

The apsidal precession could be excited when the tidal torque exists \citep{Ragozzine2009}. Distinguishing the difference between tidal dissipation and precession needs to model timings of occultation \citep{2017wasp12b,2020wasp12b, 2021wasp12bTESS}. XO-3b is also expected to be a candidate presenting precession in previous work \citep{Jordan2008, Nascimbeni2021}. We note that the period changing originating from precession and R$\o$mer effect should be unsigned as the same as from systematic underestimation.

The relation between the planet period derivative and host star acceleration rate is well modeled \citep{wasp-4b2020}. In our sample, KELT-19Ab shows a maximum stellar acceleration at 4 m s$^{-1}$ yr$^{-1}$ originating from binary companion \citep{KELT-19}. This acceleration would cause a period derivative of 5.32 ms yr$^{-1}$, according to the calculation from \citep{wasp-4b2020}. We generate the TESS timings in both 2019 and 2020. TOI catalog gives the timing at 2020 which is only 0.14 minutes different from our result (as shown in Figure \ref{image:lc} and caption therein). We find timings can be fitted with both a linear and a quadratic function (as shown in Figure \ref{image:timingbinary}). The fitting result of the quadratic function indicates a period derivative of 112$\pm$94 ms yr$^{-1}$. Therefore, we conclude that combining TESS and archival timings do not present a significant TTV dominated by stellar acceleration for KELT-19Ab. We regard the R$\o$mer effect beyond the detection limit in this work.

Further investigation requires long-term measurements with both photometric and spectroscopic instruments. The trend of radial velocity curve if presents indicate stellar companions \citep{wasp-4b2020}. Modeling timing evolution reveals TTV evidence \citep{Holman2010,2017wasp12b, yangxo3b}. Approved telescope proposals have proved to be effective in analyzing the timing offsets of hot Jupiter \citep{Ragozzine2009,2017wasp12b}. Sky surveys, e.g., Kepler, TESS, provide more light curves for timing analysis \citep{Borucki2011, Ivshina2022}. Moreover, the sample for relevant analysis can be potentially extended by upcoming time-domain surveys, e.g., Large Synoptic Survey Telescope \citep[LSST;][]{LSST, LSSTplanet}, SiTian \citep{SiTian, Yang2022}.

\section{Summary}

We discuss the ephemeris of 31 hot Jupiters, of which TOI timings show offsets. We refine the ephemeris of the sample by jointly fitting TESS timings and archival times from previously published papers. The TESS timings are obtained by our self-generated pipeline. The pipeline obtains the light curve from the raw TESS images and fits the light curve with the planet transit model. The result from our pipeline gives consistent results compared to TOI catalog.

Among the sample, TOI timings present a median offset of 17.8$\pm$4.9 minutes, equivalent to an SNR of 3.6$\sigma$ when compared to the previous ephemeris. WASP-161b and XO-3b give the most significant timing offsets. The ephemeris refinement serves the potential follow-up observations for equipment, e.g., CHEOPS, ongoing James Webb Space Telescope, and Ariel Space Telescope. The refined timing reaches a precision within 0.82 minutes in the next 5 years and 1.21 minutes in the next ten years.

WASP-161b, XO-3b, and KELT-18b present timing offsets larger than 10 $\sigma$. These three targets all have an earlier observed timing than the predictions from the previous ephemeris under a constant period assumption. We find WASP-161b and XO-3b present evidence of period decaying \citep{Yang1612022,yangxo3b}. Apsidal precession could be an alternative explanation to the TTVs. Interestingly, all four targets (WASP-161, XO-3b, WASP-12b, WASP-4b) reported with observed TTVs, show earlier timing than the prediction in a constant period model. Apsidal precession could not explain this since the timing variation caused by precession should be unsigned. Further observations, e.g., occultation timing monitoring, are helpful for confirmation.

\section*{acknowledgements}
This work made use of the NASA Exoplanet Archive \footnote{\url{https://exoplanetarchive.ipac.caltech.edu/index.html}} \citep{ExoplanetArchive} and PyAstronomy\footnote{\url{https://github.com/sczesla/PyAstronomy}} \citep{pya}. We would like to thank for Ranga-Ram Chary for helpful discussions. Su-Su Shan, Fan Yang, and Ji-Feng Liu acknowledge fundings from the National Key Research and Development Program of China (No.2016YFA0400800), the National Natural Science Foundation of China (NSFC; No.11988101), the CSST MilkyWay and Nearby Galaxies Survey on Dust and Extinction Project
CMS-CSST-2021-A09 and the Cultivation Project for LAMOST Scientific Payoff and Research Achievement of CAMS-CAS. Hai-Yan Zhang acknowledges NSFC (No.12041301, U1831128). Xing Wei is supported by NSFC (No.11872246, 12041301), and the Beijing Natural Science Foundation (No. 1202015).

\clearpage
\newpage
\section*{Appendix}

\figsetstart
\figsetnum{1}
\figsettitle{Timing differences of Type I targets of which the timings can be fitted by a linear function. The symbols are the same as Figure \ref{image:timing} while the legend inside the image is dismissed for clarity.
}
\figsetgrpstart
\figsetgrpnum{1.1}
\figsetgrptitle{WASP-161b}
\figsetplot{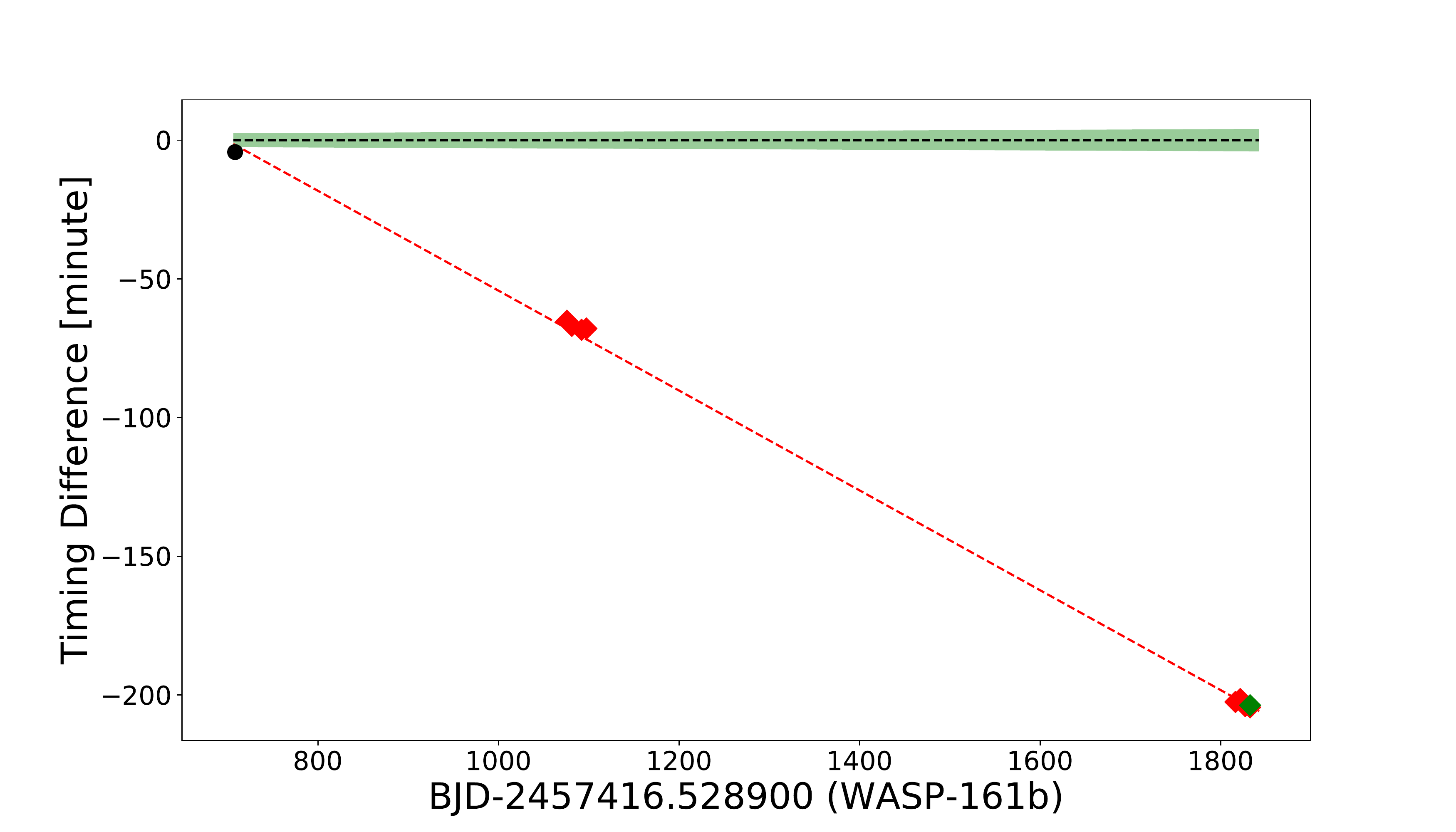}
\figsetgrpnote{WASP-161b}
\figsetgrpend
\figsetgrpstart
\figsetgrpnum{1.2}
\figsetgrptitle{KELT-18b}
\figsetplot{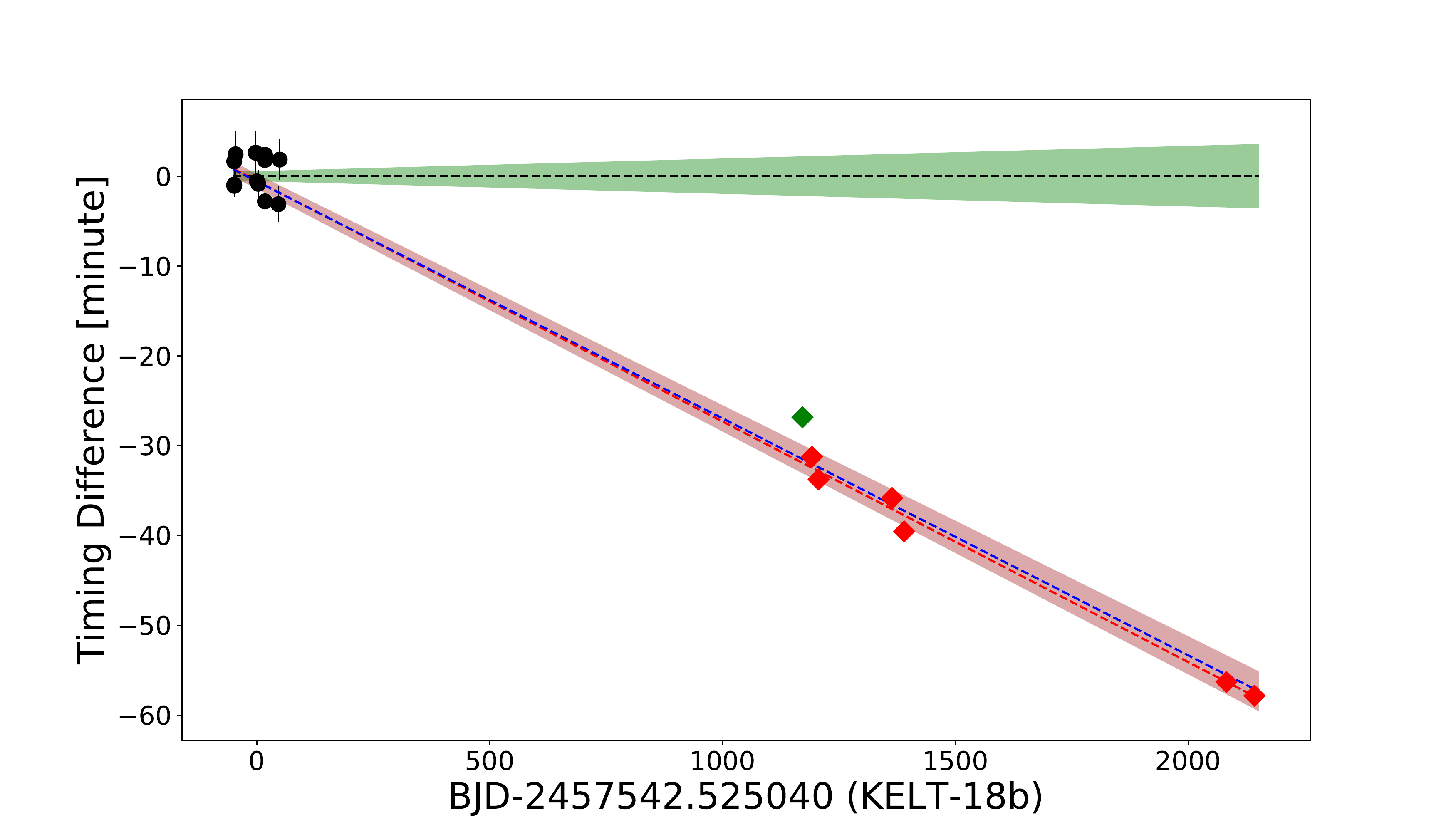}
\figsetgrpnote{KELT-18b}
\figsetgrpend
\figsetgrpstart
\figsetgrpnum{1.3}
\figsetgrptitle{WASP-54b}
\figsetplot{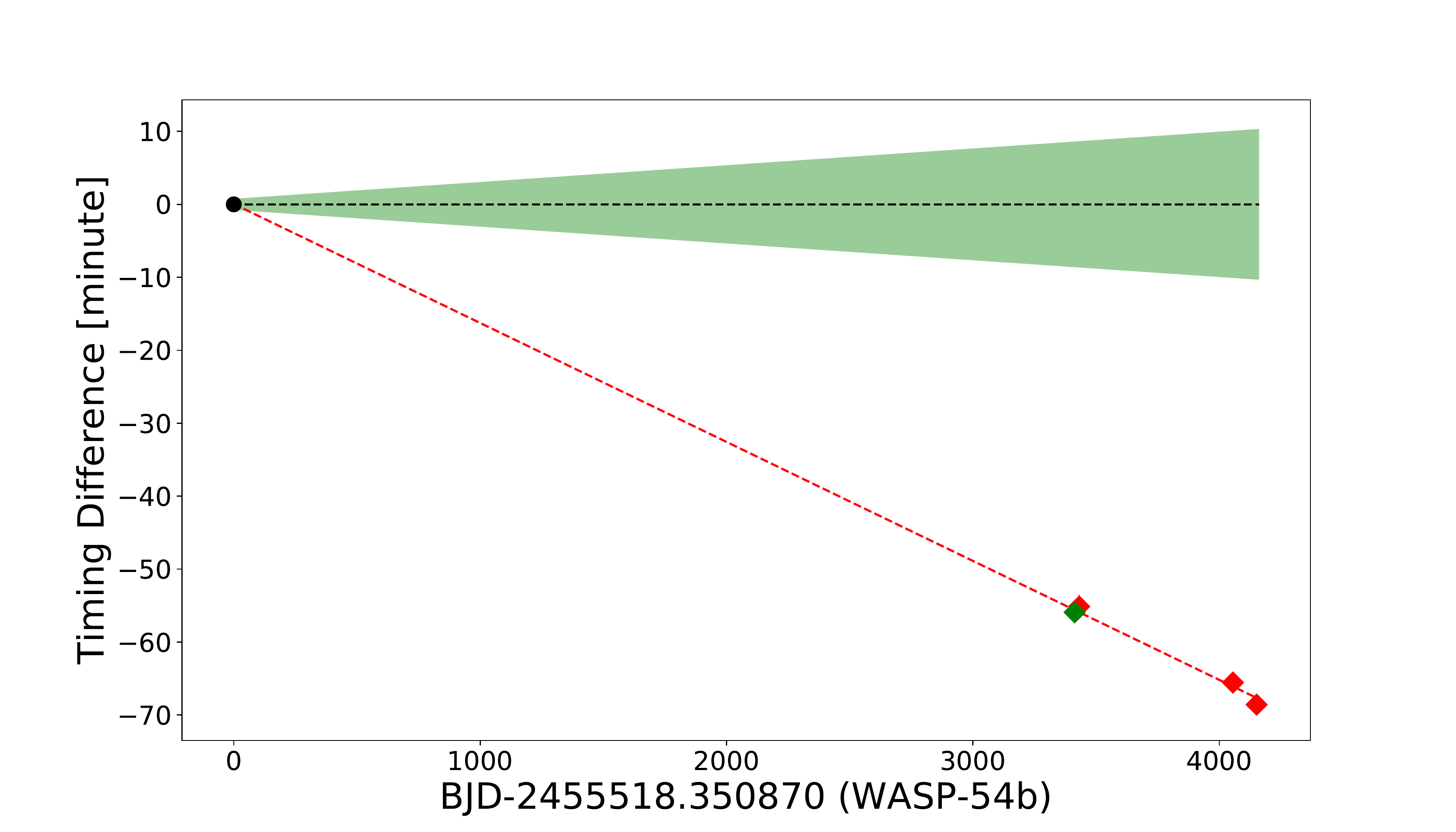}
\figsetgrpnote{WASP-54b}
\figsetgrpend
\figsetgrpstart
\figsetgrpnum{1.4}
\figsetgrptitle{K2-237b}
\figsetplot{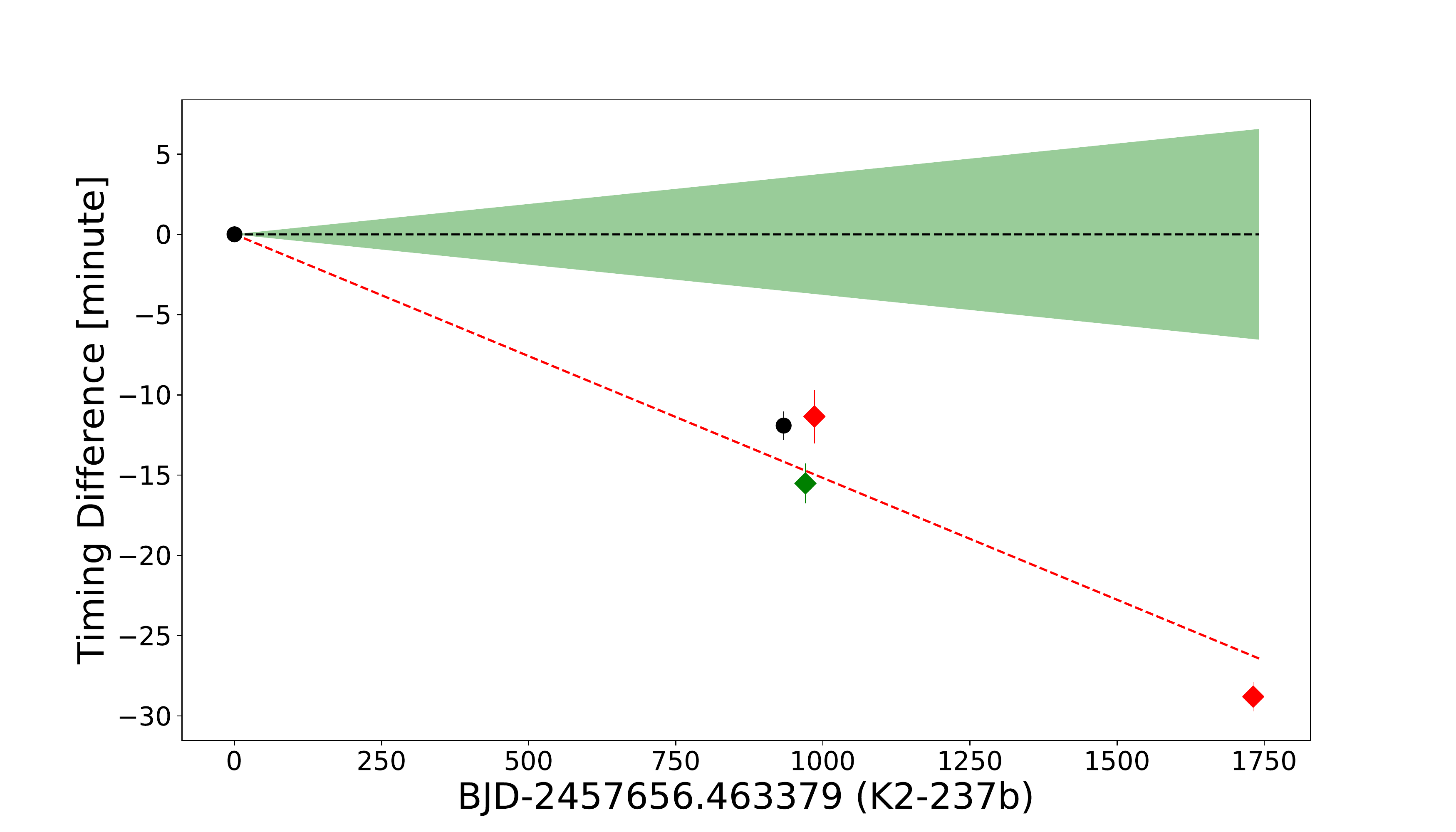}
\figsetgrpnote{K2-237b}
\figsetgrpend
\figsetgrpstart
\figsetgrpnum{1.5}
\figsetgrptitle{WASP-76b}
\figsetplot{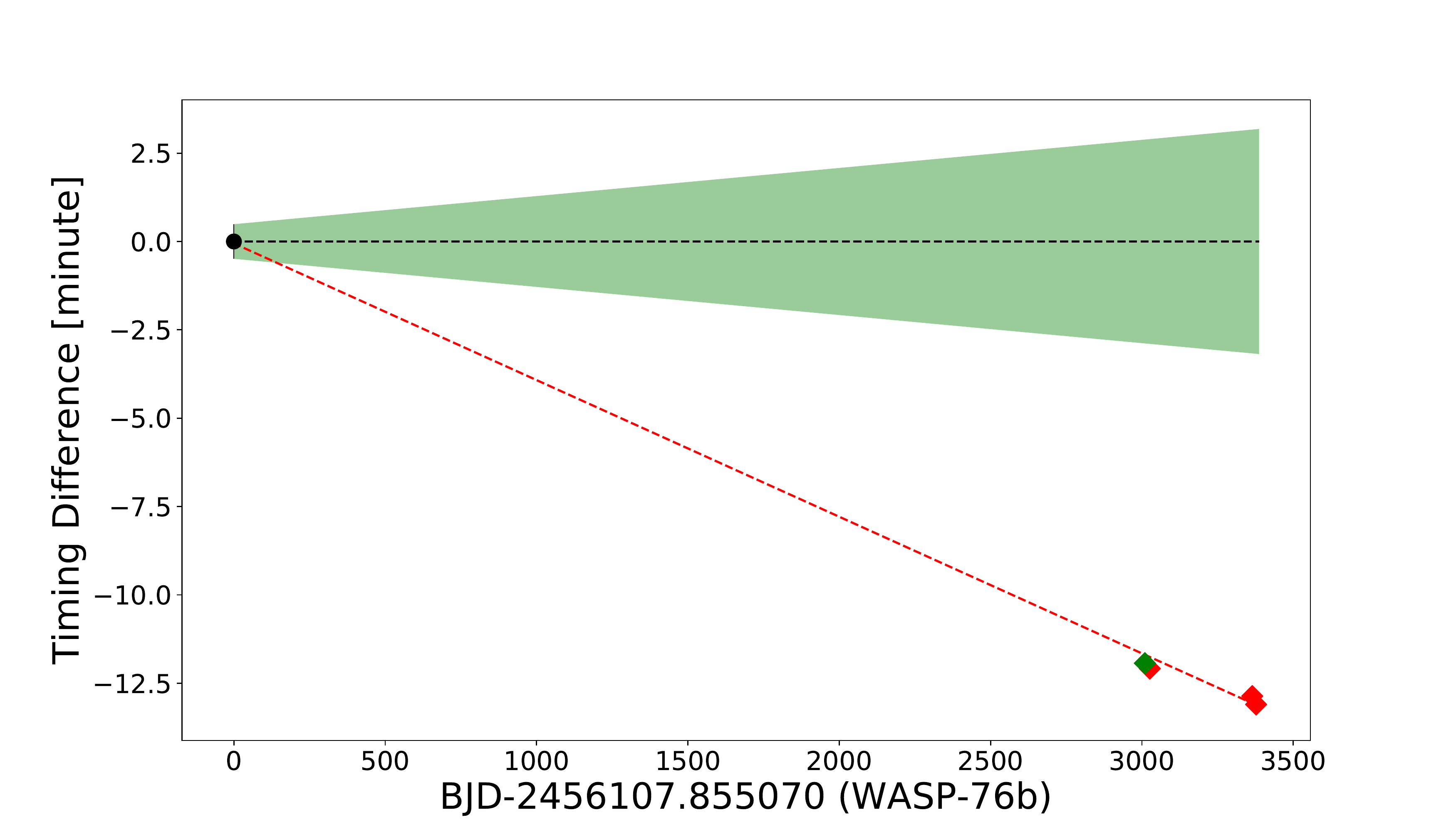}
\figsetgrpnote{WASP-76b}
\figsetgrpend
\figsetgrpstart
\figsetgrpnum{1.6}
\figsetgrptitle{WASP-95b}
\figsetplot{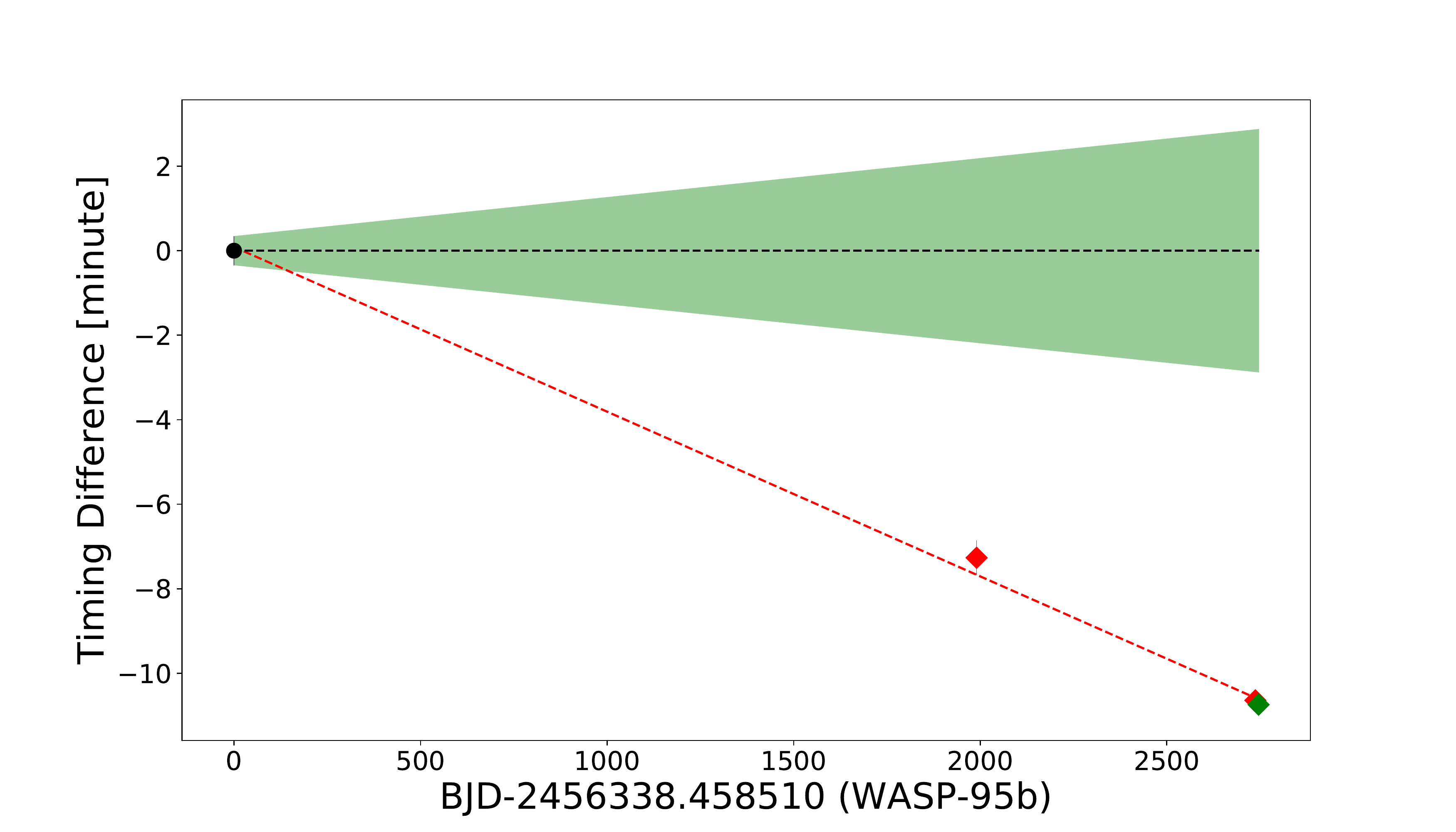}
\figsetgrpnote{WASP-95b}
\figsetgrpend
\figsetgrpstart
\figsetgrpnum{1.7}
\figsetgrptitle{WASP-101b}
\figsetplot{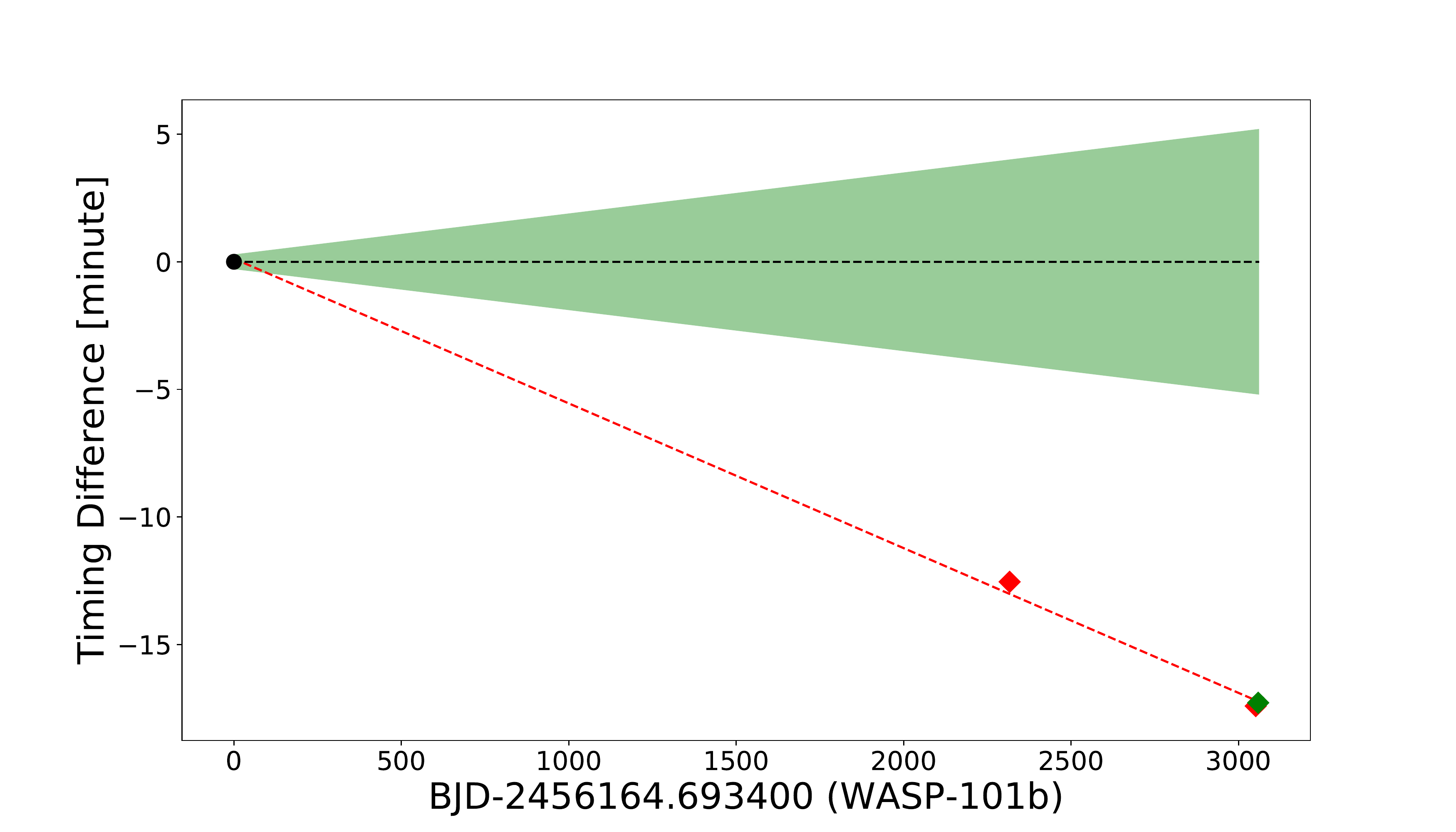}
\figsetgrpnote{WASP-101b}
\figsetgrpend
\figsetgrpstart
\figsetgrpnum{1.8}
\figsetgrptitle{WASP-35b}
\figsetplot{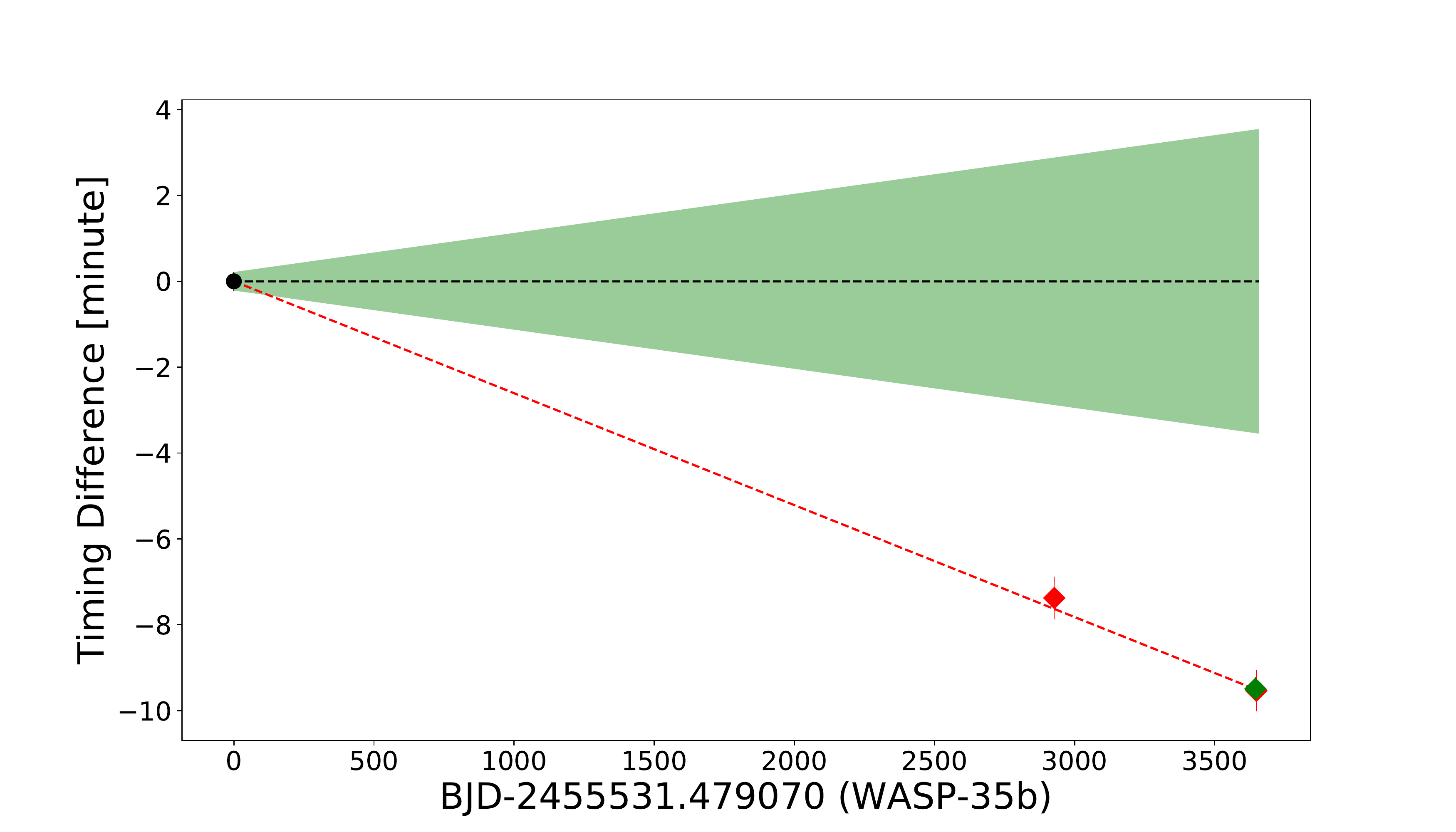}
\figsetgrpnote{WASP-35b}
\figsetgrpend
\figsetgrpstart
\figsetgrpnum{1.9}
\figsetgrptitle{TOI-163b}
\figsetplot{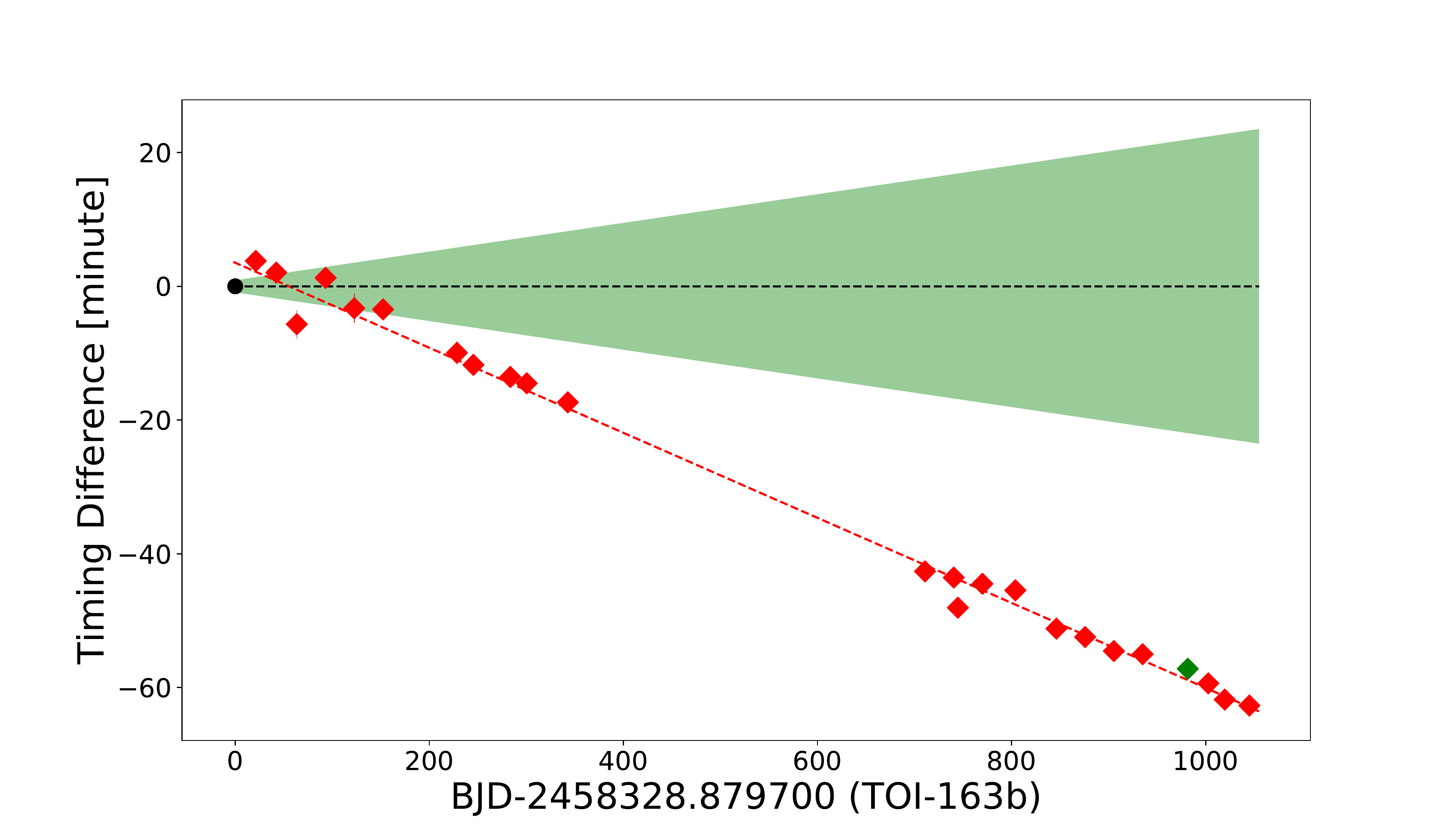}
\figsetgrpnote{TOI-163b}
\figsetgrpend
\figsetgrpstart
\figsetgrpnum{1.10}
\figsetgrptitle{KELT-14b}
\figsetplot{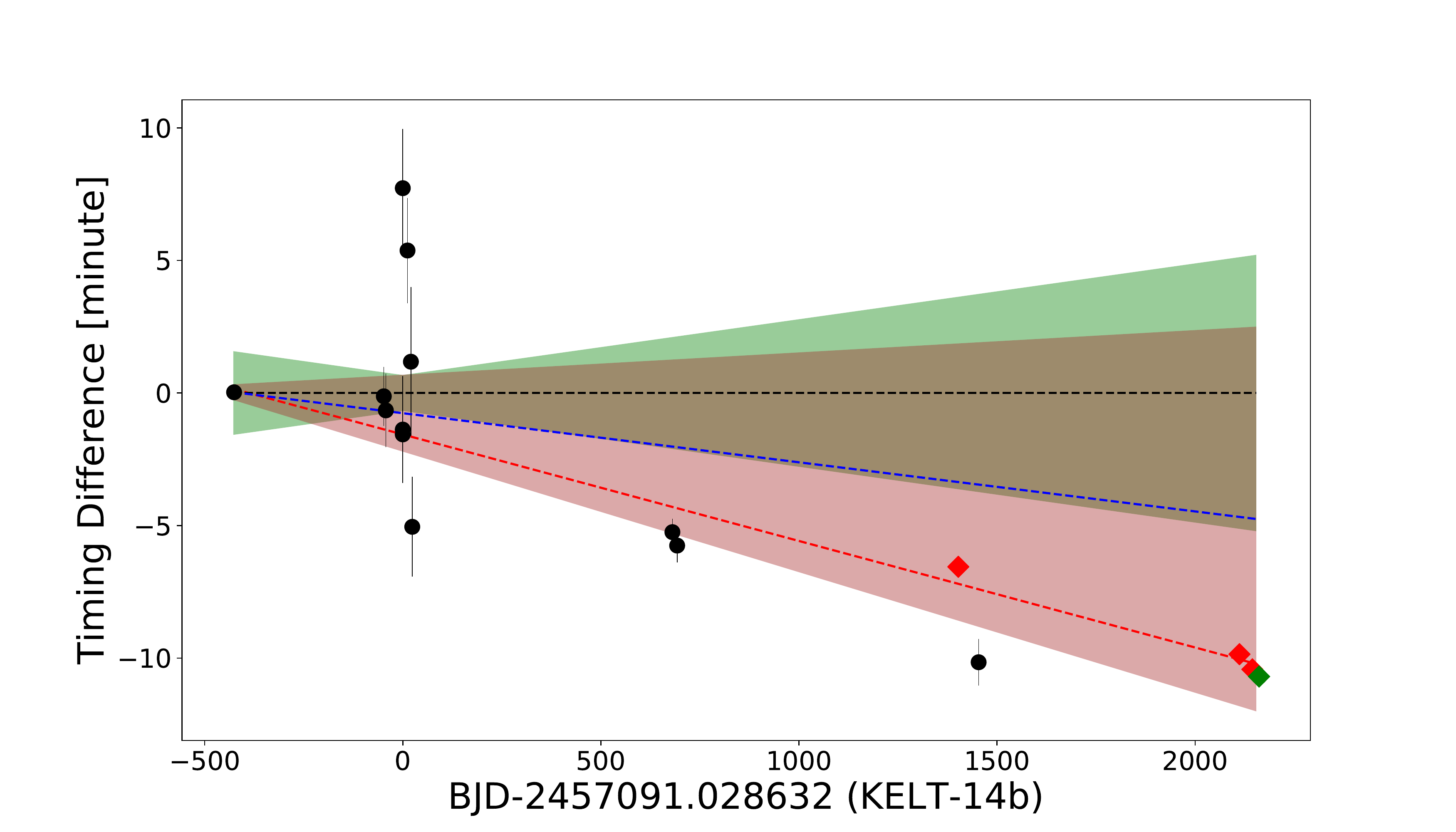}
\figsetgrpnote{KELT-14b}
\figsetgrpend
\figsetgrpstart
\figsetgrpnum{1.11}
\figsetgrptitle{KELT-7b}
\figsetplot{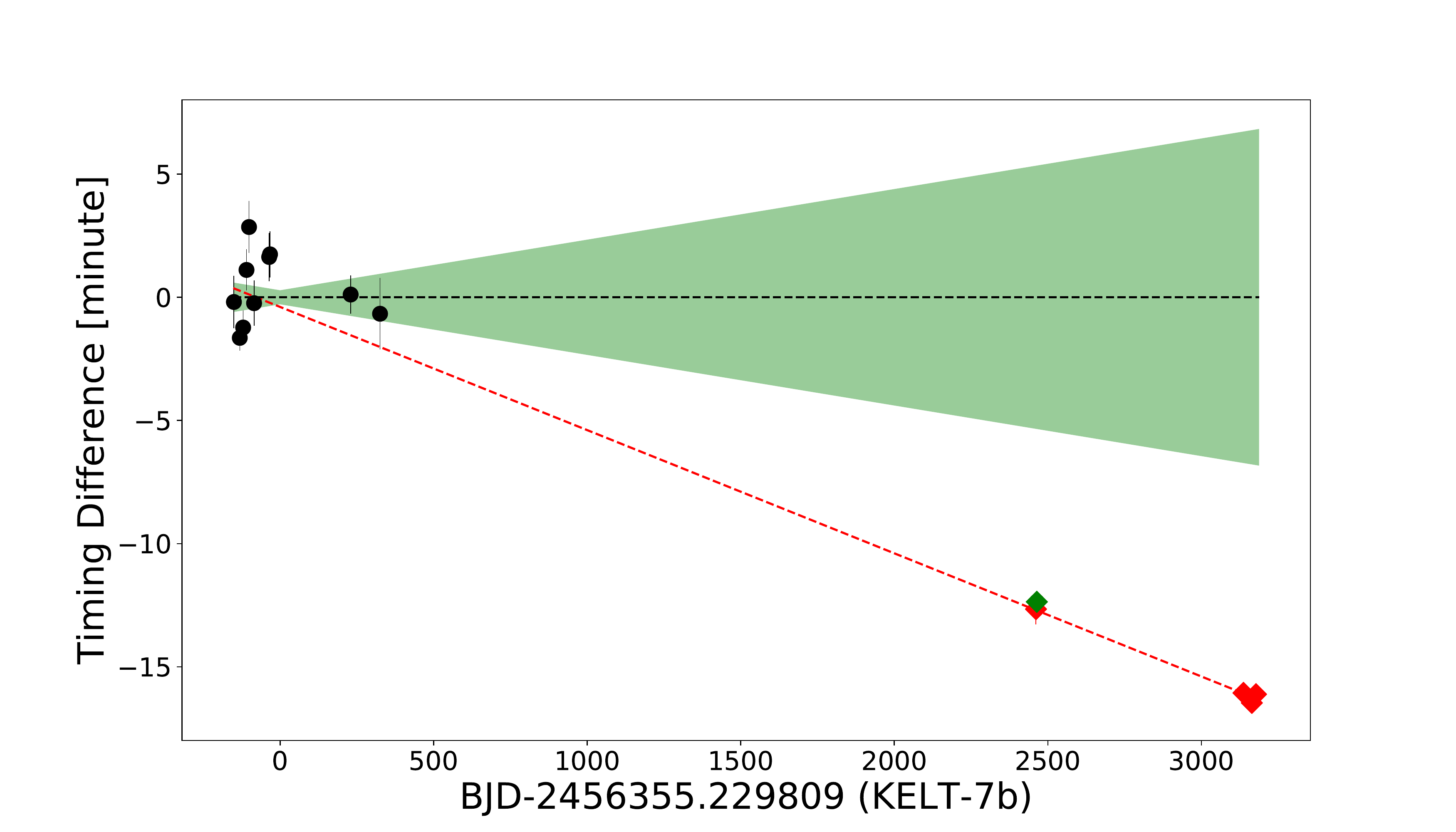}
\figsetgrpnote{KELT-7b}
\figsetgrpend
\figsetgrpstart
\figsetgrpnum{1.12}
\figsetgrptitle{HAT-P-31b}
\figsetplot{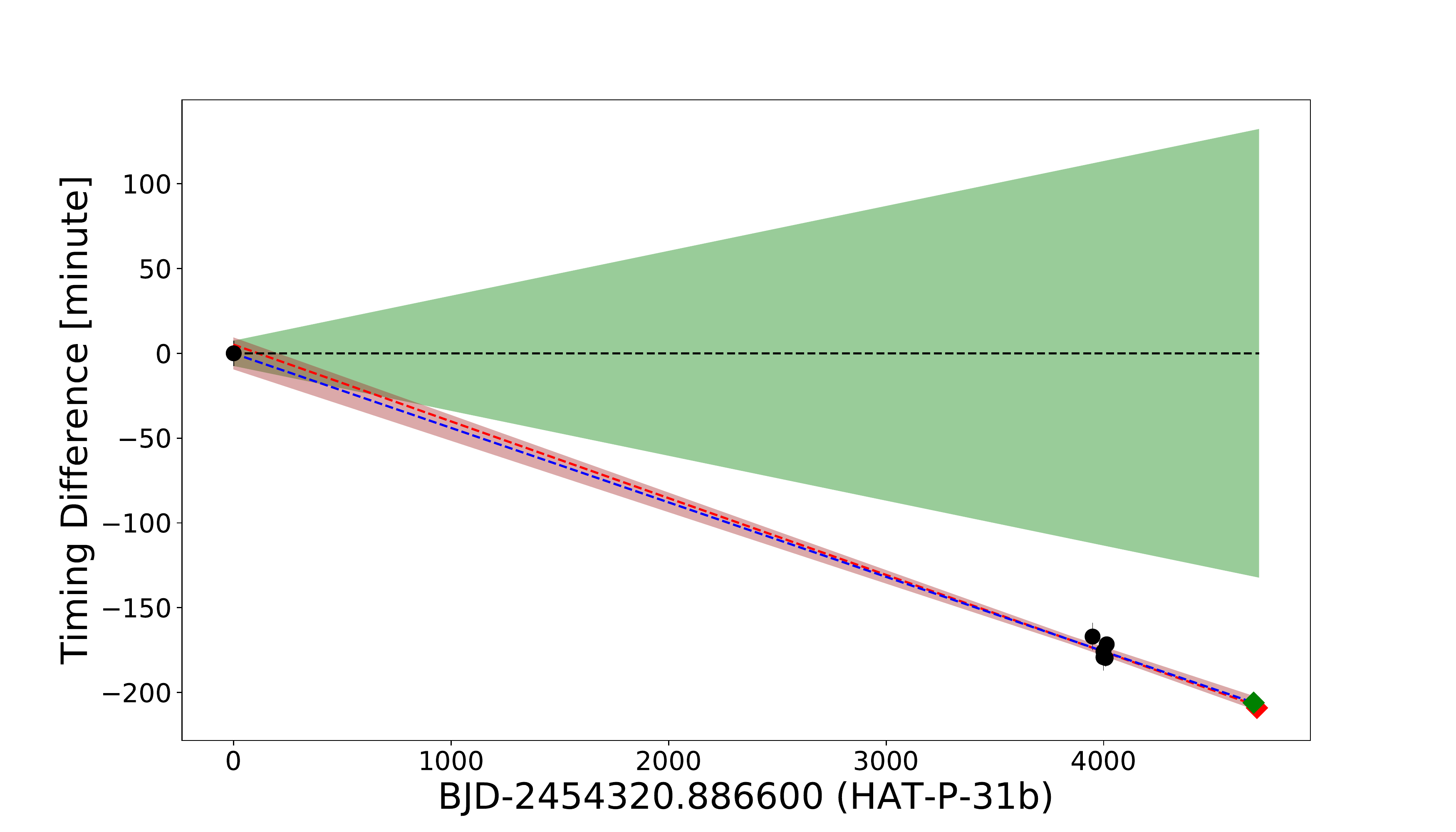}
\figsetgrpnote{HAT-P-31b}
\figsetgrpend
\figsetgrpstart
\figsetgrpnum{1.13}
\figsetgrptitle{KELT-1b}
\figsetplot{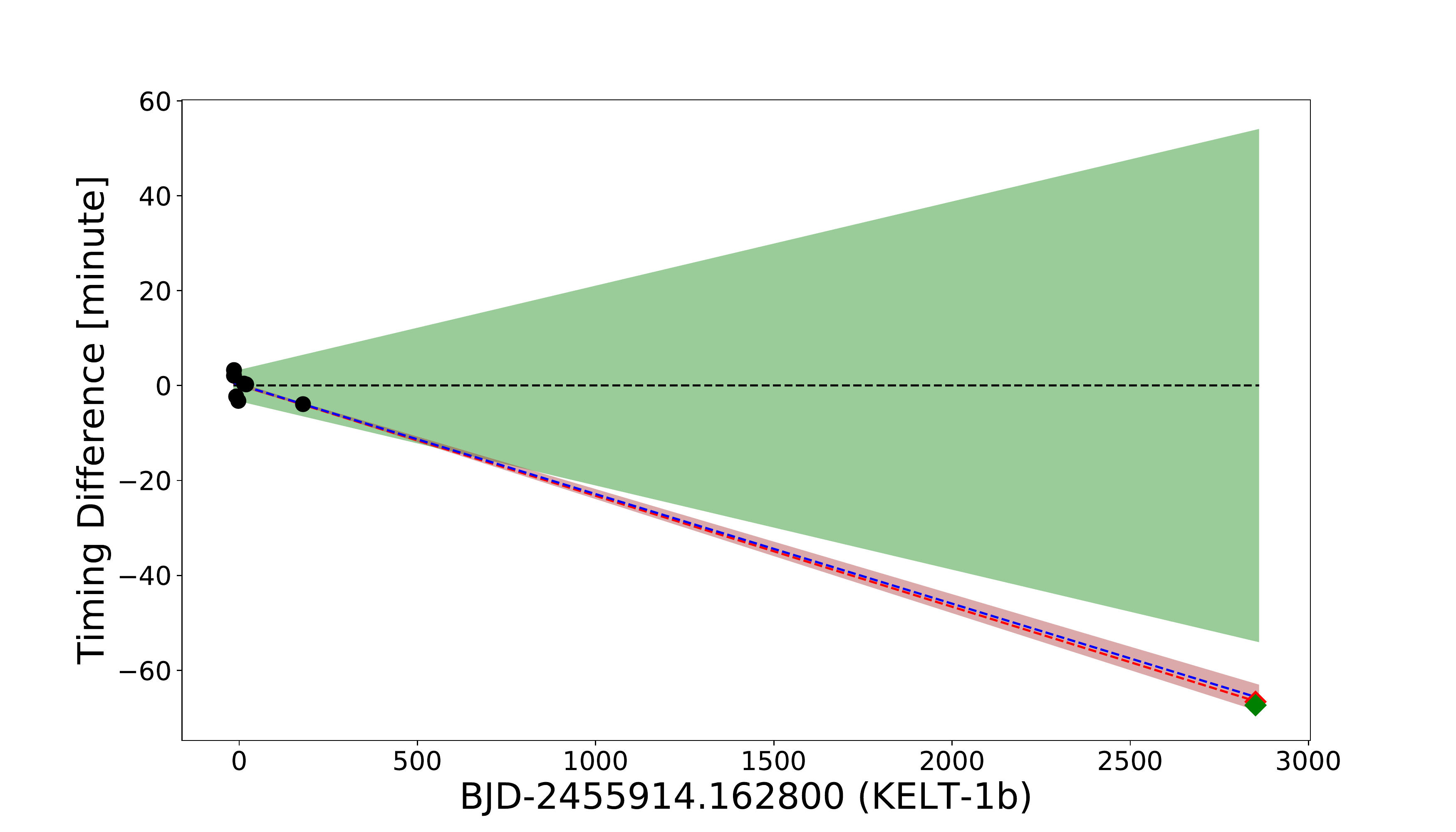}
\figsetgrpnote{KELT-1b}
\figsetgrpend
\figsetgrpstart
\figsetgrpnum{1.14}
\figsetgrptitle{KELT-21b}
\figsetplot{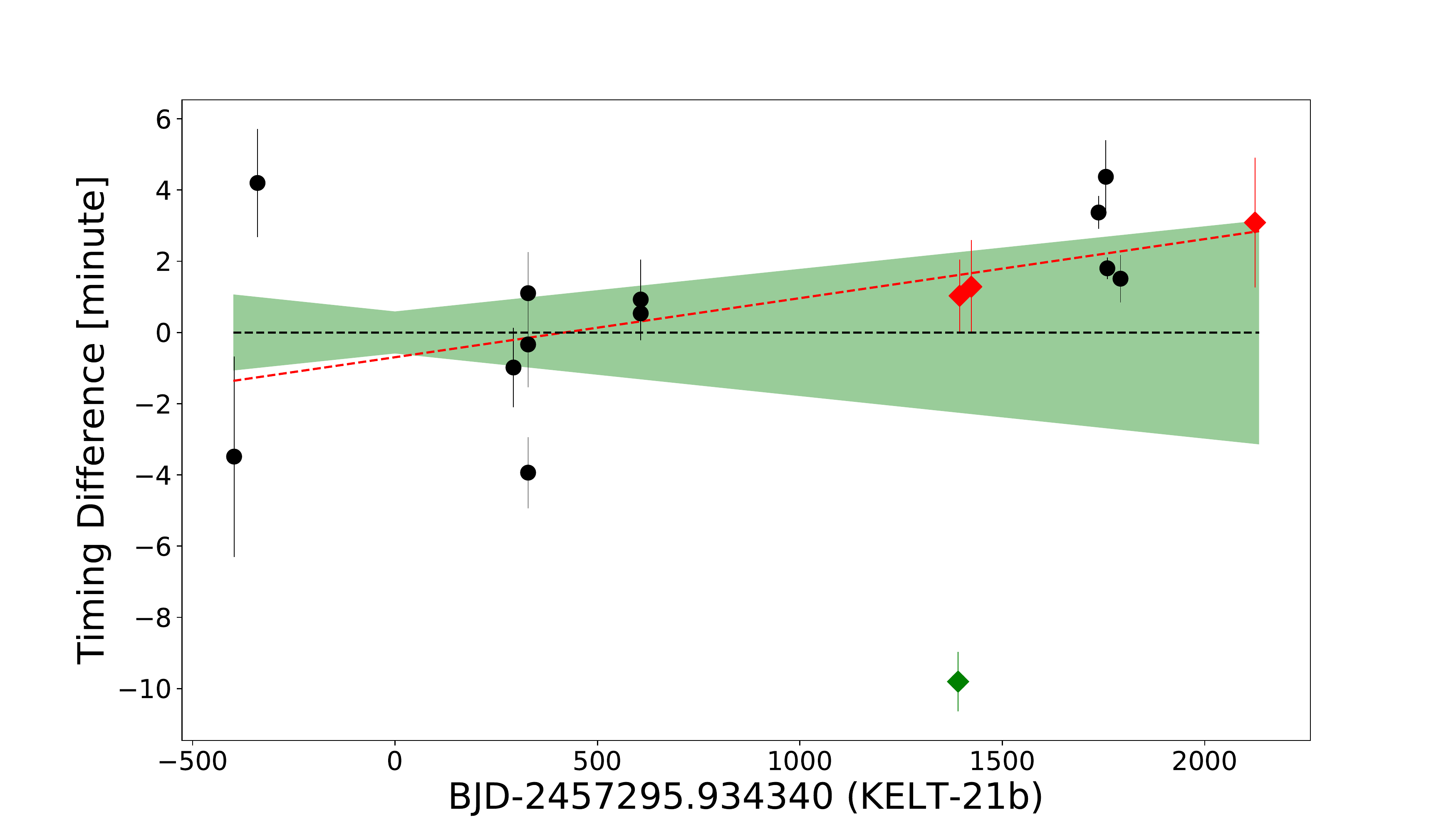}
\figsetgrpnote{KELT-21b}
\figsetgrpend
\figsetgrpstart
\figsetgrpnum{1.15}
\figsetgrptitle{WASP-17b}
\figsetplot{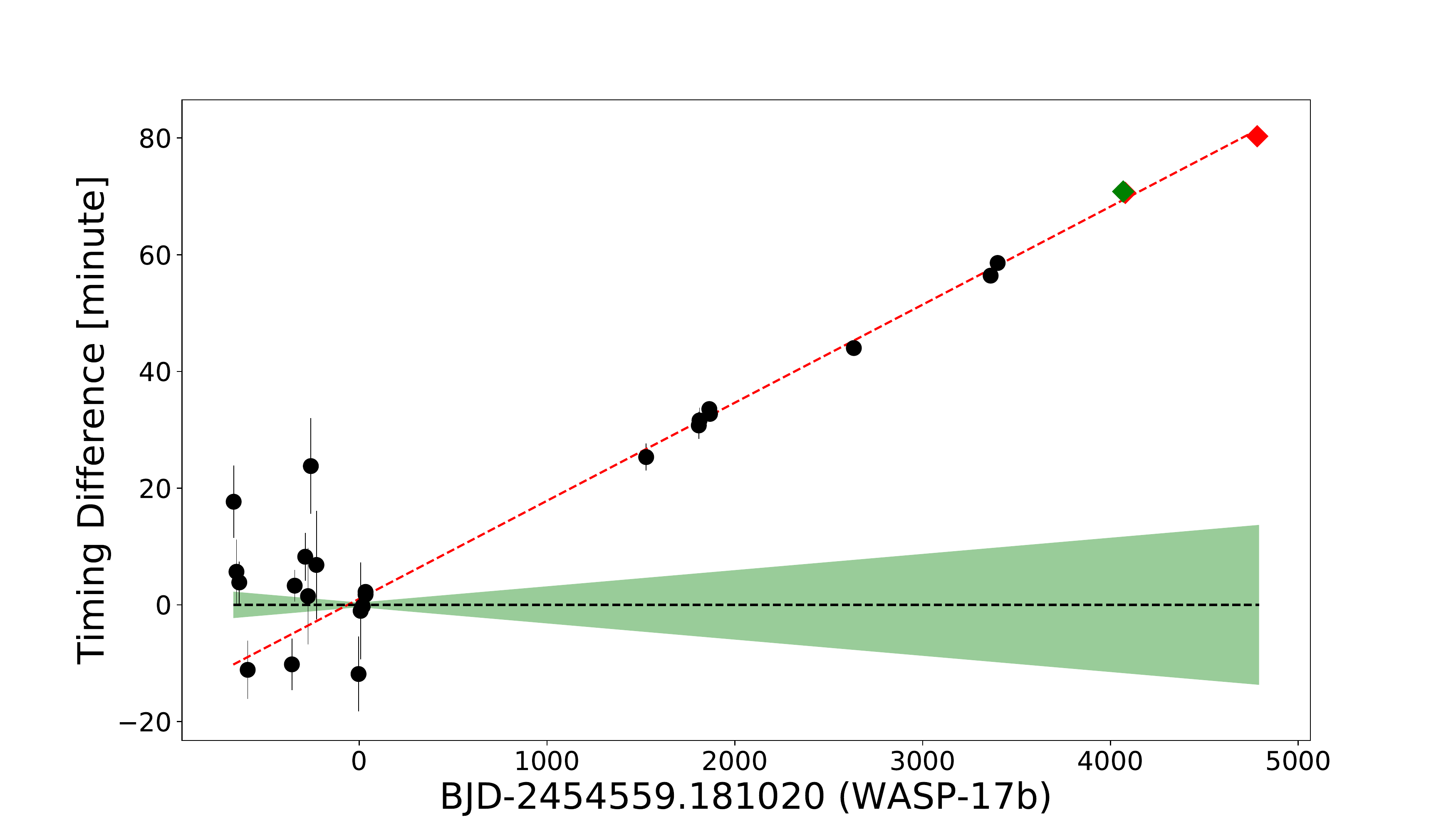}
\figsetgrpnote{WASP-17b}
\figsetgrpend
\figsetgrpstart
\figsetgrpnum{1.16}
\figsetgrptitle{WASP-178b}
\figsetplot{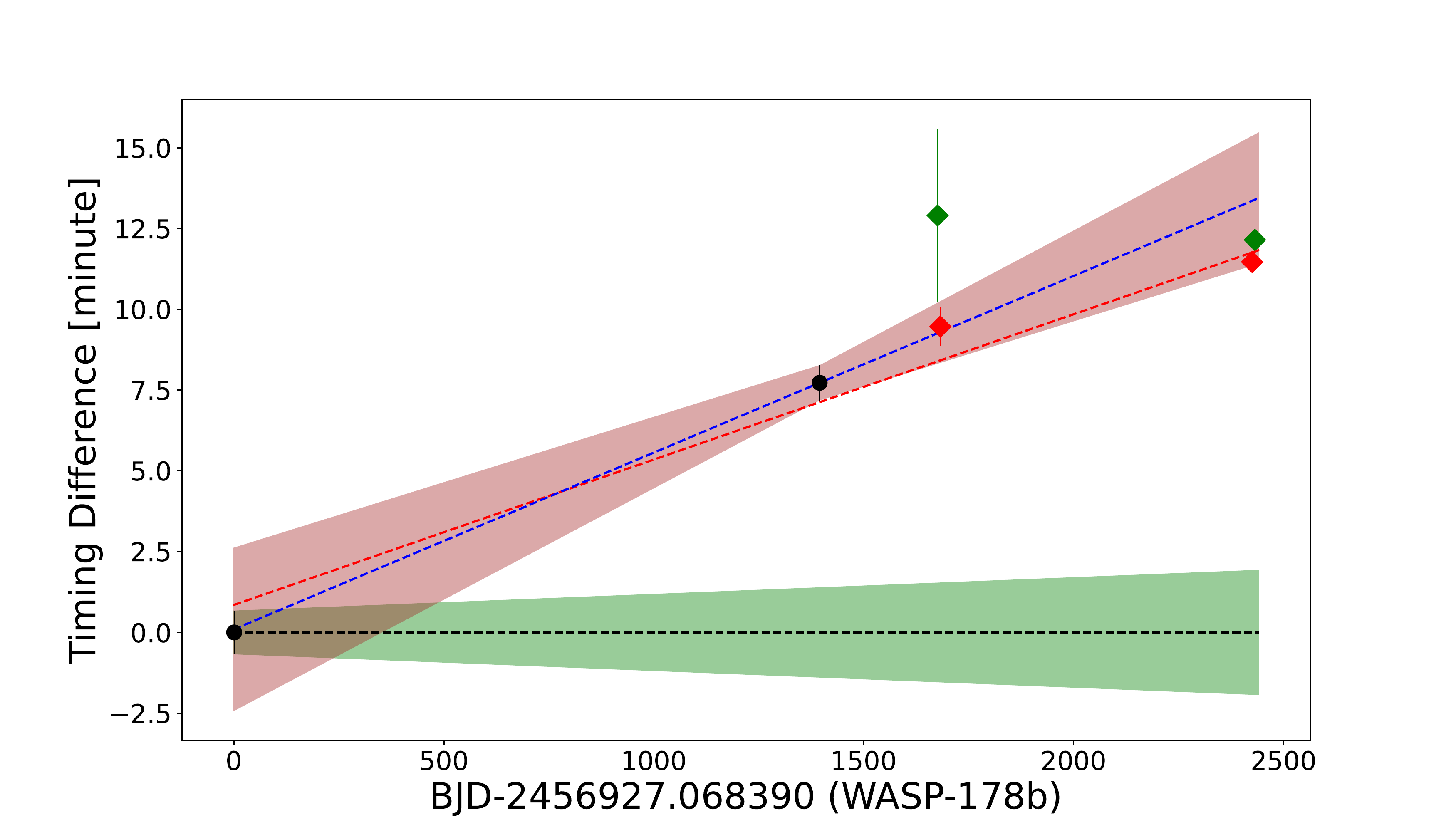}
\figsetgrpnote{WASP-178b}
\figsetgrpend
\figsetgrpstart
\figsetgrpnum{1.17}
\figsetgrptitle{KELT-19Ab}
\figsetplot{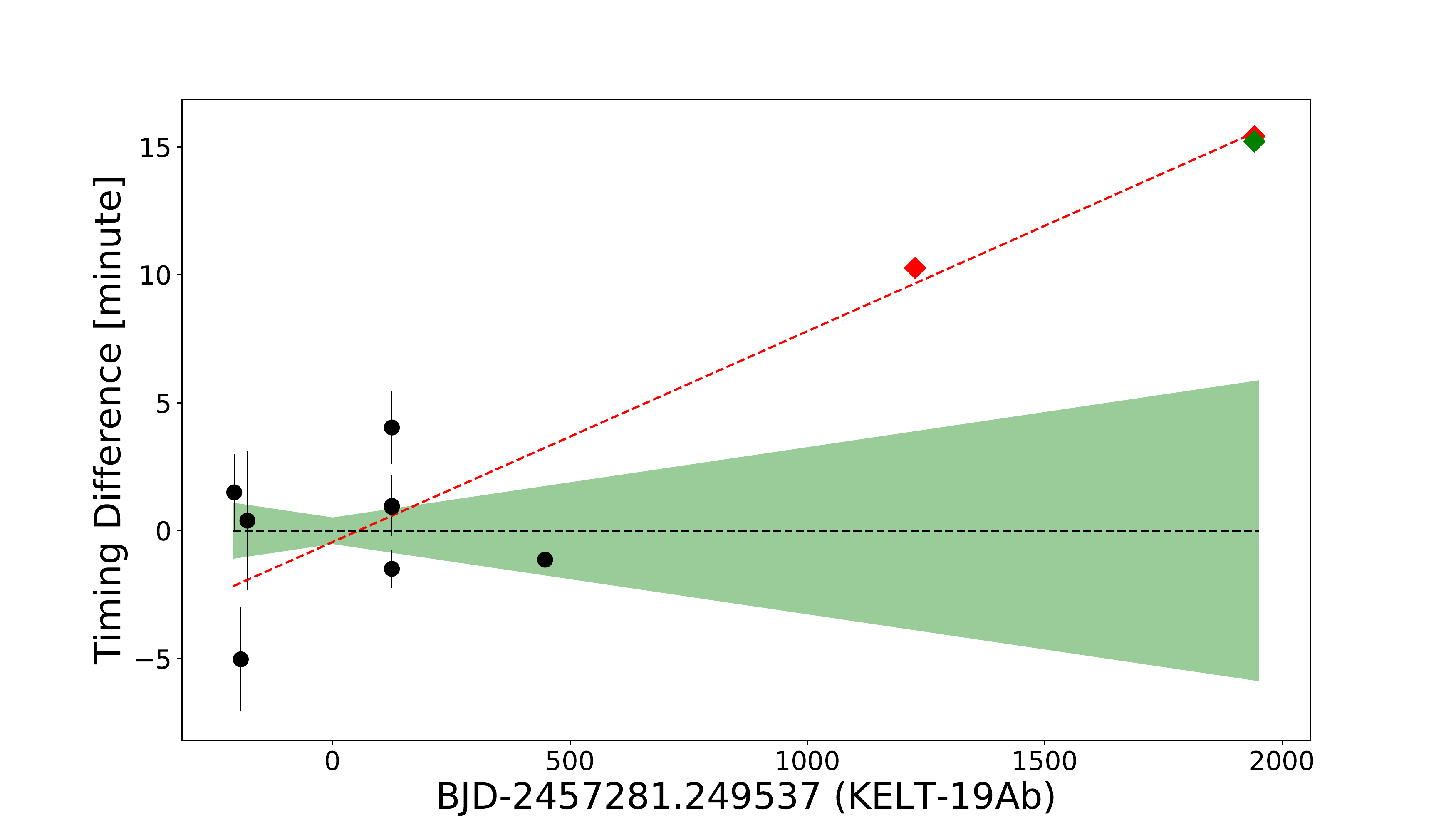}
\figsetgrpnote{KELT-19Ab}
\figsetgrpend
\figsetgrpstart
\figsetgrpnum{1.18}
\figsetgrptitle{WASP-33b}
\figsetplot{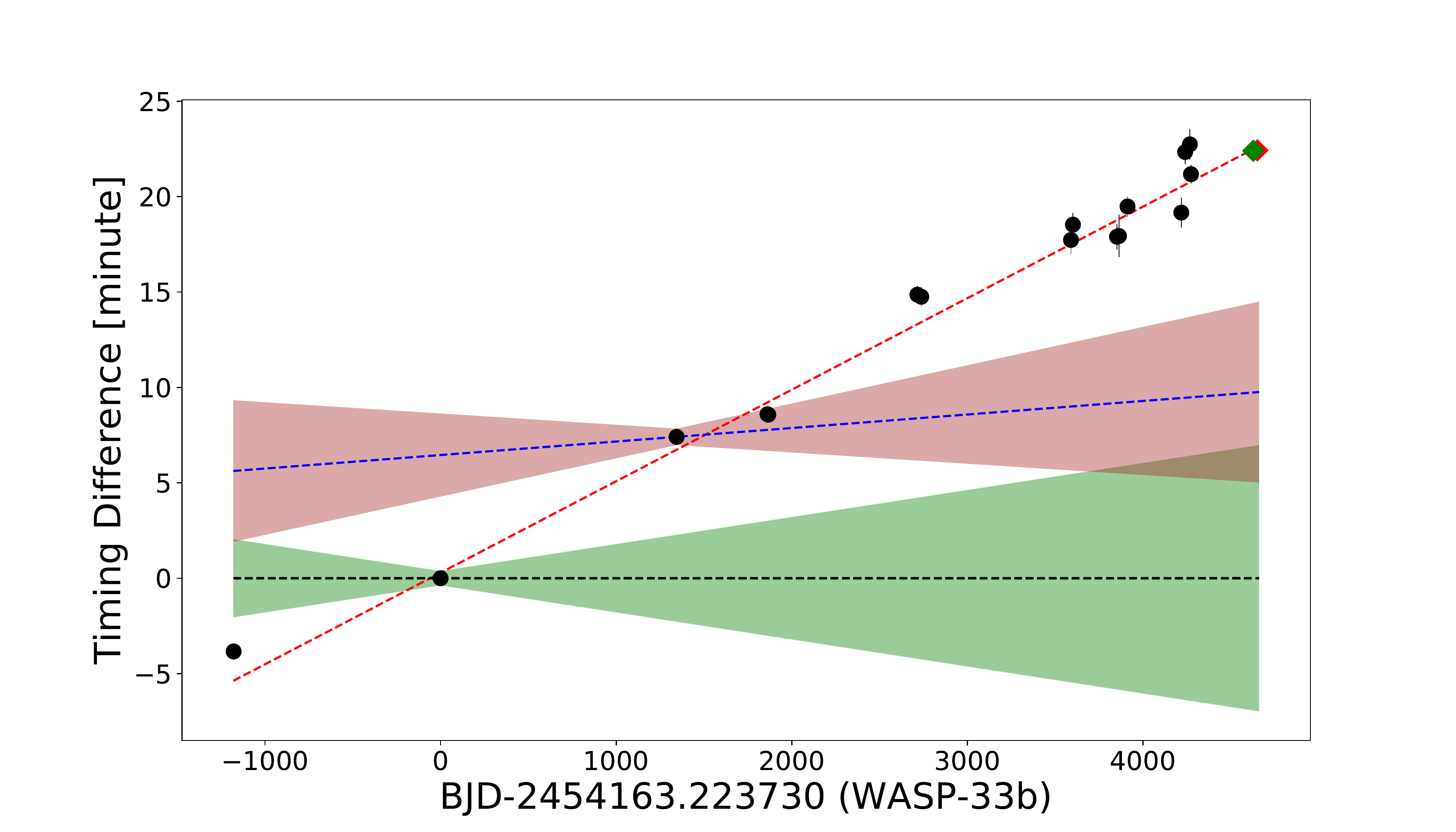}
\figsetgrpnote{WASP-33b}
\figsetgrpend
\figsetgrpstart
\figsetgrpnum{1.19}
\figsetgrptitle{HAT-P-6b}
\figsetplot{HAT-P-6b.pdf}
\figsetgrpnote{HAT-P-6b}
\figsetgrpend
\figsetend
\figsetgrpnum{1.20}
\figsetgrptitle{WASP-94Ab}
\figsetplot{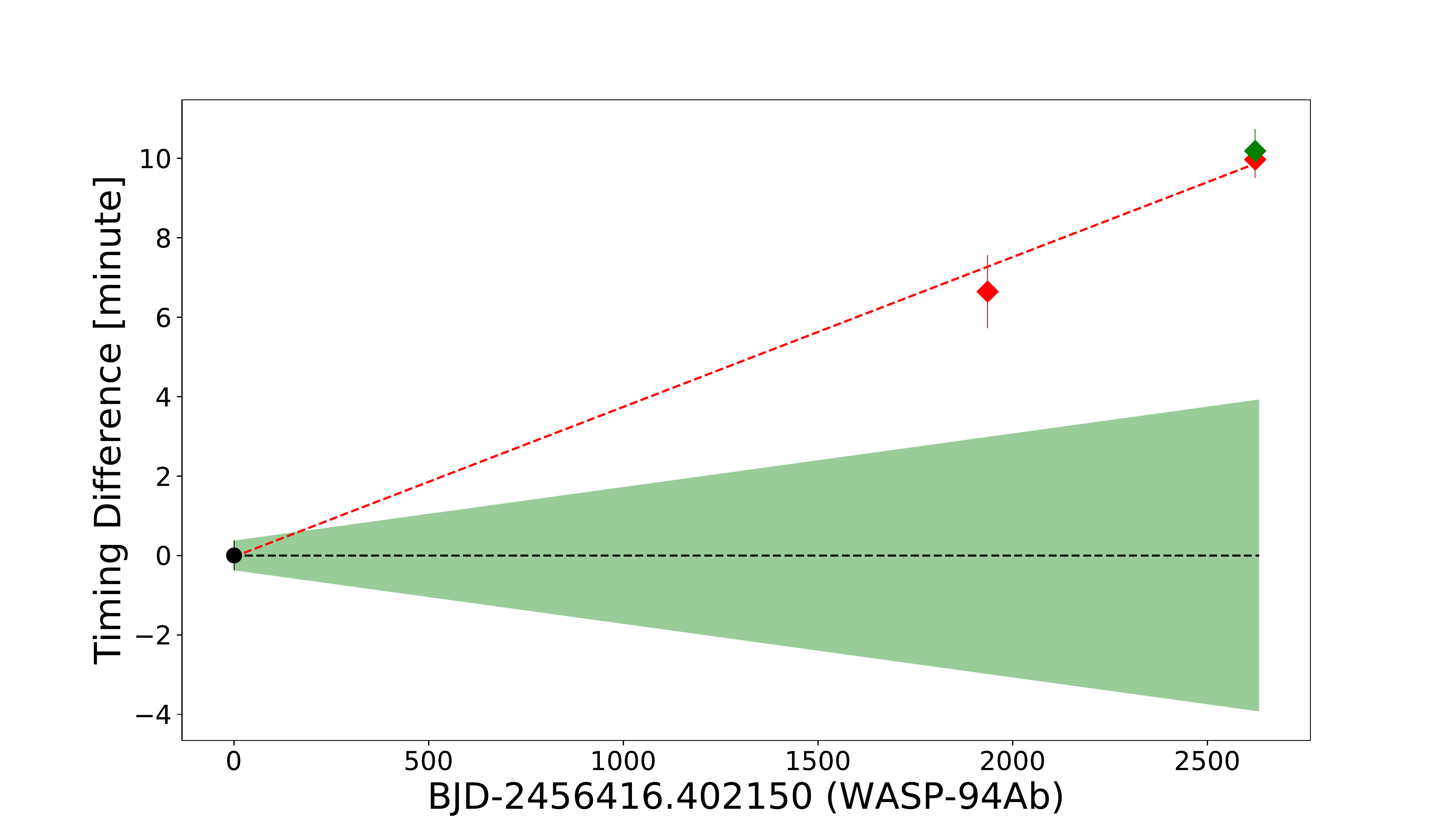}
\figsetgrpnote{WASP-94Ab}
\figsetgrpend
\figsetgrpstart
\figsetgrpnum{1.21}
\figsetgrptitle{WASP-58b}
\figsetplot{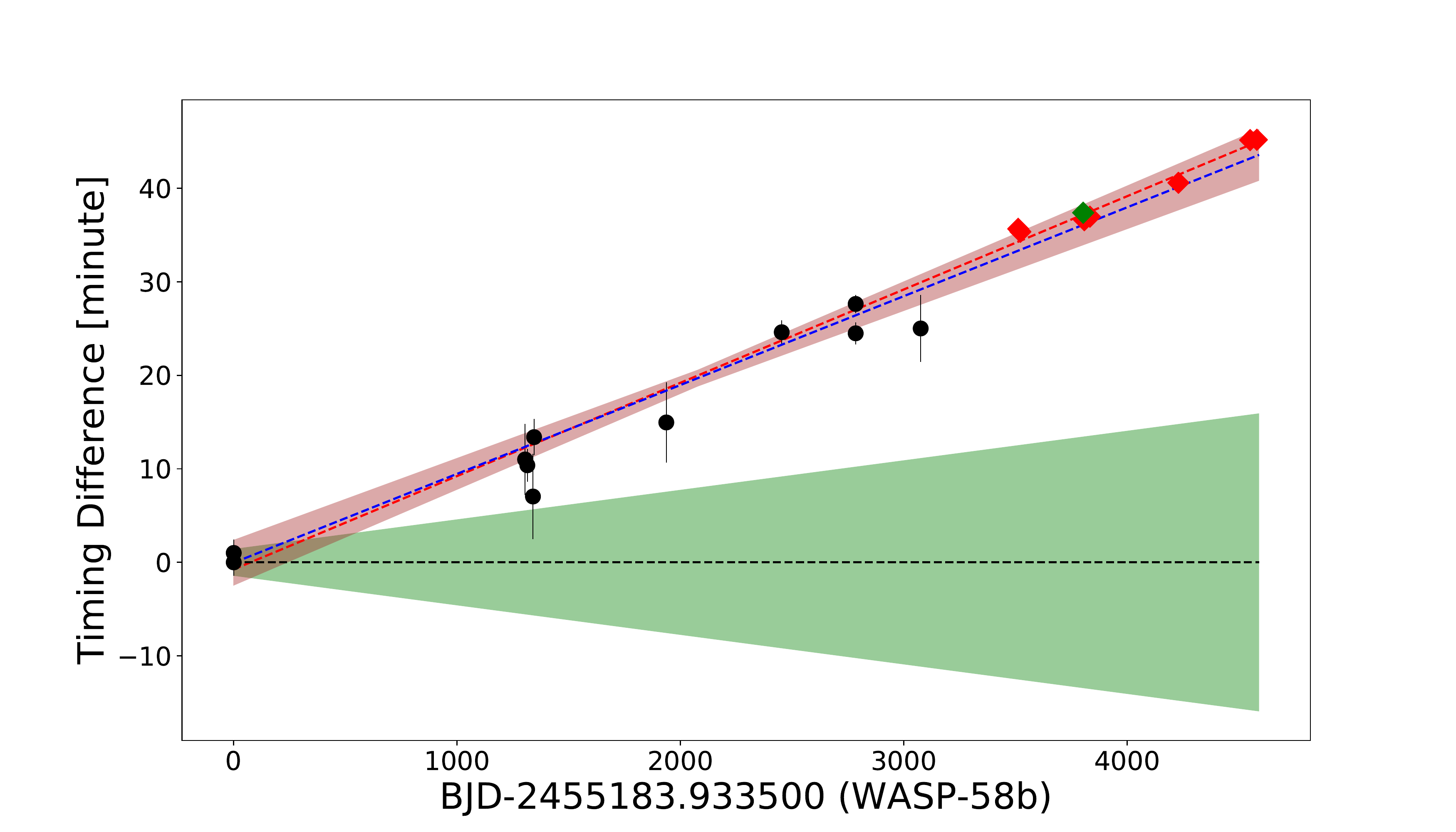}
\figsetgrpnote{WASP-58b}
\figsetgrpend
\figsetgrpstart
\figsetgrpnum{1.22}
\figsetgrptitle{WASP-99b}
\figsetplot{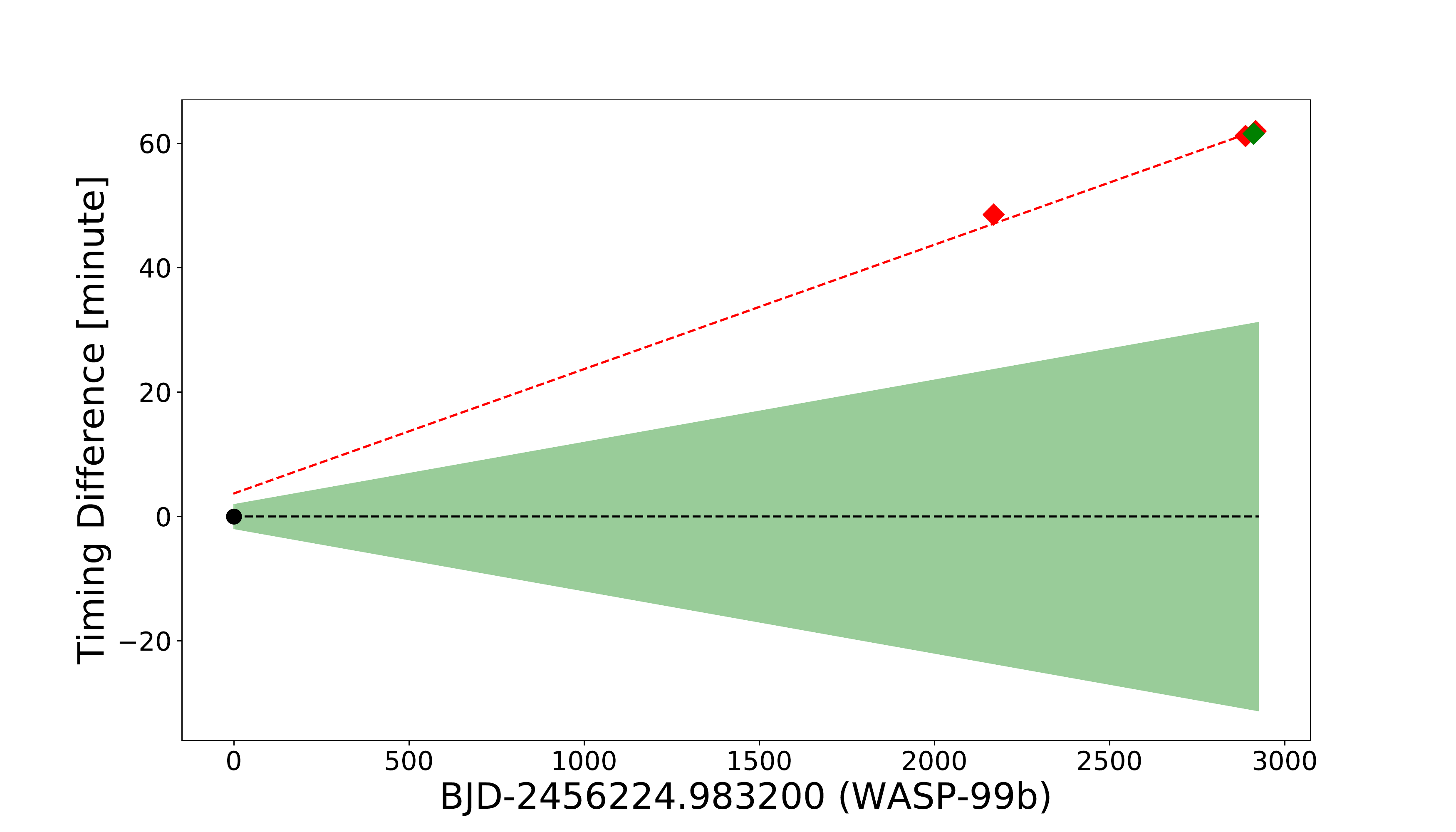}
\figsetgrpnote{WASP-99b}
\figsetgrpend
\figsetgrpstart
\figsetgrpnum{1.23}
\figsetgrptitle{TOI-1333b}
\figsetplot{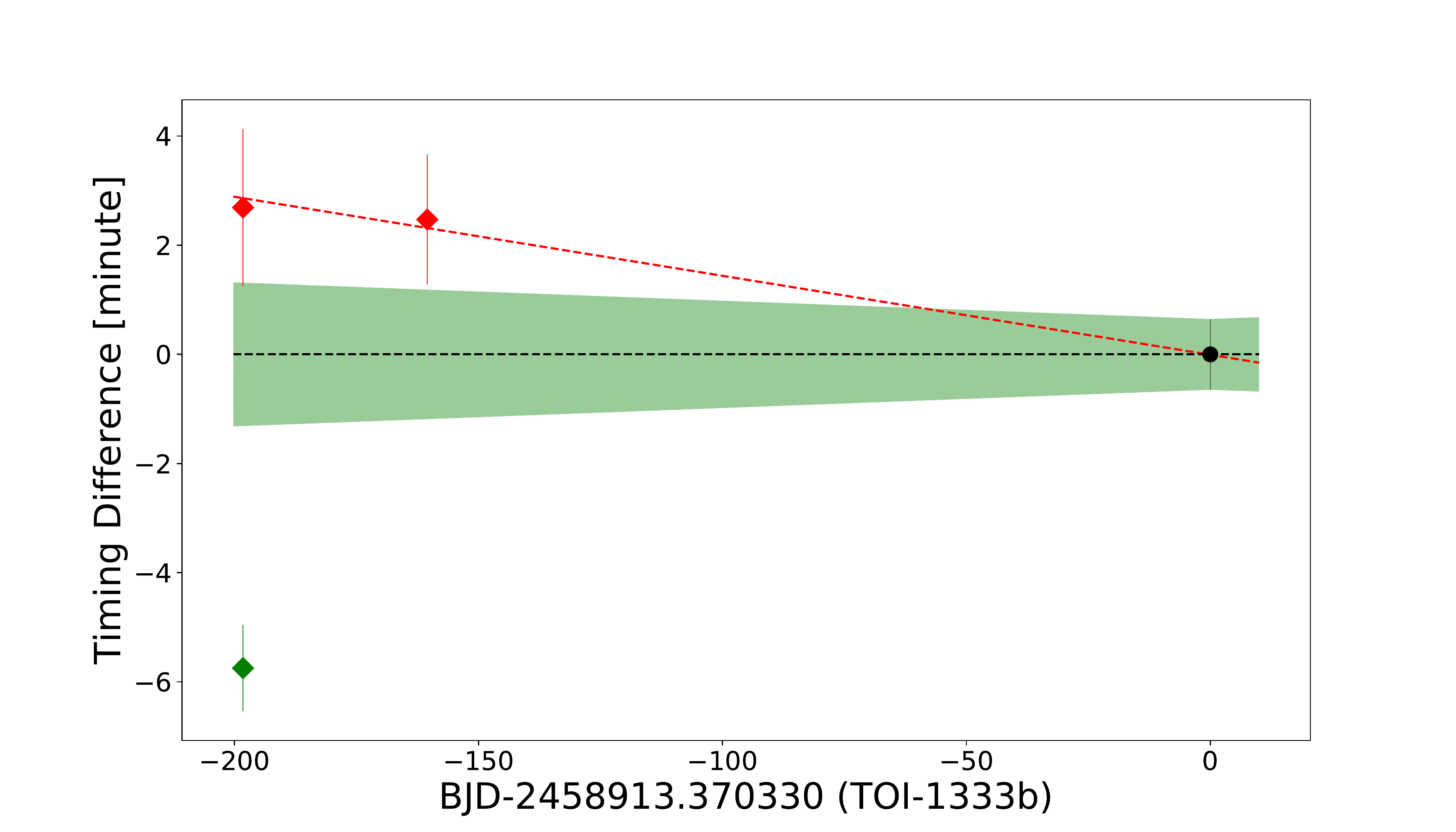}
\figsetgrpnote{TOI-1333b}
\figsetgrpend
\figsetgrpstart
\figsetgrpnum{1.24}
\figsetgrptitle{WASP-78b}
\figsetplot{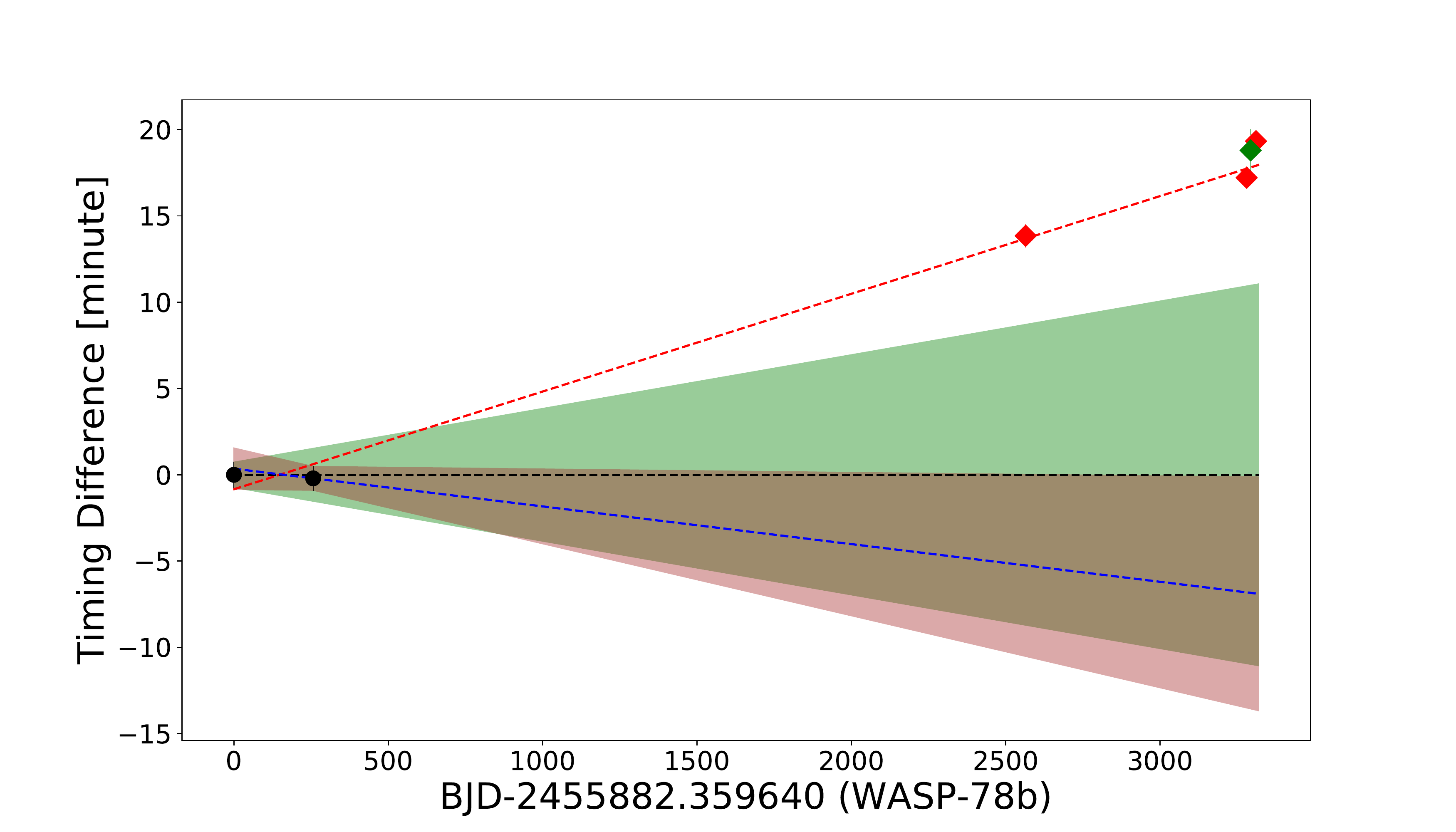}
\figsetgrpnote{WASP-78b}
\figsetgrpend
\figsetgrpstart
\figsetgrpnum{1.25}
\figsetgrptitle{WASP-173Ab}
\figsetplot{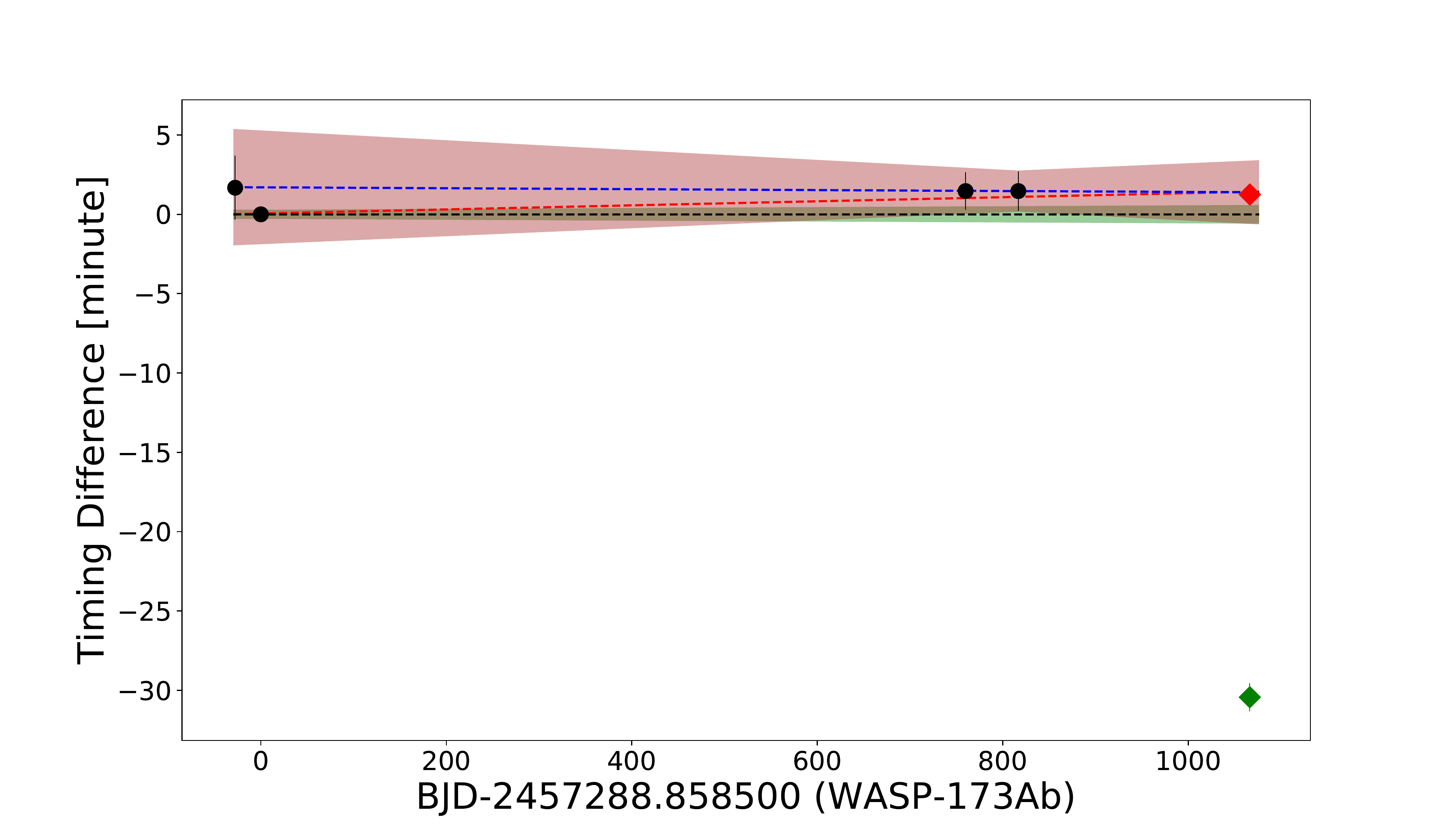}
\figsetgrpnote{WASP-173Ab}
\figsetgrpend
\figsetgrpstart
\figsetgrpnum{1.26}
\figsetgrptitle{TOI-628b}
\figsetplot{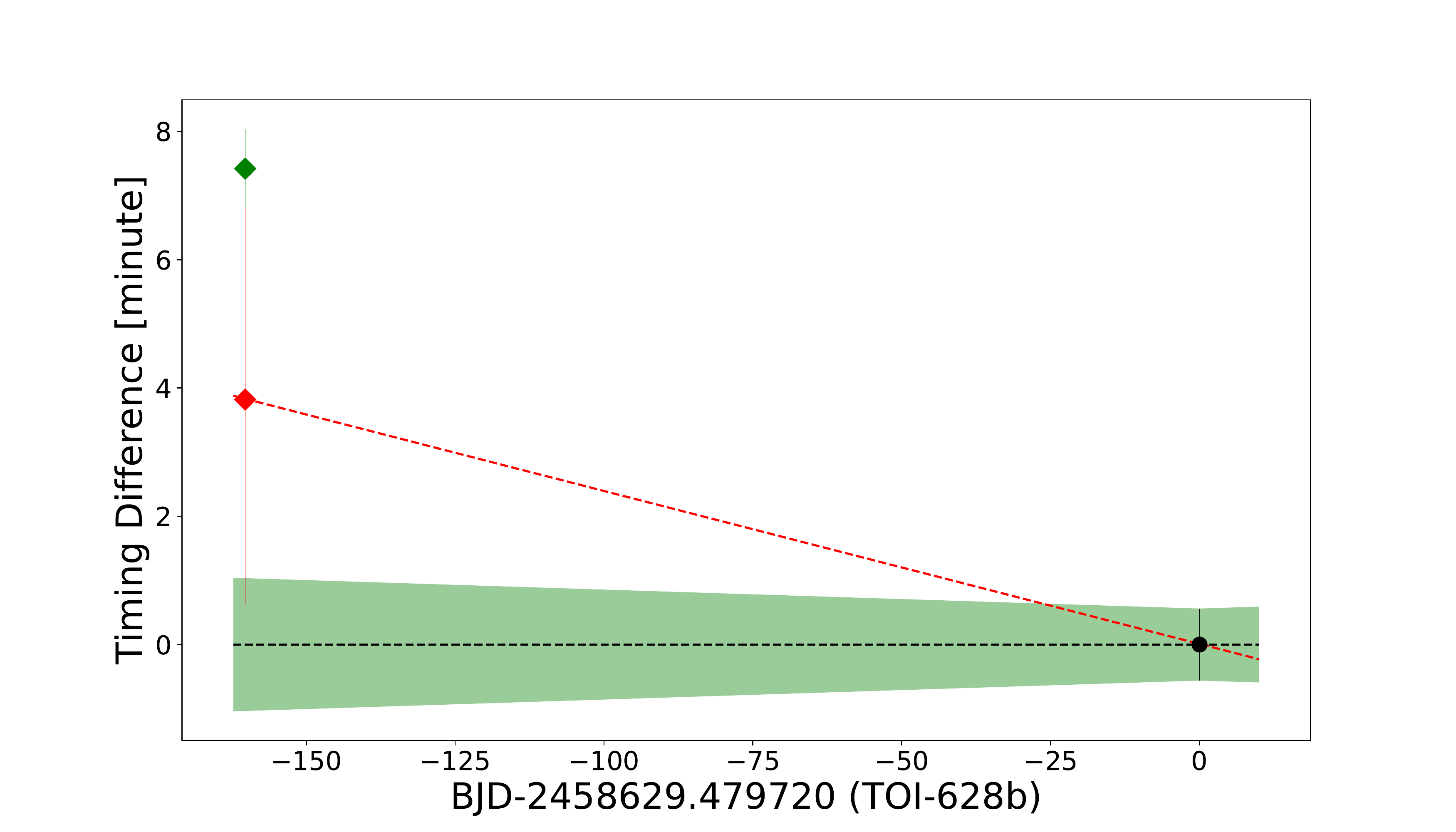}
\figsetgrpnote{TOI-628b}
\figsetgrpend
\figsetgrpstart
\figsetgrpnum{1.27}
\figsetgrptitle{KELT-24b}
\figsetplot{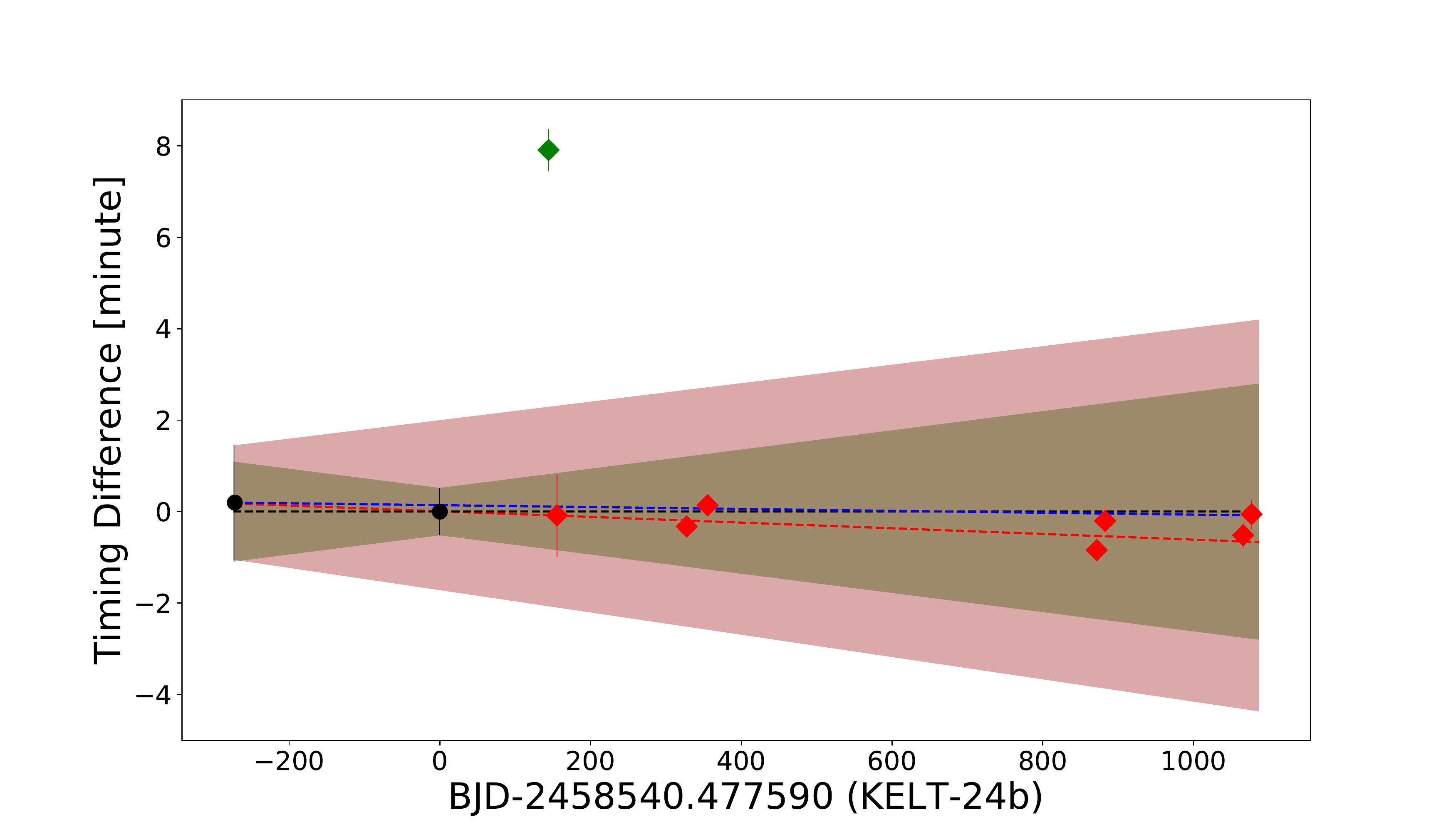}
\figsetgrpnote{KELT-24b}
\figsetgrpend
\figsetgrpstart
\figsetgrpnum{1.28}
\figsetgrptitle{WASP-187b}
\figsetplot{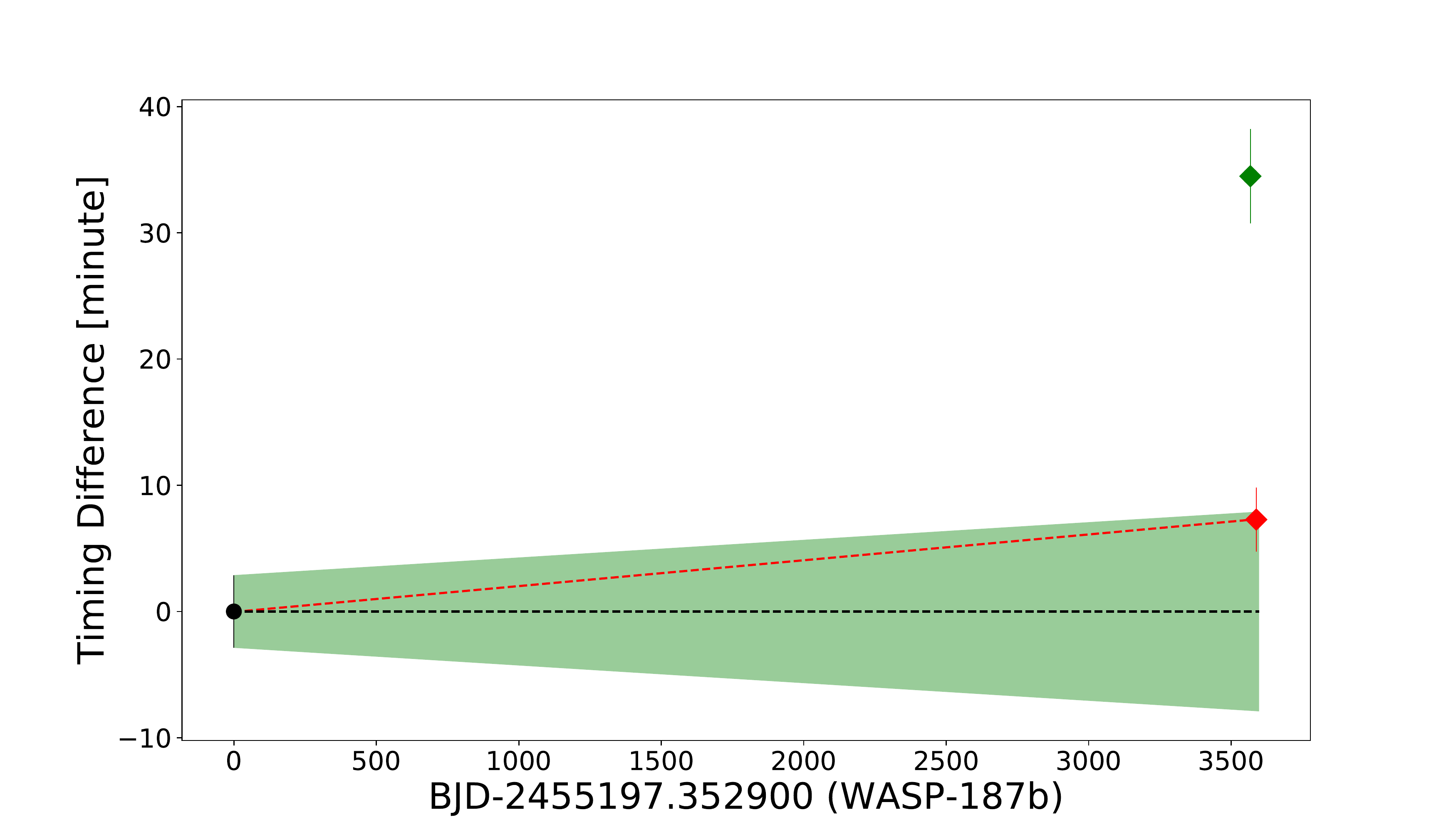}
\figsetgrpnote{WASP-187b}
\figsetgrpend
\figsetend

\figsetstart
\figsetnum{2}
\figsettitle{Timing differences of Type II targets of which timings would be modeled by a quadratic function.}
\figsetgrpstart
\figsetgrpnum{2.1}
\figsetgrptitle{XO-3b}
\figsetplot{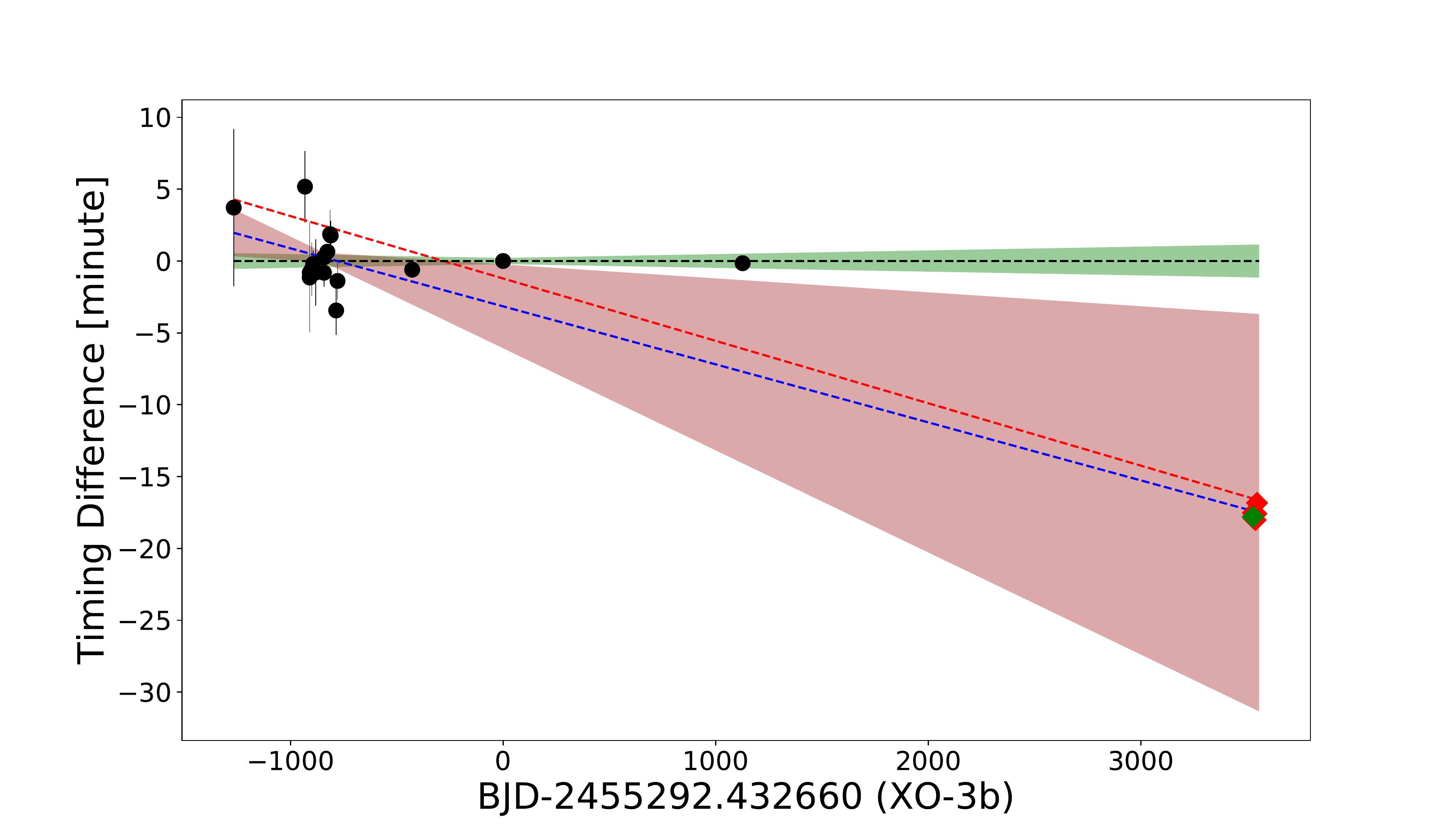}
\figsetgrpnote{XO-3b}
\figsetgrpend
\figsetend

\figsetstart
\figsetnum{3}
\figsettitle{Timing differences of Type III targets of which the timings can not be fitted with any linear or quadratic functions. The symbols are the same as Figure \ref{image:timing}.}
\figsetgrpstart
\figsetgrpnum{3.1}
\figsetgrptitle{KELT-23Ab}
\figsetplot{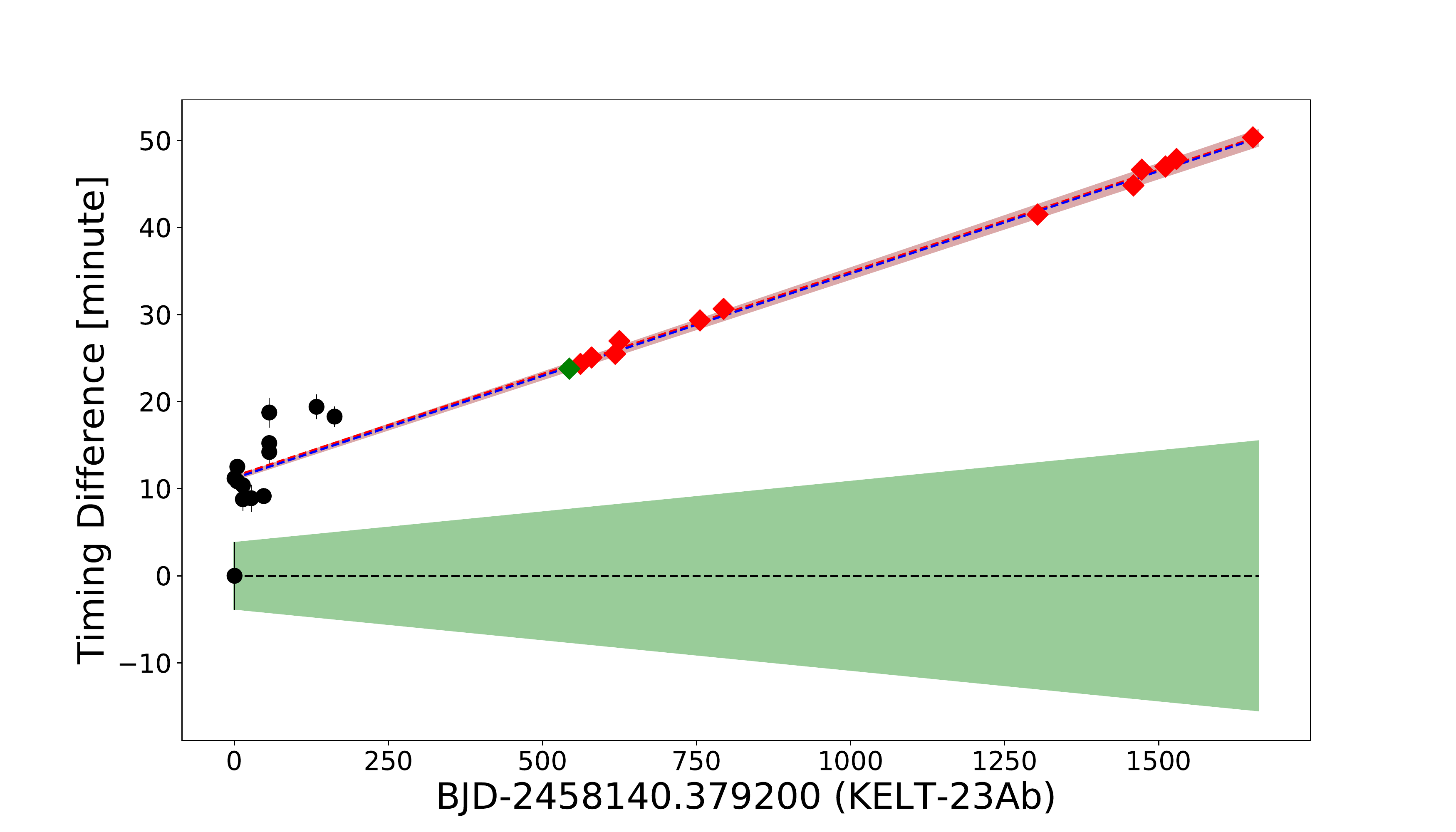}
\figsetgrpnote{KELT-23Ab}
\figsetgrpend
\figsetgrpstart
\figsetgrpnum{3.2}
\figsetgrptitle{HAT-P-69b}
\figsetplot{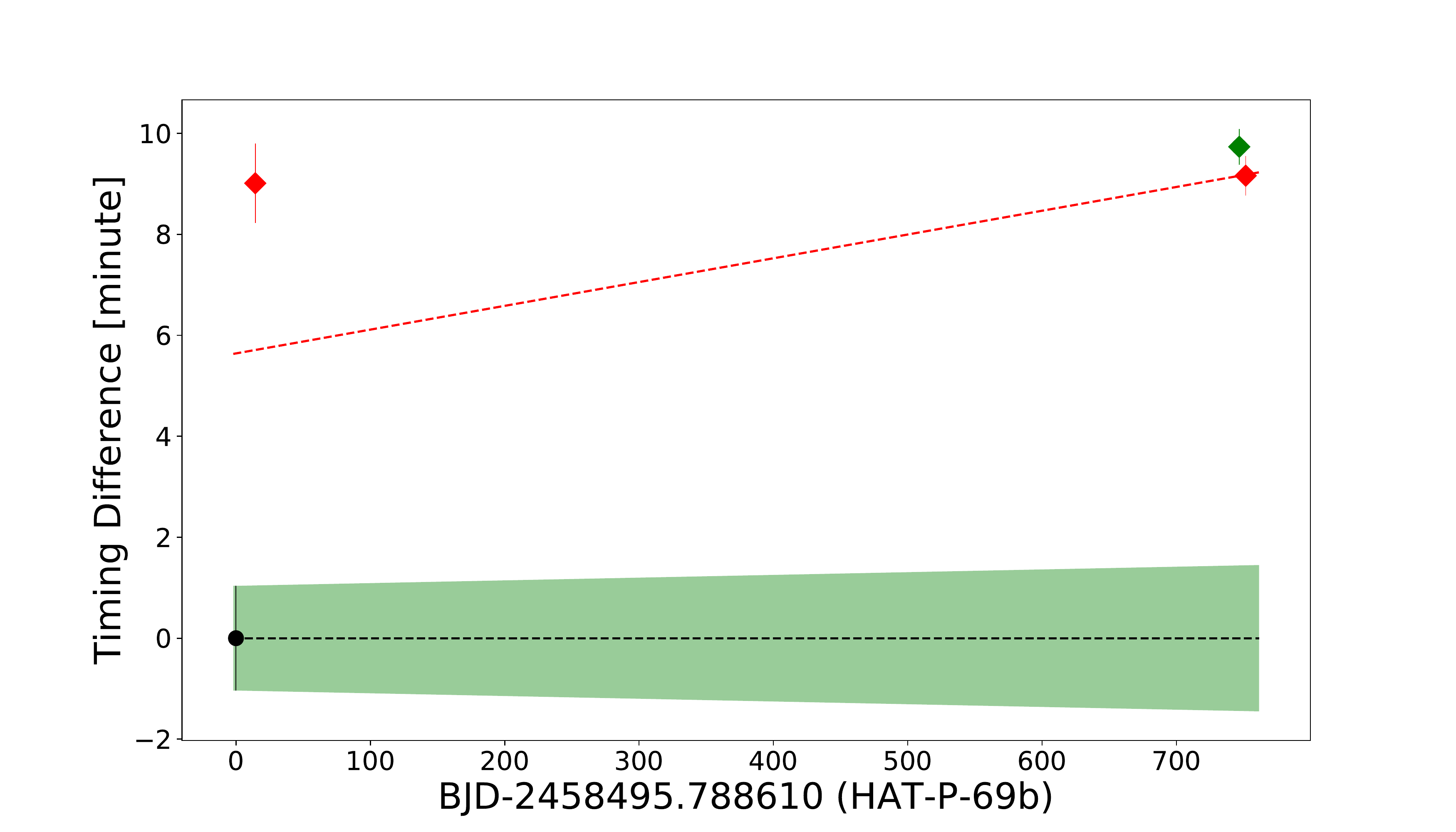}
\figsetgrpnote{HAT-P-69b}
\figsetgrpend
\figsetend

Figure A1-A3 shows the timing differences of 31 targets classified by three types. 

\renewcommand{\thefigure}{A\arabic{figure}}
\setcounter{figure}{0}

\begin{figure*}[ht]
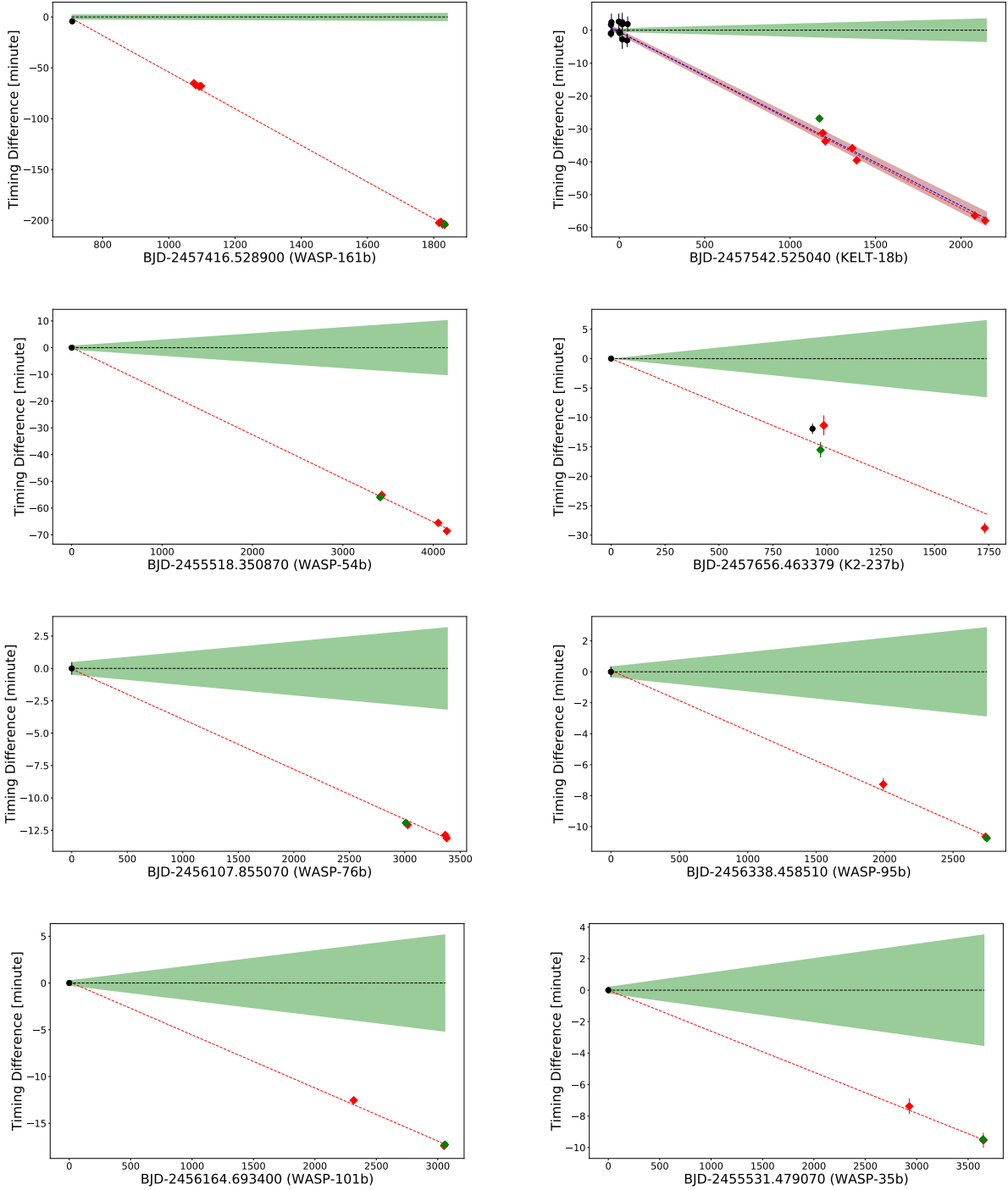

  \centering
  \caption{Timing differences of Type I targets of which the timings can be fitted by a linear function. The symbols are the same as Figure \ref{image:timing} and
  the legend inside the image is dismissed for clarity.}
   \includegraphics[width=3.3in]{appendix/WASP-161b.pdf}
    \includegraphics[width=3.3in]{appendix/KELT-18b.pdf}
    \includegraphics[width=3.3in]{appendix/WASP-54b.pdf} 
    \includegraphics[width=3.3in]{appendix/K2-237b.pdf}
    \includegraphics[width=3.3in]{appendix/WASP-76b.pdf}
    \includegraphics[width=3.3in]{appendix/WASP-95b.pdf}
    \includegraphics[width=3.3in]{appendix/WASP-101b.pdf}
    \includegraphics[width=3.3in]{appendix/WASP-35b.pdf}
   \label{image:timing appendix}
\end{figure*}

  \addtocounter{figure}{-1}
 \begin{figure*}[ht]
  \begin{tabular}{cc}
   \includegraphics[width=3.3in]{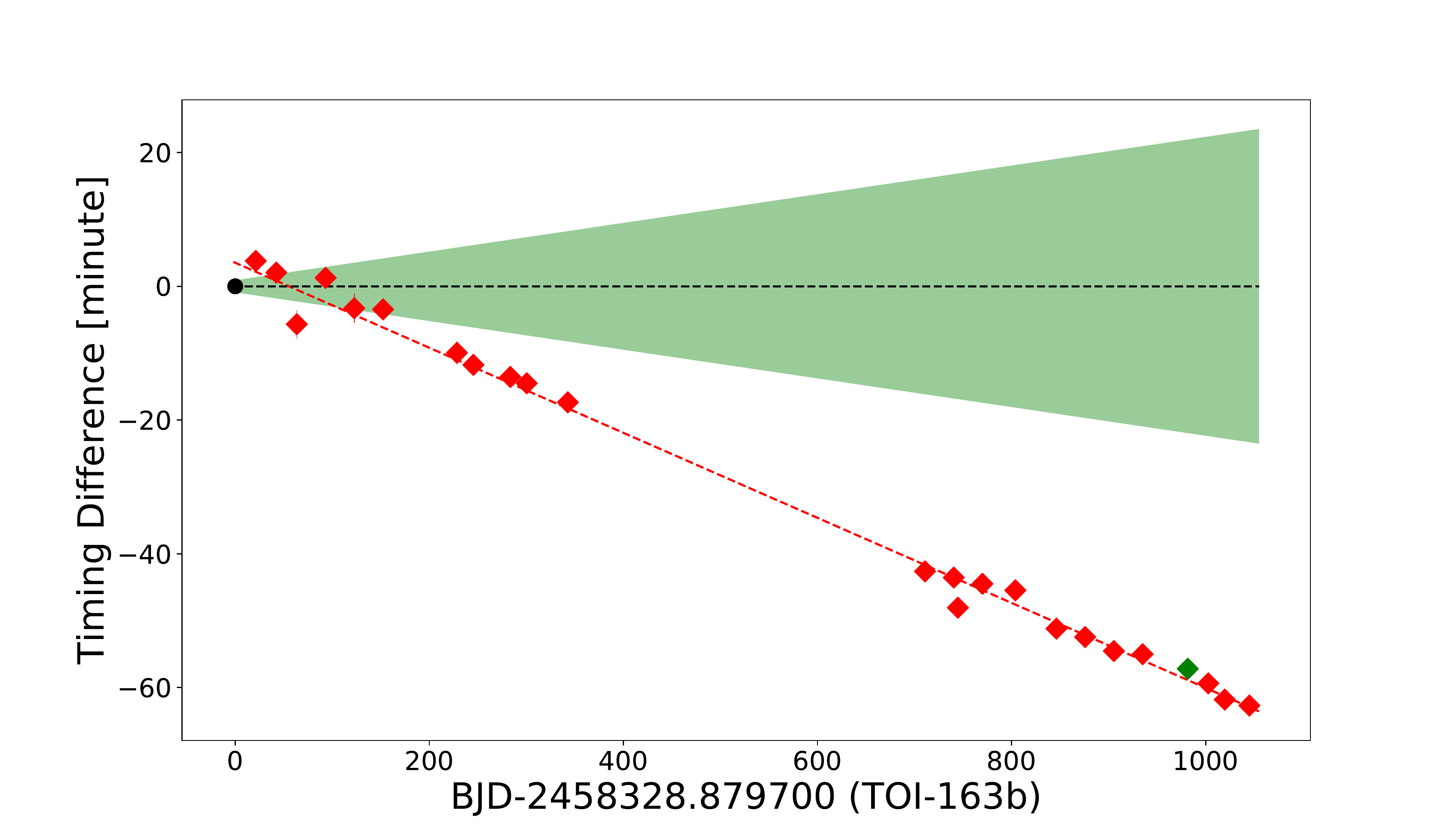}&
  \includegraphics[width=3.3in]{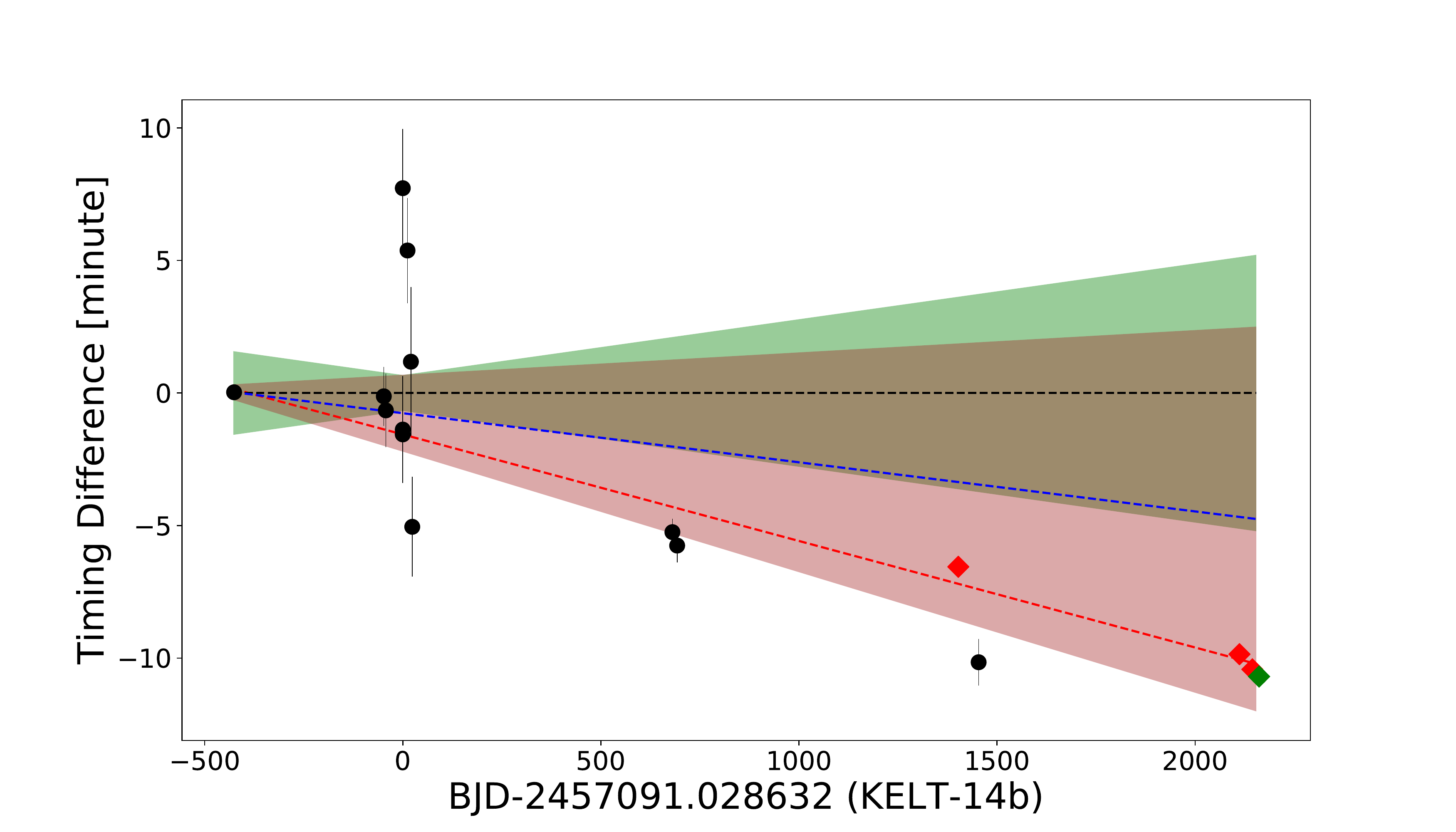}\\
 \includegraphics[width=3.3in]{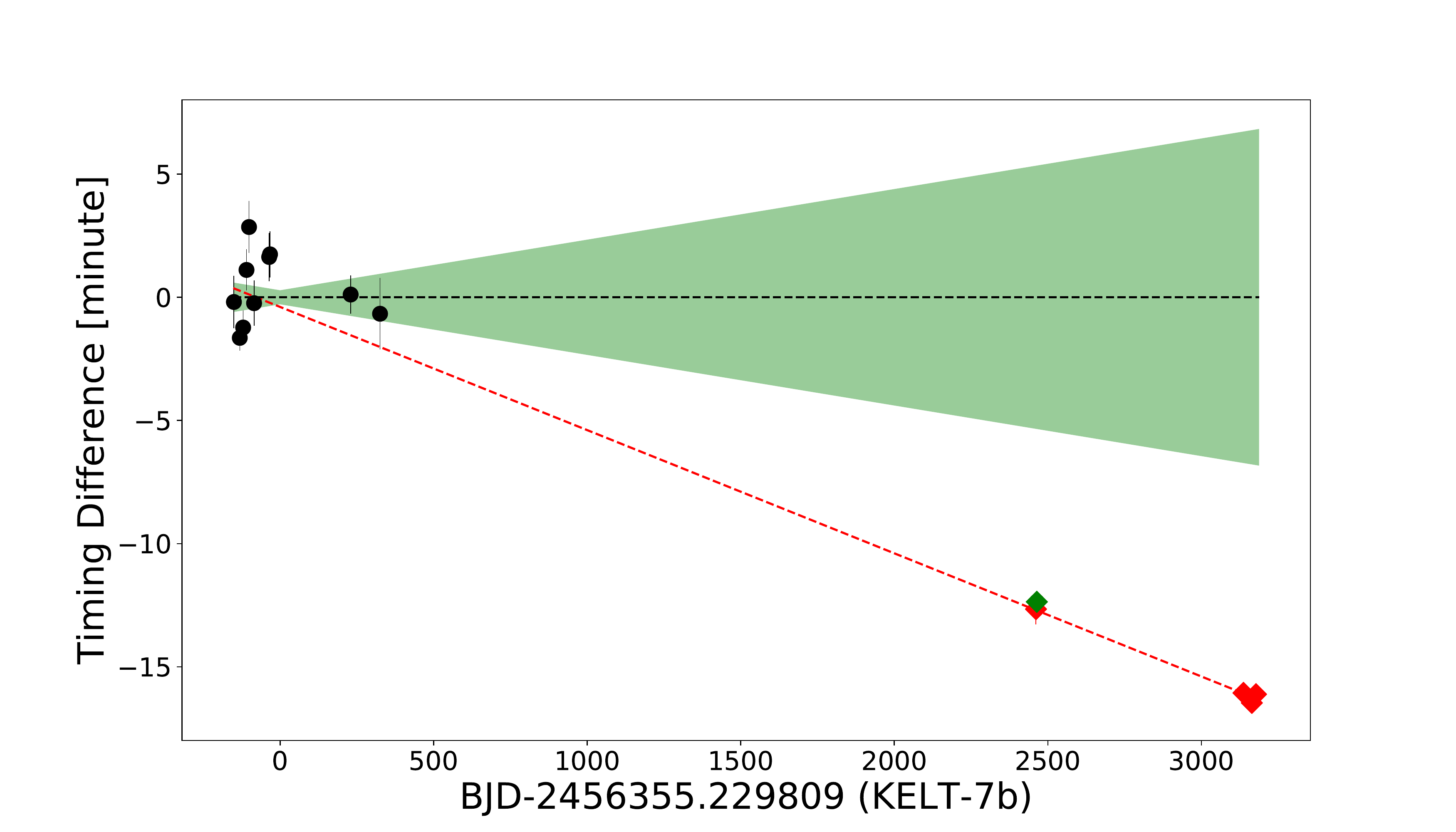}&
  \includegraphics[width=3.3in]{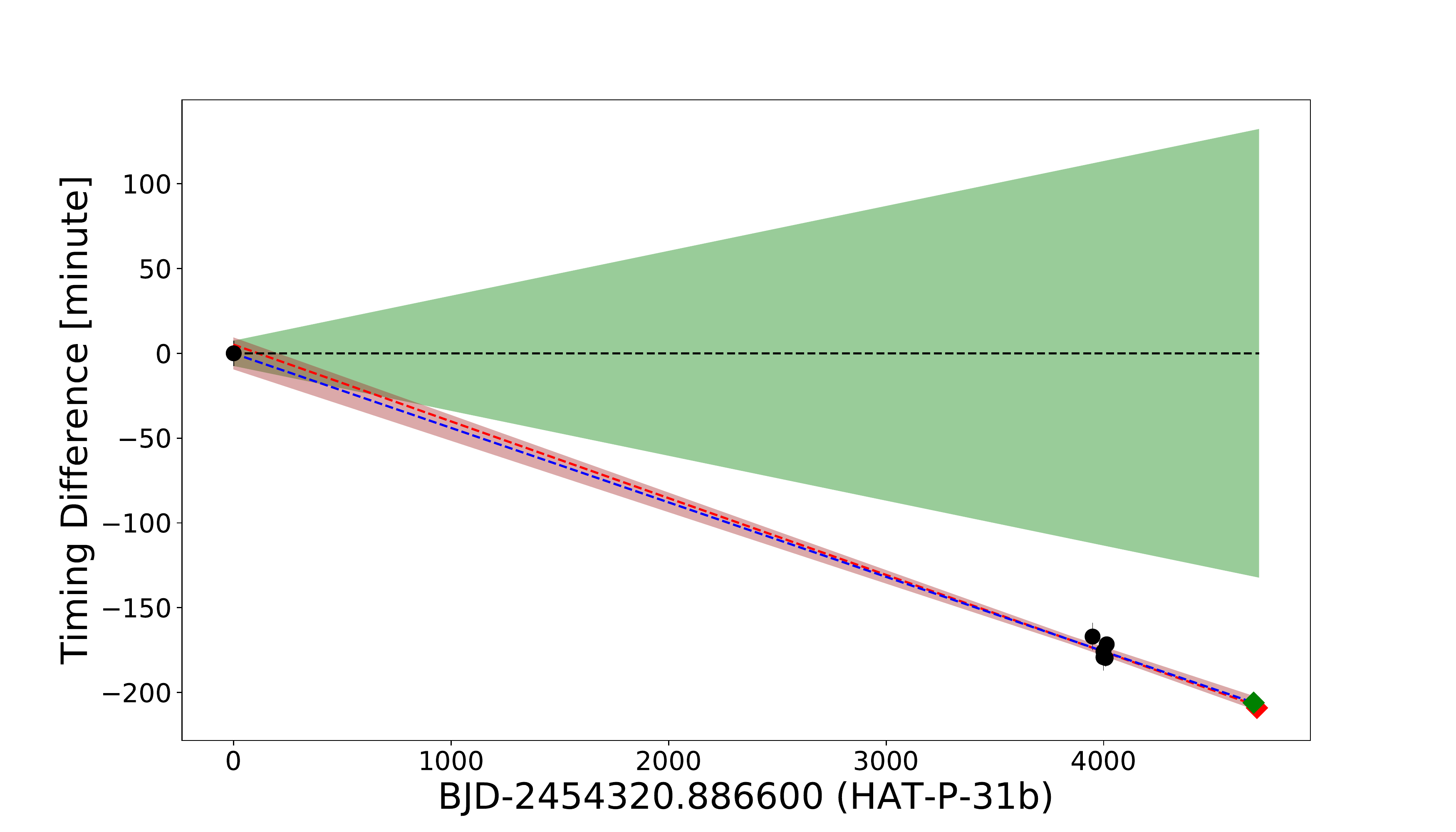}\\
  \includegraphics[width=3.3in]{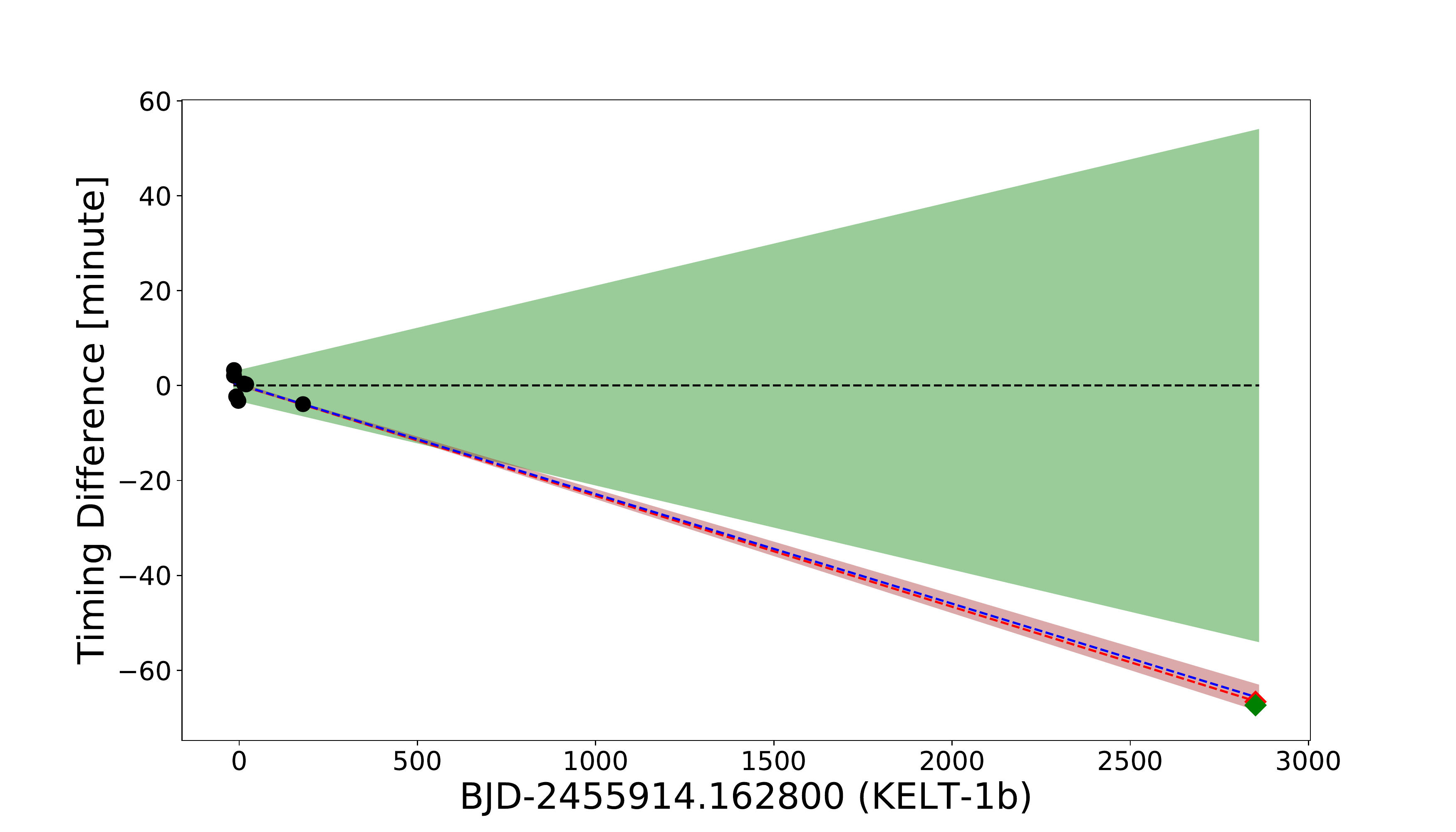}&
  \includegraphics[width=3.3in]{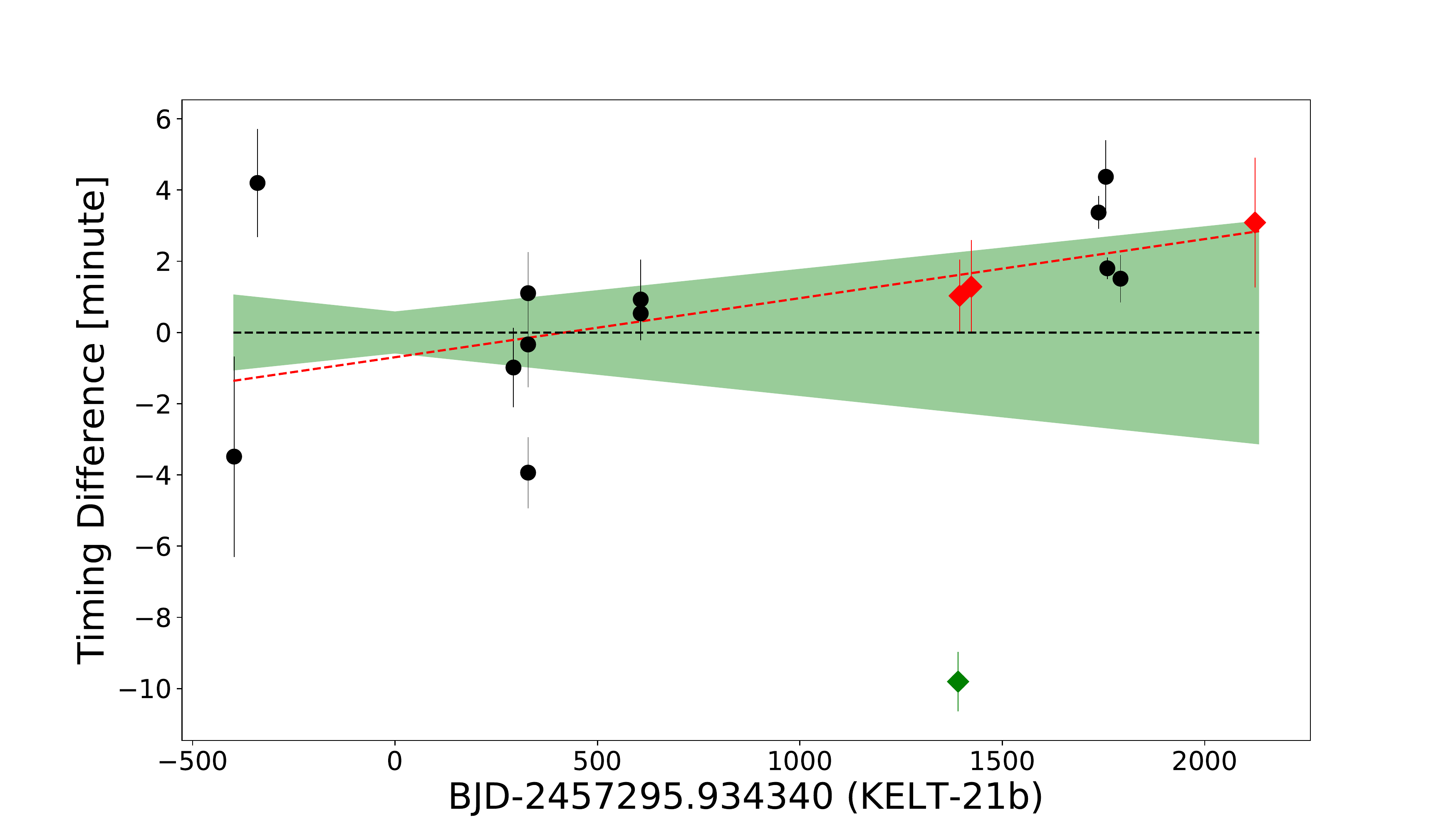}\\
   \includegraphics[width=3.3in]{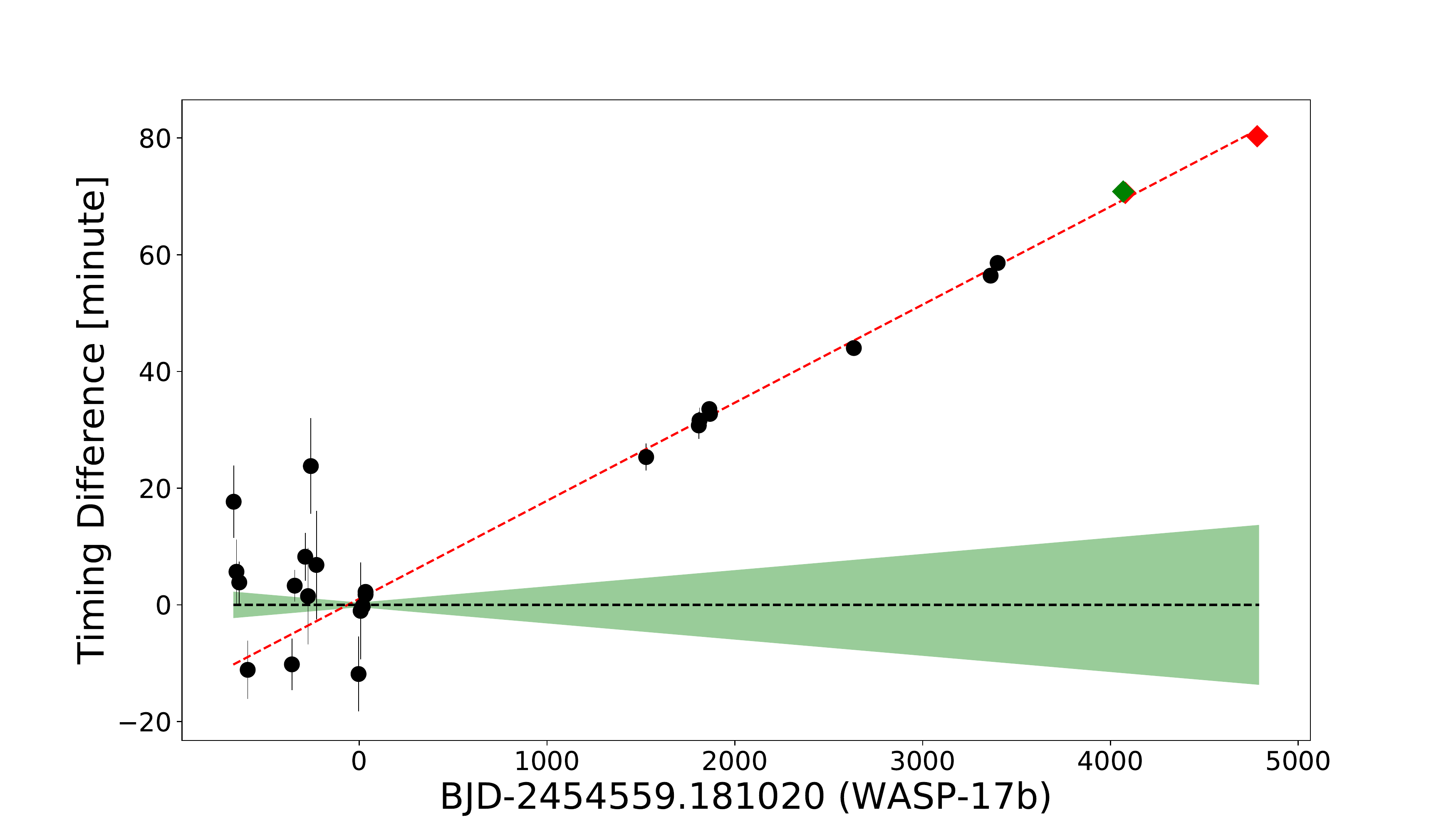}&
   \includegraphics[width=3.3in]{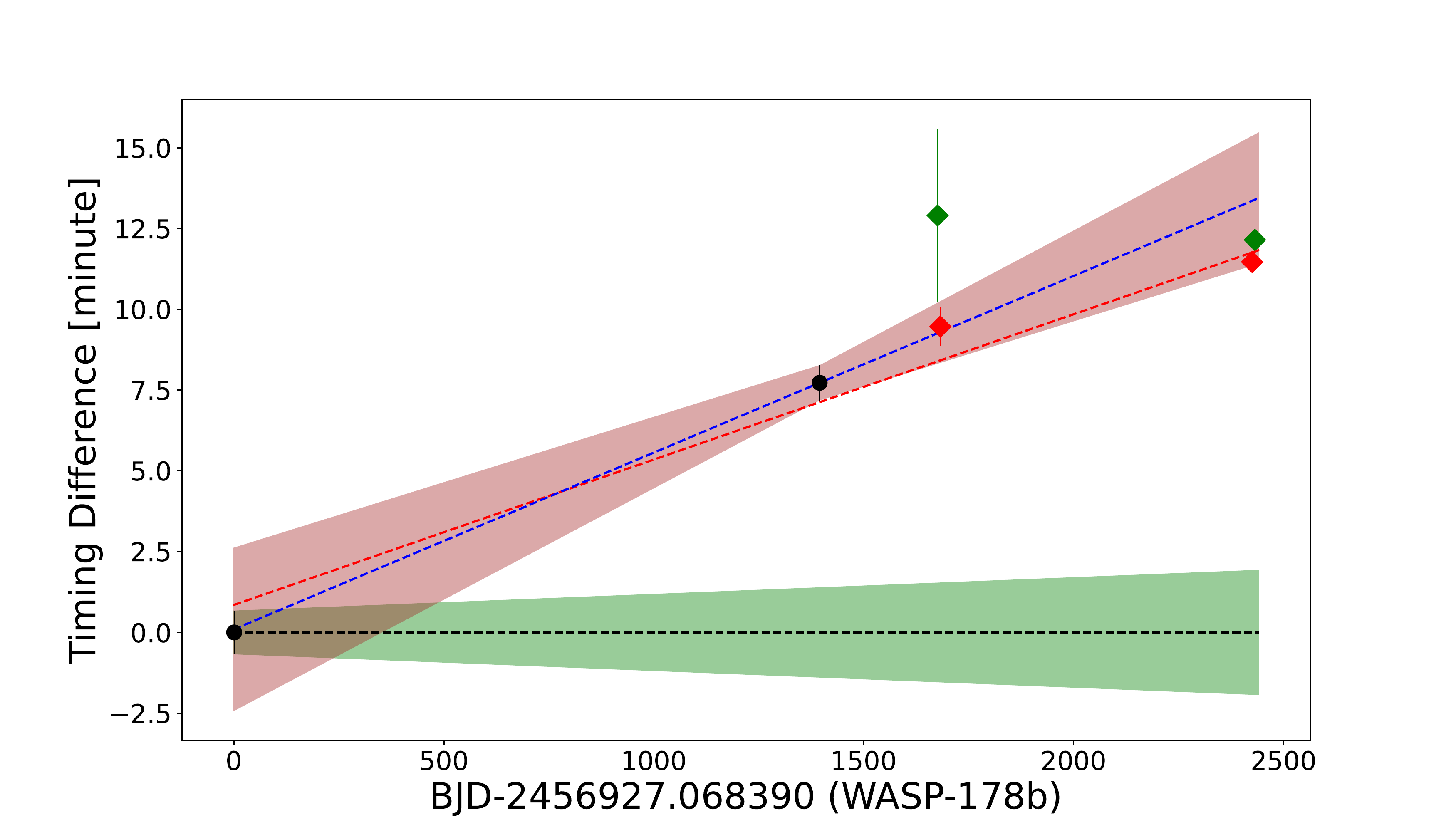}\\
   \end{tabular}
  \caption{(Continued) }
  \end{figure*} 

   \addtocounter{figure}{-1}
 \begin{figure*}[ht]
  \begin{tabular}{cc}
   
   \includegraphics[width=3.3in]{appendix/WASP-33b.pdf}&
   \includegraphics[width=3.3in]{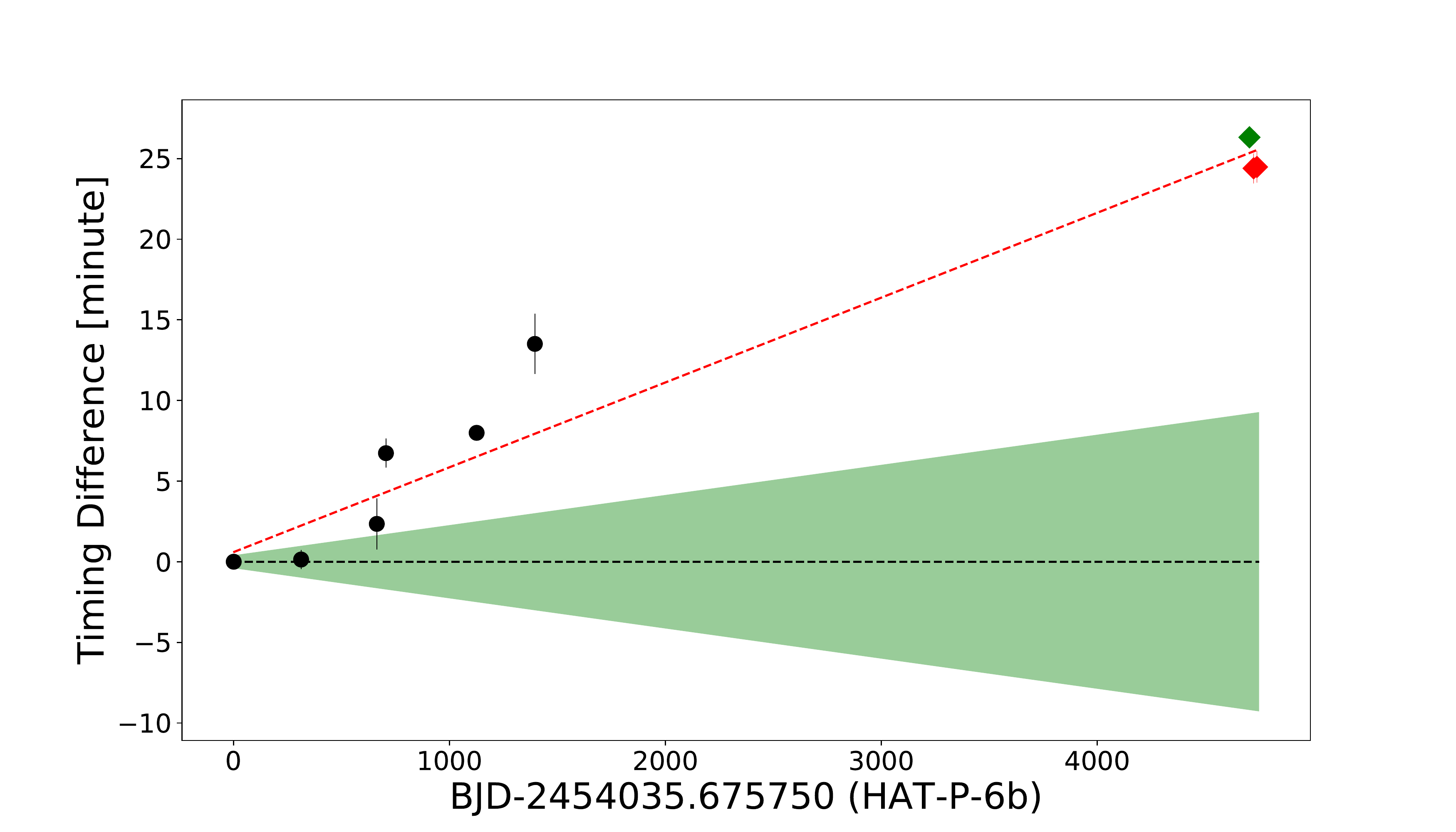}\\
   \includegraphics[width=3.3in]{appendix/KELT-19Ab.pdf}&
   \includegraphics[width=3.3in]{appendix/WASP-94Ab.pdf}\\
   \includegraphics[width=3.3in]{appendix/WASP-58b.pdf}&
   \includegraphics[width=3.3in]{appendix/WASP-99b.pdf}\\
   \includegraphics[width=3.3in]{appendix/TOI-1333b.pdf}&
   \includegraphics[width=3.3in]{appendix/WASP-78b.pdf}\\
   \end{tabular}
  \caption{(Continued) }
  \end{figure*} 
  
 \addtocounter{figure}{-1}
 \begin{figure*}[ht]
  \begin{tabular}{cc}
   \includegraphics[width=3.3in]{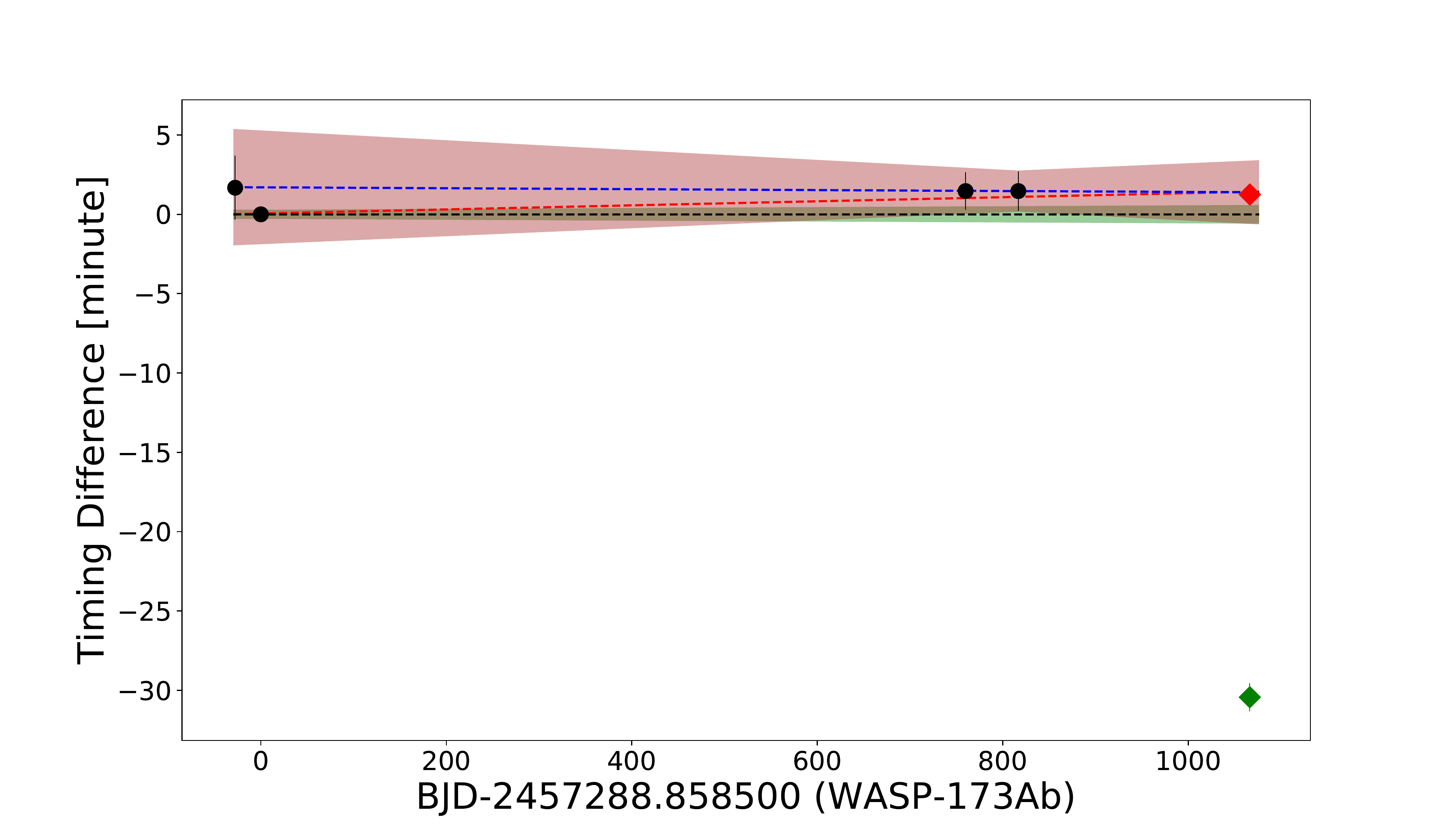}&
   \includegraphics[width=3.3in]{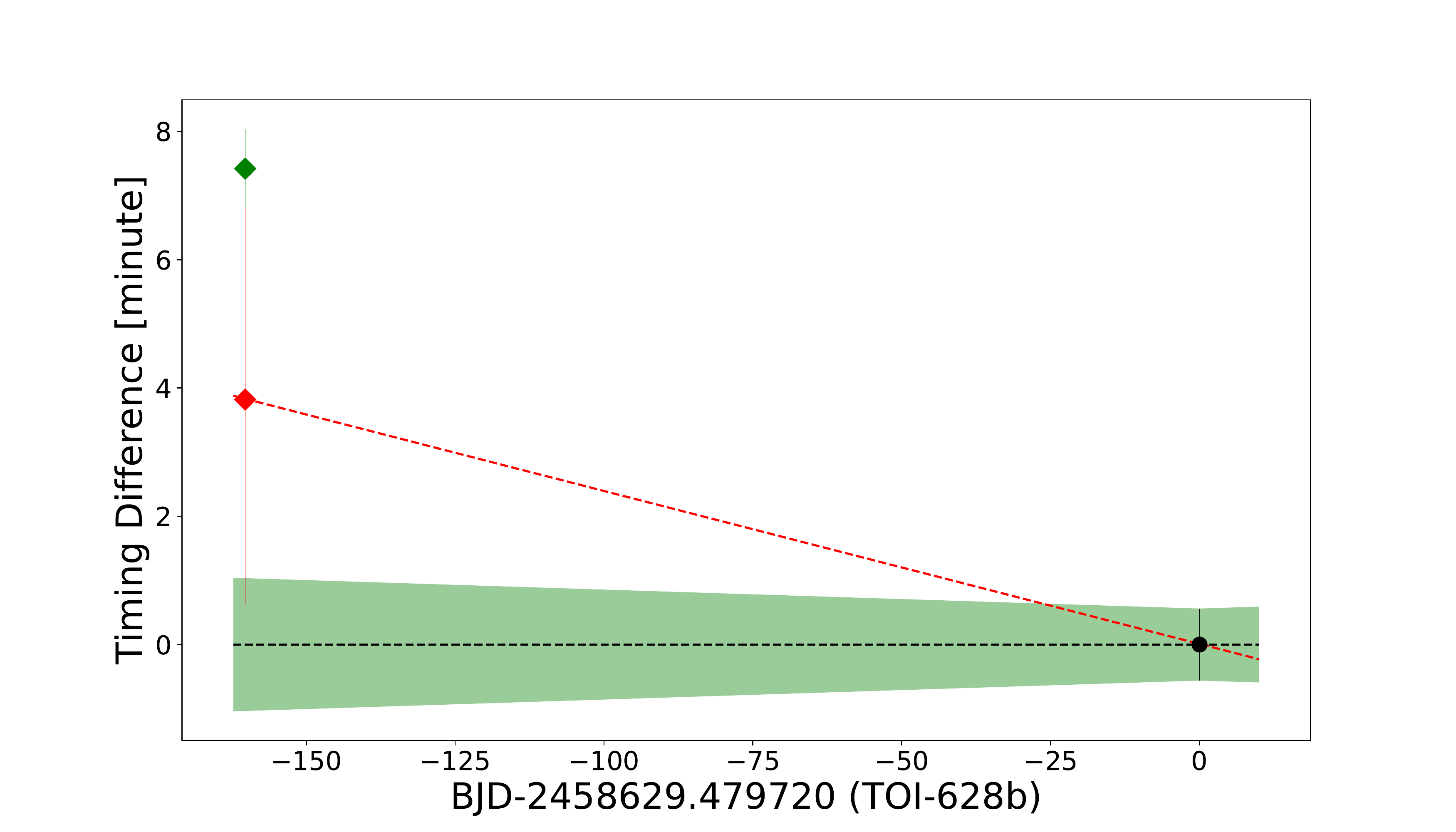}\\
   \includegraphics[width=3.3in]{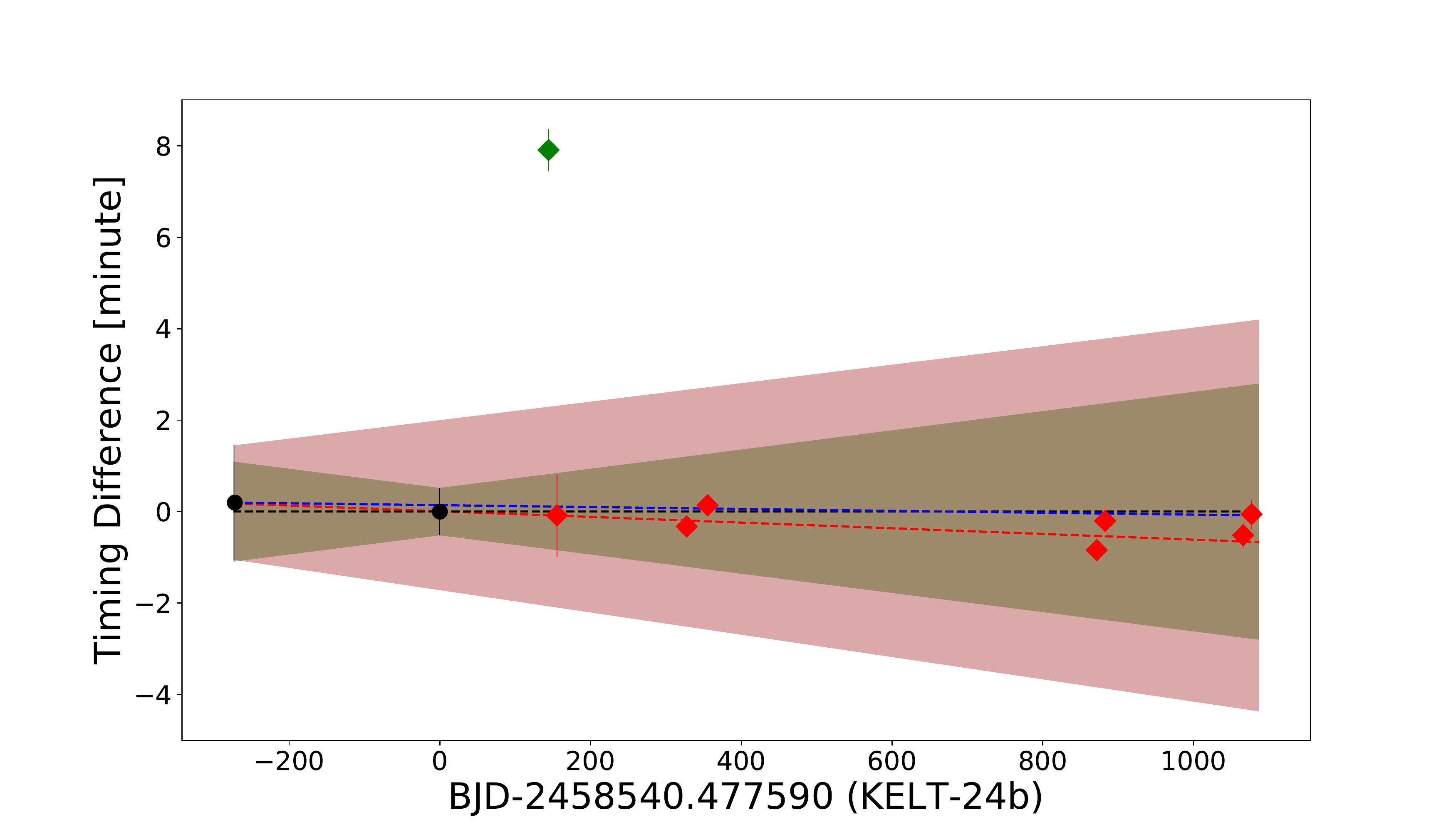}&
   \includegraphics[width=3.3in]{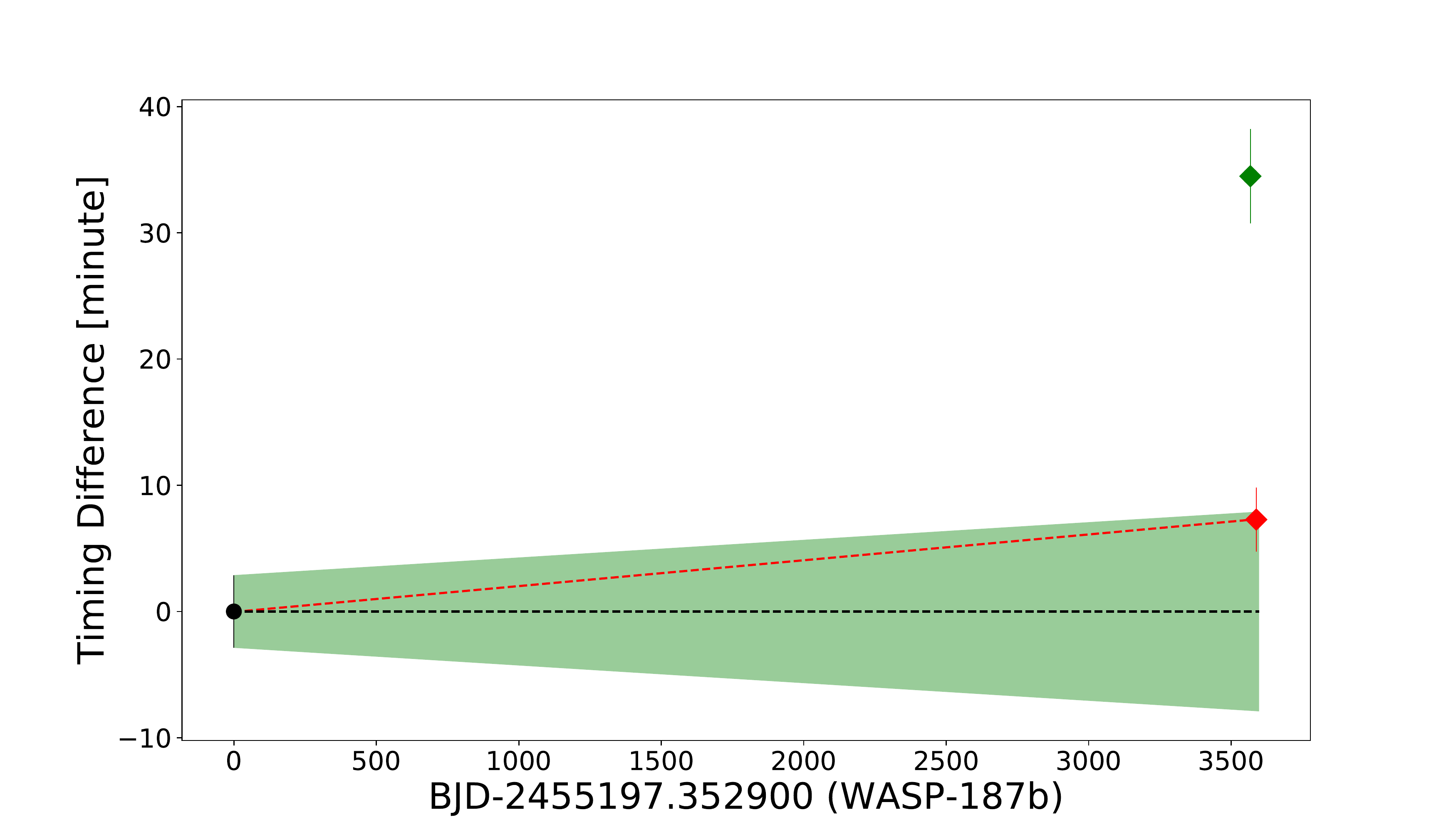}\\

   \end{tabular}
  \caption{(Continued) }
  \end{figure*}

\begin{figure*}[t]
  \centering 
   \caption{Timing differences of Type II targets of which timings would be modeled by a quadratic function.}
   \includegraphics[width=3.3in]{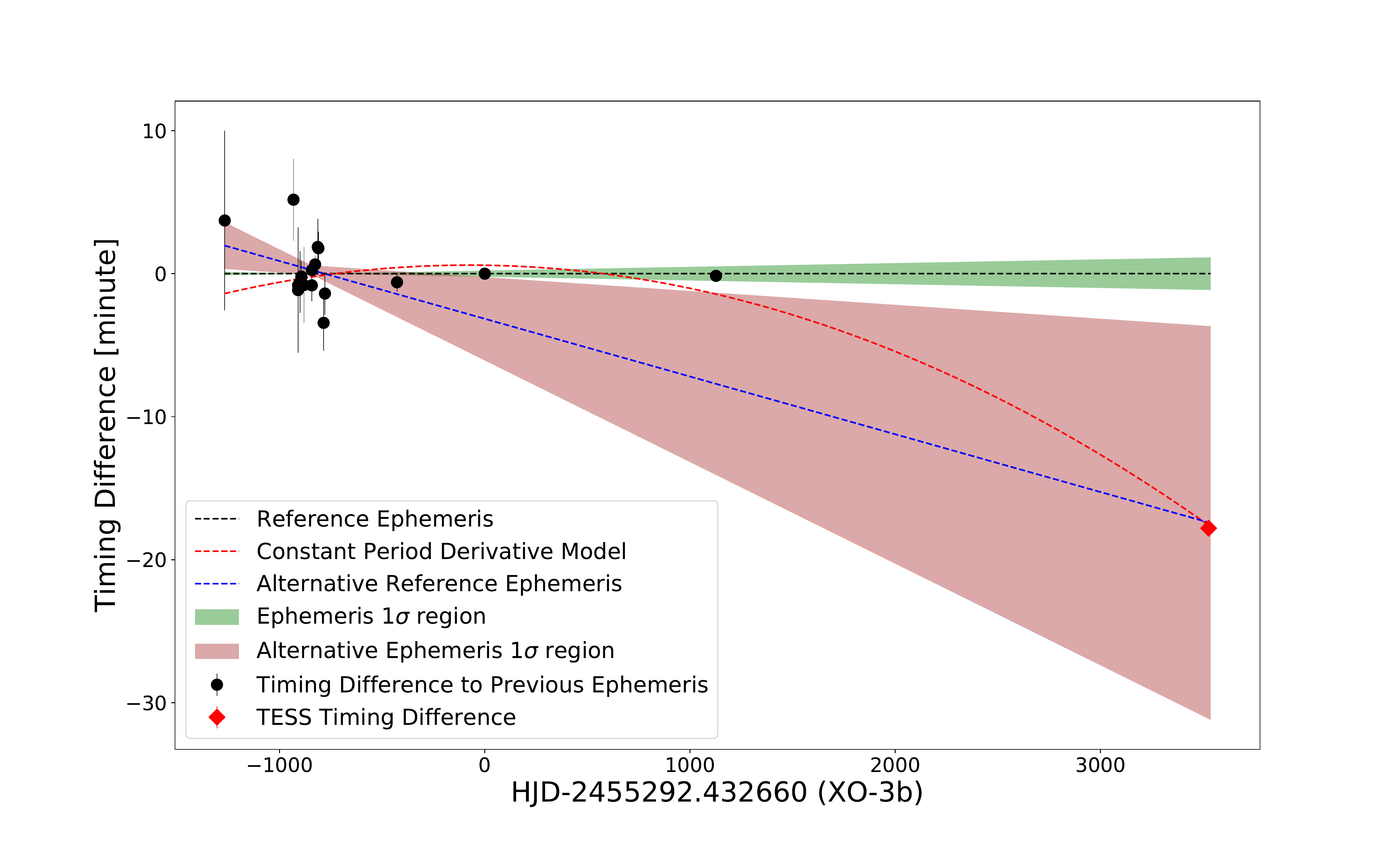}
   
\end{figure*}

\begin{figure*}[b]
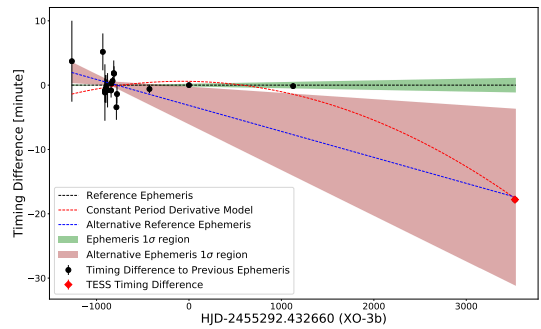

  \centering
  \caption{Timing differences of Type III targets of which the timings can not be fitted with any linear or quadratic functions. }
     \includegraphics[width=3.3in]{appendix/HAT-P-69b.pdf}
     \includegraphics[width=3.3in]{appendix/KELT-23Ab.pdf}
  \end{figure*}

\clearpage

\newpage
{
\setlength{\tabcolsep}{2pt}
\setlength\LTcapwidth{\textwidth} 
%\begin{landscape}
\begin{longtable*}{|l|l|l|}
\caption{The single mid-transit times of each target from the literature if available.} 
%Category flag \uppercase\expandafter{\romannumeral1} 
\label{table:timingsupplement}\\

\hline

Planet ID    &     Mid-Transit time $T$  &  Reference  \\
%\hline
              &     BJD$_{TDB} $              &               \\

\endfirsthead
\multicolumn{3}{c}%
{\tablename\ \thetable\ -- \textit{Continued from previous page}} \\
\hline
Planet ID   &  $T$    & Reference  \\
%\hline
              &             BJD$_{TDB} $              &  \\
\hline
\endhead
\hline \multicolumn{3}{c}{\textit{Continued on next page}} \\
\endfoot
\hline
\endlastfoot
\hline % inserts single horizontal line
WASP-161b &2458492.286046$\pm$0.00140  &  \cite{Yang1612022} \\
  &2458497.690811$\pm$0.00140  &  \cite{Yang1612022} \\
  &2458508.501901$\pm$0.00140  &  \cite{Yang1612022} \\
   &2458513.908266$\pm$0.00140  &  \cite{Yang1612022} \\
   &2459232.818367$\pm$0.00094  &  \cite{Yang1612022} \\
   &2459238.225141$\pm$0.00094  &  \cite{Yang1612022} \\
   &2459243.629420$\pm$0.00094  &  \cite{Yang1612022} \\
   &2459249.035140$\pm$0.00094  &  \cite{Yang1612022} \\
\hline
 XO-3b   &  2458819.06428$\pm$0.00035 &  \cite{yangxo3b}  \\ 
          &2458822.25556$\pm$0.00034 &  \cite{yangxo3b}  \\   
          &  2458825.44732$\pm$0.00037&  \cite{yangxo3b}  \\ 
          &2458831.83008$\pm$0.00035 &  \cite{yangxo3b}  \\ 
          &  2458835.02191$\pm$0.00034&  \cite{yangxo3b}  \\ 
          &2458838.21397$\pm$0.00042 &  \cite{yangxo3b}  \\
           &  2454864.76684$\pm$0.00040&  \cite{yangxo3b}  \\ 
          &2454025.3967$\pm$0.0038  &  \cite{yangxo3b}  \\    &2454360.50866$\pm$0.00173 &  \cite{Winn2008}\\ 
           &2454382.84500$\pm$0.00265&  \cite{Winn2008}\\
           &2454382.84523$\pm$0.00112  &  \cite{Winn2008}\\
           &2454392.41999$\pm$0.00130&  \cite{Winn2008}\\
           &2454395.61179$\pm$0.00167&  \cite{Winn2008}\\
          & 2454398.80332$\pm$0.00066 &  \cite{Winn2008}\\
           &2454411.56904$\pm$0.00161&  \cite{Winn2008}\\ 
           &2454449.86742$\pm$0.00067 &  \cite{Winn2008}\\
           &2454465.82610$\pm$0.00038
           &  \cite{Winn2008}\\
           &2454478.59308$\pm$0.00119&  \cite{Winn2008}\\
           &2454481.78455$\pm$0.00070 
           &  \cite{Winn2008}\\
           &2454507.31319$\pm$0.00118
           &  \cite{Winn2008}\\
           &2454513.69768$\pm$0.00090
           &  \cite{Winn2008}\\

     \hline
   KELT-18b      & 2457493.70451$_{-0.00084}^{+0.00082}$ &  \cite{McLeod2017}  \\
        & 2457493.7064$\pm$0.0011 &  \cite{McLeod2017}  \\        
      &  2457493.7046$_{-00.00087}^{+0.00086}$&  \cite{McLeod2017}  \\  
      & 2457496.5787$_{-0.0018}^{+0.0017}$ &  \cite{McLeod2017}  \\
        & 2457539.6551$\pm$0.0017 &  \cite{McLeod2017}  \\        
      & 2457545.3962$\pm$0.0011 &  \cite{McLeod2017}  \\
      & 2457559.7568$\pm$0.0011         &  \cite{McLeod2017}  \\        
      &  2457559.7572$\pm$0.0020         &  \cite{McLeod2017}  \\  
      &2457559.7536$_{-0.0020}^{+0.0019}$ &  \cite{McLeod2017}  \\
        & 2457588.4709$_{-0.0013}^{+0.0014}$ &  \cite{McLeod2017}  \\        
      & 2457591.3461$_{-0.0016}^{+0.0015}$ &  \cite{McLeod2017}  \\      
    \hline
    
   K2-237b   & 2458589.73380$\pm$0.00061 &  \cite{Edwards2021MNRAS.504.5671E} \\ 
   \hline
   
   KELT-14b   &   2457043.146899$\pm$0.000775 &  \cite{Rodriguez2016}\\
           &     2457048.276707$\pm$0.000961&  \cite{Rodriguez2016}\\
            &    2457091.027548$\pm$0.001076&  \cite{Rodriguez2016}\\
           &    2457091.033997$\pm$0.001551&  \cite{Rodriguez2016}\\
           &    2457091.027674$\pm$0.001400&  \cite{Rodriguez2016}\\
           &    2457103.002776$\pm$0.001377&  \cite{Rodriguez2016}\\
           &    2457111.550157$\pm$0.001956&  \cite{Rodriguez2016}\\
            &   2457114.965950$\pm$0.001308&  \cite{Rodriguez2016}\\
            &2457771.62839$\pm$0.00035 &\cite{Edwards2021MNRAS.504.5671E} \\
            &2457783.59845$\pm$0.00044 &\cite{Edwards2021MNRAS.504.5671E} \\
            &2458544.57156$\pm$0.00061 &\cite{Edwards2021MNRAS.504.5671E} \\
\hline

  KELT-7b   &2456204.817057$\pm$0.000741&  \cite{Bieryla2015}\\
      &2456223.959470$\pm$0.000358&  \cite{Bieryla2015}\\
      &2456234.898861$\pm$0.000486&  \cite{Bieryla2015}\\
      &2456245.839584$\pm$0.000579&  \cite{Bieryla2015}\\
      &2456254.045118$\pm$0.000730&  \cite{Bieryla2015}\\
      &2456270.451621$\pm$0.000637&  \cite{Bieryla2015}\\
      &2456319.678871$\pm$0.000683&  \cite{Bieryla2015}\\
      &2456322.413721$\pm$0.000648&  \cite{Bieryla2015}\\
      &2456584.950978$\pm$0.000544&  \cite{Bieryla2015}\\
      &2456680.667558$\pm$0.001007&  \cite{Bieryla2015}\\
      \hline
HAT-P-31b & 2458270.05094$\pm$0.00564 & \cite{Mallonn2019} \\                     & 2458320.09907$\pm$0.00131 & \cite{Mallonn2019} \\ 
          & 2458320.09673$\pm$0.00550 & \cite{Mallonn2019} \\  
          & 2458330.10726$\pm$0.00340 &  \cite{Mallonn2019} \\ 
          & 2458335.11829$\pm$0.00213 &  \cite{Mallonn2019} \\ 
\hline
 KELT-1b & 2455899.5549$\pm$0.0010 &   \cite{Siverd2012}    \\
         & 2455899.55408$\pm$0.00044 &   \cite{Siverd2012}    \\
         &2455905.63860$^{+0.00084}_{-0.00082}$ & \cite{Siverd2012}    \\
         & 2455911.72553$\pm$0.00045 &   \cite{Siverd2012}    \\
         & 2455927.55574$^{+0.00040}_{-0.00042}$ &\cite{Siverd2012}    \\
         & 2455933.64320$^{+0.00041}_{-0.0003}$ & \cite{Siverd2012}    \\
          \hline  
    KELT-21b   &  2456898.527802$\pm$0.001956 &  \cite{Johnson2018}  \\  &2456956.337374$\pm$0.001053&  \cite{Johnson2018}     \\
&2457588.567597$\pm$0.000775 &  \cite{Johnson2018}     \\
&2457624.696694$\pm$0.000799&  \cite{Johnson2018}     \\
&2457624.695694$\pm$0.000833 &  \cite{Johnson2018}     \\
&2457624.693194$\pm$0.000694 &  \cite{Johnson2018}     \\
&2457902.879452$\pm$0.000775 &  \cite{Johnson2018}     \\
&2457902.879181$\pm$0.000521 &  \cite{Johnson2018}     \\
&2459033.67650$\pm$0.00032& \cite{Garai2022MNRAS.tmp.1071G}\\
&2459051.74102$\pm$0.00071& \cite{Garai2022MNRAS.tmp.1071G}\\ 
&2459055.35200$\pm$0.00021& \cite{Garai2022MNRAS.tmp.1071G}\\ 
&2459087.86668$\pm$0.00046& \cite{Garai2022MNRAS.tmp.1071G}\\
\hline
WASP-17b   &  2453890.549230$\pm$0.004306  &  \cite{Anderson2010}      \\
&2453905.482660$\pm$0.003819  &  \cite{Anderson2010}\\
 &  2453920.423160$\pm$0.0025 &  \cite{Anderson2010}\\
 &  2453965.238059$\pm$0.003472&  \cite{Anderson2010}\\
 &  2454200.571537$\pm$0.003056&  \cite{Anderson2010}\\
  & 2454215.522667$\pm$0.001875&  \cite{Anderson2010}\\
  & 2454271.557737$\pm$0.002847&  \cite{Anderson2010}\\
 &  2454286.494817$\pm$0.005764&  \cite{Anderson2010}\\
  & 2454301.452058$\pm$0.005694&  \cite{Anderson2010}\\
  & 2454331.323827$\pm$0.006458&  \cite{Anderson2010}\\
  & 2454555.437350$\pm$0.004444&  \cite{Anderson2010}\\
 &  2454566.651190$\pm$0.005764&  \cite{Anderson2010}\\
 & 2454592.801221$\pm$0.000382&  \cite{Anderson2010}\\
&2456423.18973$\pm$0.00023&  \cite{Alderson2022MNRAS.512.4185A}    \\
&2456426.9246$\pm$0.0003&  \cite{Alderson2022MNRAS.512.4185A}    \\
&2457921.1177278$\pm$0.000775&  \cite{Alderson2022MNRAS.512.4185A}    \\
&2457958.473652$\pm$0.000775 &  \cite{Alderson2022MNRAS.512.4185A}    \\
&2456367.15615529$\pm$0.001615 &  \cite{Alderson2022MNRAS.512.4185A}    \\
&2456086.99426107$\pm$0.001615 &\cite{Alderson2022MNRAS.512.4185A}    \\
&2456370.8921914$\pm$0.001499 &  \cite{Alderson2022MNRAS.512.4185A}    \\
\hline

WASP-33b   &  2452984.82964$\pm$0.00030 &  \cite{Turner2016MNRAS.459..789T}      \\
 & 2456029.62604$\pm$0.0001624& \cite{Zhang2018AJ....155...83Z}     \\
&2456024.74659$\pm$0.00014 & \cite{Zhang2018AJ....155...83Z} \\
&2456878.65777$\pm$0.00033&\cite{Maciejewski2018AcA....68..371M}\\
&2456900.61530$\pm$0.00036 &\cite{Maciejewski2018AcA....68..371M}\\
&2457753.30433$\pm$0.00052 &\cite{Maciejewski2018AcA....68..371M}\\
&2457764.28369$\pm$0.00043 &\cite{Maciejewski2018AcA....68..371M}\\
&2458015.57583$\pm$0.00046 &\cite{Maciejewski2018AcA....68..371M}\\ 
&2458026.55466$\pm$0.00077&\cite{Maciejewski2018AcA....68..371M}\\ 
&2458075.35041$\pm$0.00037 &\cite{Maciejewski2018AcA....68..371M}\\
&2458381.53678$\pm$0.00055&\cite{Maciejewski2018AcA....68..371M}\\
&2458403.49659$\pm$0.00045 &\cite{Maciejewski2018AcA....68..371M}\\
&2458430.33394$\pm$0.00056&\cite{Maciejewski2018AcA....68..371M}\\ 
&2458436.43219$\pm$0.00034&\cite{Maciejewski2018AcA....68..371M}\\
\hline

KELT-23Ab &  2458144.898400$\pm$0.000463  &     \cite{Johns2019}\\
  &  2458144.897240$\pm$0.000440  &     \cite{Johns2019}\\
  &  2458153.917930$\pm$0.000590  &     \cite{Johns2019}\\
  &  2458153.916810$\pm$0.000949  &     \cite{Johns2019}\\
  &  2458167.448400$\pm$0.001100  &     \cite{Johns2019}\\
  &  2458187.745830$\pm$0.000637  &     \cite{Johns2019}\\
  &  2458196.770350$\pm$0.001100  &     \cite{Johns2019}\\
  &  2458196.771060$\pm$0.000625  &     \cite{Johns2019}\\
  &  2458196.773500$\pm$0.001192  &     \cite{Johns2019}\\
  &  2458273.452490$\pm$0.000984  &     \cite{Johns2019}\\
  &  2458302.769970$\pm$0.000810  &     \cite{Johns2019}\\

\hline

HAT-P-6b &  2454347.76763$\pm$0.00042  &    \cite{Szabo2010}     \\
 &    2454698.3908$\pm$0.0011    &  \cite{Szabo2010}     \\
 &2454740.77668$\pm$0.00063&\cite{Todorov2012ApJ...746..111T}\\
 &2455160.75292$\pm$0.00034&\cite{Todorov2012ApJ...746..111T}\\
 &2455430.4657$\pm$0.0013&\cite{Todorov2012ApJ...746..111T}\\
\hline

KELT-19Ab    &  2457073.723660$\pm$0.001042  &     \cite{Siverd2018}\\
  &  2457087.554255$\pm$0.001412  &     \cite{Siverd2018}\\
  &  2457101.393149$\pm$0.001887  &     \cite{Siverd2018}\\
  &  2457405.764653$\pm$0.000521  &     \cite{Siverd2018}\\
  &  2457405.766335$\pm$0.000683  &     \cite{Siverd2018}\\
  &  2457405.768490$\pm$0.000995  &     \cite{Siverd2018}\\
  &  2457405.766362$\pm$0.000822  &     \cite{Siverd2018}\\
  &  2457728.584553$\pm$0.001042  &     \cite{Siverd2018}\\
\hline

WASP-58b   &  2455183.9342$\pm$0.0010&    \cite{Mallonn2019}   \\
&2456488.40790$\pm$.00264 &    \cite{Mallonn2019}   \\
&2456498.44187$\pm$0.00121 &    \cite{Mallonn2019}   \\
&2456523.52545$\pm$0.00316&    \cite{Mallonn2019}   \\
&2456528.54704$\pm$0.00134 &    \cite{Mallonn2019}   \\
&2457120.57537$\pm$0.00297&    \cite{Mallonn2019}   \\ 
&2457637.35161$\pm$0.0008975&    \cite{Mallonn2019}   \\
&2457968.48759$\pm$0.00068141&    \cite{Mallonn2019}   \\
&2457968.48541$\pm$0.00082141&    \cite{Mallonn2019}   \\
&2458259.48221$\pm$0.00249199&    \cite{Mallonn2019}   \\
\hline

WASP-173Ab & 2457261.1266$^{+0.0013}_{-0.0014}$  & \cite{Labadie2019}  \\
       & 2458048.74546$^{+0.00084}_{-0.00078}$  &  \cite{Labadie2019}  \\
       & 2458105.59824$^{+0.00090}_{-0.00084}$  &  \cite{Labadie2019}  \\
 \hline

%\multicolumn{this is a try of the version}
\end{longtable*}
%\end{landscape*}
}

\clearpage
\newpage
\bibliographystyle{aasjournal}
\bibliography{ref}
\end{document}